\documentclass[11pt]{article}

\usepackage[utf8]{inputenc}
\usepackage{amssymb}
\usepackage{tabularx}
\usepackage{adjustbox}
\usepackage[table]{xcolor}
\usepackage{multirow}
\usepackage{bbold}
\usepackage{amsmath}
\usepackage{amsthm}  
\usepackage[left=3cm,right=3cm,top=2.5cm,bottom=2.5cm]{geometry}
\usepackage{mathrsfs}
\usepackage[framemethod=tikz]{mdframed} 
\usepackage{comment}
\usepackage{orcidlink} 

\usepackage{authblk} 
\usepackage[style=numeric-comp,sorting=none]{biblatex} 
\usepackage{url}
\usepackage{hyperref}
\usepackage[hyphenbreaks]{breakurl}
\mdfdefinestyle{defaultstyle}{%
  linecolor=black,
  middlelinewidth =0.5pt,
  innertopmargin=0pt,
  innerbottommargin=8pt,
}

\mdfdefinestyle{definitionstyle}{
    backgroundcolor=black!2.5,
        middlelinewidth     =2pt,
        innertopmargin      =0pt,
        innerbottommargin   =8pt,
        topline=false,
        bottomline=false,
            }

\newmdtheoremenv[style=defaultstyle]{assumption}{Assumption} 
\newmdtheoremenv[style=definitionstyle]{definition}{Definition}
\newmdtheoremenv[style=defaultstyle]{summary}{Summary}

\newcommand{\tens}[1]{%
  \mathbin{\mathop{\otimes}\limits_{#1}}%
}

\theoremstyle{definition}

\newtheorem{example} {Example}

\DeclareUnicodeCharacter{0301}{} 

\begin{document}

\title{A Theoretical Framework for Self-Gravitating \\ k-Form Boson Stars with Internal Symmetries}

\author[1,a]{Jakob Hoffmann \thanks{\href{mailto:jhoffmann@itp.uni-frankfurt.de}{jhoffmann@itp.uni-frankfurt.de}}}
\author[1,2,a]{Cédric Jockel\,\orcidlink{0009-0007-7617-7178} \thanks{\href{mailto:cedric.jockel@aei.mpg.de}{cedric.jockel@aei.mpg.de}}}
\affil[1]{Institute for Theoretical Physics, Goethe University, Max von Laue Straße 1, 60438 Frankfurt am Main, Germany, European Union}
\affil[2]{Max Planck Institute for Gravitational Physics (Albert Einstein Institute), Am Mühlenberg 1, 14476 Potsdam, Germany, European Union}
\affil[a]{Both authors are first authors}

\date{\today}

\maketitle

\begin{abstract}
\noindent 
Current boson star models are largely restricted to global symmetries and lower spin fields. In this work, we generalize these systems of self-gravitating bosonic fields to allow for arbitrary totally antisymmetric tensor fields and arbitrary internal gauge symmetries. We construct a generalized formalism for Yang-Mills-like theories, which allows for arbitrary k-form fields, instead of just vector fields. The k-form fields have gauge symmetries described by semisimple, compact Lie groups. We further derive equations of motion for the k-form fields and connection coefficients of the Lie group. Extensions and applications are also discussed. We present a novel way to fix the group connection using a spacetime connection. As an example, we derive explicitly the connection coefficients for SU(2) in a spherically symmetric spacetime using rectangular vielbeins. The combination of methods presented leads to a powerful, adaptable and practical framework. As a proof of concept, we derive ordinary differential equations for a 0-form field with a SU(2) symmetry. Our framework can be used to model self-gravitating (multi) particle states with internal symmetries, such as pion condensates or dark matter. It is also suited as a tool to approach open problems in modified gravity and string theory.
\end{abstract}

\tableofcontents

\section{Introduction} \label{sec:introduction}
Gauge theories are the central paradigm of all fundamental physical theories. All fundamental interactions of elementary particle physics mediated through bosons can be consistently described by gauge fields. This includes gravity which is described by general relativity. It is built upon a gauge symmetry which reflects invariance under coordinate transformations (see e.g. \cite{Tecchiolli:2019hfe,Gronwald:1995em,HehlReview76}). This however seems less apparent, judging purely from its common tensorial description. \\

\noindent
The descriptions of the gauge theories used to describe the standard model of particle physics and general relativity appear to be very different. Works in particle physics are mostly based on using gauge fields as a tool for computing different observables \cite{DeGrand:2003xu,Briceno:2014tqa}. The mathematical structure of general relativity on the other hand emphasizes the use of concepts from differential geometry, such as parallel transport, connection coefficients and covariant derivatives \cite{Misner:1973prb}. Even though both approaches may seem very different, their inherent mathematical structure is very similar. Both, particle physics and general relativity can be formulated on a classical level using an action that is invariant under certain (gauge) symmetries. From a mathematical perspective, gauge theories in particle physics, as well as general relativity, can be understood in the context of differential forms with internal Lie group symmetries (for the case of gravity, see e.g. \cite{Tecchiolli:2019hfe,Cattaneo:2023cnt}). Both theoretical descriptions then contain covariant derivatives which are constructed via a connection. The distinction between the gauge sectors in the standard model arises through the choice of the internal symmetry group and the connection. The strong nuclear force for example is described by a local $SU(3)$ gauge theory. There the connection is chosen such that the connection coefficients correspond to the gauge field components. Quantum electrodynamics in turn features a local $U(1)$ symmetry. General relativity can be understood as a gauge theory with a local Lorentz symmetry that is also invariant under diffeomorphisms (coordinate transformations) and uses the Levi-Civita connection (i.e. the unique metric compatible connection with zero torsion). \\

\noindent
Making use of the different descriptions of gauge theories is important because they reveal different aspects of the theories. In order to benefit from the different approaches, it is necessary to bridge the gap between physically motivated and mathematical formulations. The physics approach makes heavy use of index notation. This makes it easy to do concrete calculations and to obtain physical intuition. However, this can also obscure the fundamental structure of the theory. The mathematical approach uses differential forms and is more abstract. This makes it easy to see the bigger picture and allows to explore a broader variety of theories. But at the same time, it is more difficult to perform useful explicit computations or to provide concrete solutions. Ideally, a joint mathematical and physical treatment should allow us to learn more about which mathematical structures are useful to describe nature, and at the same time to gain a more holistic understanding of physical theories. \\

\noindent
All the gauge fields in the standard model and general relativity are described by vector fields, i.e. $1$-form fields. But there are also a number of applications for gauge fields of higher differential-form-degree $k$. One motivation for our work comes from generalizing the concept of boson stars and from theories of modified gravity. Boson stars are self-gravitating objects that were first proposed by \cite{Kaup:1968zz}. They are modeled using classical scalar fields \cite{Kaup:1968zz,Liebling:2012fv} or vector fields \cite{Brito:2015pxa,Cardoso:2021ehg,Aoki:2022woy,Brihaye:2017inn,Lazarte:2024jyr}. The fields are usually taken to be complex fields with a global $U(1)$ symmetry that allows for a definition of a conserved Noether current. They can be used to describe e.g. accumulations of dark matter at galactic scales \cite{Lee:2008ab,Sharma:2008sc,Urena-Lopez:2010zva,Rindler-Daller:2011afd,Marsh:2015wka,DellaMonica:2022kow} or as black hole mimickers \cite{Guzman:2009zz,Vincent:2015xta,Rosa:2022tfv,Herdeiro:2021lwl,Rosa:2022toh}. Boson stars constructed from fields with internal symmetries have also been studied \cite{daRocha:2020jdj,Bartnik:1988am,Schunck:2003kk,Brihaye:2004nd,Dzhunushaliev:2006sy,Brihaye:2009hf,Soni:2016gzf,Martinez:2022wsy,Jain:2022kwq,Soni:2016yes}. In the context of dark matter, these fields can for example describe a fundamental particle. Thus, mixing various types of boson stars with ordinary relativistic matter has been studied intensively \cite{Henriques:1989ez,Kouvaris:2010vv,Liebling:2012fv,Bertone:2007ae,Zurek:2013wia,Kouvaris:2013awa,Diedrichs:2023trk,DiGiovanni:2021ejn,Jockel:2023rrm}. As for modified gravity, modifications using different gauge symmetries have also been studied e.g. in \cite{Partanen:2023dkt}. For these cases, $k$-form fields with internal symmetries can serve as a useful tool to generalize models and probe more physical scenarios. \\
Another motivation comes from string theory. In 10-dimensional superstring theory, particularly type IIB superstring theory \cite{Sen_2016}, massless string excitations on configurations of $D$-branes can be described by differential $k$-form fields with internal symmetries. In Type IIB superstring theory, which contains only closed strings, these include a $2$-form field which is often referred to as the Kalb-Ramond field. Other $k$-form fields called Ramond-Ramond fields are also included. They are described using $0$-form, $2$-form and $4$-form fields (for more details we refer to \cite{Maharana_1997,Polchinski_1995,Polchinski_1996,Witten_1995,Witten_1996,Witten_1997,Polyakov_1996,maldacena1996black,Brodie_1997,Hohm_2011,Schwarz_2013,hamilton2016field,Sen_2016}). All these fields can be grouped into representations of $SO(8) \times SO(8)$. To understand possible physical implications of these fields it is crucial to first study simpler setups of differential $k$-form fields with internal symmetries. This work may contribute to a better understanding of the nature of such objects. \\

\noindent
In this work, we aim to combine concepts from differential geometry and group theory to derive an algorithmic framework with which one can study a wide range of physically relevant applications. We use differential forms to derive a classical action that can describe the systems introduced in the previous paragraphs. We start with a mathematical model which is motivated by Yang-Mills theory. We then proceed to generalize Yang-Mills theory using Lie algebra valued differential $k$-forms. This generalization is then coupled to gravity via a classical Palatini-Cartan term. We then derive equations of motion for this framework and apply them to concrete examples. In this way, we use methods from mathematics and physics in tandem. \\

\noindent 
This work is structured as follows. In section \ref{sec:mathematical-background}, we provide a pedagogical introduction to the mathematical background needed to understand this work. It is written in a way to make it easy to understand for a broad audience of mathematicians and physicists. We show how certain mathematical structures are used in physics and make clear how physical principles are described using mathematical structures. We also provide interesting notes aiming to build both mathematical and physical intuition. In section \ref{sec:formulation-and-motivation-of-action}, we introduce our main model mathematically. We use the Yang-Mills action as a motivational starting point to construct a generalized action that works for arbitrary $k$-form fields instead of exclusively vector ($1$-form) fields. This action is subsequently coupled to gravity using a Palatini-Cartan term. We then discuss possible extensions and limitations of our model. We proceed by deriving the associated equations of motion in section \ref{sec:derivation-of-EOM}. In section \ref{sec:connection-for-Lie-groups}, we discuss the choice of the group connection, which is a central building block of our theory. For this, we use non-square vielbeins to compute coefficients of the connection of a Lie group. The construction of those vielbeins and this connection is described in detail. This method can be useful to those working with gauge theories in curved spacetimes and any theoretical constructions that involve spin-connections. In section \ref{sec:solving-the-equations-of-motion}, we solve the equations of motion explicitly using these methods. Finally in section \ref{sec:conclusions-and-outlook}, we summarize our findings and discuss future research prospects. We also provide additional information and detailed calculations in the appendices \ref{sec:appendix a} to \ref{sec:appendix D}.

\newpage
\section{Mathematical Background} \label{sec:mathematical-background}
In this chapter, we introduce the necessary mathematical building blocks to understand the remaining sections of this work. This section may be skipped if the reader already has a firm mathematical understanding of differential forms, Lie groups, (local) gauge symmetries, connections and covariant derivatives, as well as the tetrad formalism. Otherwise, this section is meant as a light-weight and pedagogic mathematical introduction (without proofs) of the mathematical concepts this work is build upon. By providing this section, we aim to make our work as self-contained as possible. Concrete examples are also provided and are meant to build a bridge between pure mathematics and physics. The goal is to build both mathematical and physical intuition. We do this by connecting abstract mathematical concepts to concepts known to most physicists. We also highlight the physical use of certain mathematical structures and offer additional perspectives. \\
Most concepts and definitions presented in this section are inspired by or taken from at least one of the following works \cite{Tecchiolli:2019hfe,Berning_skript:2020,lee2009manifolds,tu2008introduction,jost2017riemannian,MITscript}.

\subsection*{Tensors}

We start the section by defining the basic building block of any physical theory: tensors. Tensors are mathematical objects which have useful transformation properties. They are used to describe various physical quantities.

\begin{definition}[Vector Spaces and Dual Spaces] \label{def:dual-space} $\:$\\
A vector space $V$ is a space over a field $\mathbb{F}$, with a basis $\left\{ e_1, ..., e_n\right\}$. The linear mappings $\lambda^i$ with the property $\lambda^i (e_j) = \delta^i_{\: j}$ then define a basis $\left\{ \lambda^1, ..., \lambda^n \right\}$ of the vector space $V^\star$, which we call the dual space of $V$. That means that all vectors in the dual space $V^\star$ can map a vector in the vector space $V$ to a scalar, and vice-versa.
\end{definition}

\noindent
Vector spaces can also be used to build more complicated mathematical structures such as tensors.

\begin{definition}[Tensors] \label{def:tensor} $\:$\\
Let $V$ be an $n$-dimensional vector space over a field $\mathbb{F}$ and let $V^*$ be its dual vector space. Then we can define an $(l,k)$-tensor as a multilinear mapping $f \colon V^k \times (V^*)^l \rightarrow \mathbb{F}$ with $V^k := \underbrace{V \times V \times ... \times V}_{k-times}$ and $(V^*)^l  := \underbrace{V^\star \times V^\star \times ... \times V^\star}_{l-times}$. We define $T^{(l,k)}$ as the space of all \textit{(l,k)}-tensors. $(0,k)-tensors$ and $(l,0)$-tensors are commonly called covariant- and contravariant tensors respectively.
\end{definition}

\noindent
Tensors can thus be understood as objects whose basis is constructed out of different elements in a vector space and the dual space. In index-notation, the components of a (1,0)-tensor (or vector) are denoted by $a^\mu$, and the components of a (0,1)-tensor (or covector) are denoted by $a_\mu$. Higher-order tensors just gain additional (lower and/or upper) indices. A large part of modern physics is build upon tensors. A special kind of tensors are alternating linear forms.

\begin{definition}[Alternating Linear Forms] \label{def:alternating-linear-forms} $\:$\\
Let $V$ be an $n$-dimensional vector space. Let further $S_k$ denote the set of all permutations of $\{1, ..., k \}$ and \\ $\sigma \in S_k$ be one specific permutation with $sgn(\sigma)$ being the sign of $\sigma$. We define the set of antisymmetric tensors as $A_k(V)=\{\: f \in T^{(0,k)}\: \: | \: \sigma \circ f=sgn(\sigma) f \: \: \forall \: \sigma \in S_k \}$. $A_k(V)$ is also called the set of all alternating linear forms of degree $k$ on $V$.
\end{definition}
\noindent
We define an anti-symmetrization operator $A(T)$ which converts any tensor $T \in T^{(0,k)}$ into an antisymmetric tensor. In index-notation, it can be written as follows:
\begin{align}
    T_{[\mu_1 ... \mu_k]} := \frac{1}{k!} \sum_{\sigma \in S_k} sgn(\sigma) \: T_{\mu_{\sigma(1)} ...\mu_{\sigma(k)}} \: . \label{eq:anti-symmetrization-operator-definition}
\end{align}
Explicitly for a $(0,2)$- and $(0,3)$-tensor, the operation reads:
\begin{align}
    T_{[\mu\nu ]} = \frac{1}{2} \left( T_{\mu\nu} - T_{\nu\mu}\right) \:\: , \:\: T_{[\mu\nu\rho]} = \frac{1}{3!} \left( T_{\mu\nu\rho} - T_{\mu\rho\nu} - T_{\nu\mu\rho} + T_{\nu\rho\mu} + T_{\rho\mu\nu} - T_{\rho\nu\mu} \right) \: .
\end{align}

\noindent
In physics, we extensively work with tensors. We therefore need to define mathematical operations between them.

\begin{definition}[Tensor Product] \label{def:tensor-product} $\:$\\
Let $f \in T^{(n,k)}$ and $g \in T^{(m,l)}$. We define a tensor product $\tens{}$ between those tensors as
\begin{align}
    f \tens{} g (a_1, ..., a_{k+l}, a^1, ..., a^{n+m}) := f(a_1, ..., a_k, a^1, ..., a^{n}) g(a_{k+1}, ..., a_{k+l}, a^1, ..., a^{m}) \: .
\end{align}
\end{definition}

\begin{definition}[Wedge Product] \label{def:wedge-product} $\:$\\
We define the wedge product $\wedge$ as an operation between two alternating tensors \\ $f \in A_k(V)$, $g \in A_l(V)$ as
\begin{align}
    f \wedge g := \frac{1}{k! l!} A(f \tens{} g) \: .
\end{align}
The wedge product between $f$ and $g$ has the following properties:
\begin{itemize}
    \item[a)] $f \wedge g = (-1)^{kl} g \wedge f$ ,
    \item[b)] $f \wedge (g \wedge h) = (f \wedge g) \wedge h = f \wedge g \wedge h$ (it is associative),
    \item[c)] $f \wedge g = 0 \:$ if \:$ k+l > dim(V)$.
\end{itemize}

\end{definition}
\noindent
The notion of the wedge product will be further expanded upon later in this section when considering it in the context of differential forms (see Def. \ref{def:wedge-product-and-differential-forms}).

\begin{definition}[Scalar Product] \label{def:scalar-product} $\:$\\
Let $V$ be an $n$-dimensional vector space and $v,w \in V$. A scalar product is a bilinear form $g: V \times V \rightarrow \mathbb{R}$ with $g(v,w) = g(w,v)$ and if $g(v,w)= 0 \:\: \forall \: w\in V $ then $v=0$. If $g(v,v) \geq 0 \: \forall \: v\in V$ and $g(v,v) = 0 \Rightarrow v=0$, then we call $g$ positive definite. We write $g(v,w)$ also as $\left<v,w \right>$.\\
Further, we define the index of a scalar product $ind_g$ as the dimension of the largest subspace $W \subset V$, where the scalar product $g|_W$ is negative definite.
\end{definition}
For example the scalar product induced by the Minkowski metric $\eta(v,w) := - v^0 w^0 + \sum_i v^i w^i$ has $ind_\eta = 1$. \\
In the case of general tensors (see Def. \ref{def:tensor}), the scalar product can be generalized as a contraction over indices of the tensors. If the degree of both tensors is equal and if we perform a contraction over all indices, the result will be a scalar.

\begin{definition}[Scalar Product Between Alternating Forms] \label{def:scalar-product-over-alternating-forms} $\:$\\
Let now $V$ be a vector space with a scalar product $g$. On $A^k(V)$ (see Def. \ref{def:alternating-linear-forms}), we define a scalar product as

\begin{align}
    \left< \alpha^1 \wedge ... \wedge \alpha^k, \beta^1 \wedge ... \wedge \beta^k \right> := det( \left< \alpha^i, \beta^j \right> )_{i,j} \: \text{ with the basis-forms } \: \alpha^i, \beta^j \in V^\star \: .
\end{align}
\end{definition}
This corresponds to contracting over all indices of the alternating tensors in $A^k(V)$, resulting in a scalar. This notion will become important later, when we construct an action using the scalar product between tensors.

\subsection*{Manifolds and Differential Forms}

Manifolds are the basic geometrical building blocks of any physical theory. For us, a manifold is a space which can be populated with various mathematical objects and structures. These include for example tensors, differential forms, scalar products, all of which we define in this section at different points.

\begin{definition}[Manifolds] \label{def:manifold} $\:$\\ 
A topological manifold (or simply manifold) $M$ is an $n$-dimensional topological space which
\begin{itemize}
    \item[a)] is hausdorff,
    \item[b)] has a countable basis,
    \item[c)] has at each point $p \in M$ a neighbourhood $U$ so that there exists a homeomorphism $\varphi \colon U \rightarrow V$, where $V \subset \mathbb{R}^n$ is open.
\end{itemize}
\end{definition}
\noindent
Manifolds can take various shapes such as spheres, tori or flat planes. For the cases studied in this work, we will assume manifolds to be arbitrarily curved $n$-dimensional surfaces. A manifold itself is just a collection of points with a certain structure. To be able to work with them, we also need a way to look at neighbourhoods of specific points on the manifold. This is done using a coordinate system.

\begin{definition}[Map and Atlas] \label{def:map-and-atlas} $\:$\\
Let $M$ be an $n$-dimensional manifold and let $U$ be an open neighbourhood around a point $p \in M$. We define a map as the tuple $(U,\varphi)$, where $\varphi$ is a homeomorphism as defined as in Def. \ref{def:manifold}. A map is also called a coordinate system. \\
A ($C^\infty$-) atlas $\mathcal{A}$ is a family of maps $\left\{ (U_i, \varphi_i) \right\}$ on a manifold $M$ such that \\ $M = \bigcup_i U_i$ and all maps are pairwise compatible. That is, for two maps $(U_i,\varphi_i)$, $(U_j,\varphi_j)$ the mapping $\varphi_i \circ \varphi_j^{-1} \colon \varphi_j (U_i \cap U_j) \rightarrow \varphi_i (U_i \cap U_j)$ (i.e. change of maps) is a smooth ($C^\infty$-) diffeomorphism. A maximal atlas is an atlas, which contains all possible atlas' on $M$ i.e. all possible atlas' on $M$ are subsets of $\mathcal{A}$.
\end{definition}

\noindent
One can think of an atlas as one system of patches with coordinates on them that cover the whole manifold. We now have a way to look at neighbourhoods of points on a manifold using a global system of coordinates. Equipping the manifold with a maximal atlas gives us the freedom to change coordinates at will. This makes such manifolds very useful in physical applications.

\begin{definition}[Smooth Manifolds] \label{def:smooth-manifold} $\:$\\
A smooth manifold is an $n$-dimensional manifold $M$ with a maximal atlas $\mathcal{A}$.
\end{definition}
\noindent
Examples of smooth manifolds are $\mathbb{R}^n$ or the general linear group $GL(n,\mathbb{R})$ which contains all real invertible $n \times n$ matrices.

\begin{definition}[Connected Manifolds] \label{def:connected-manifold} $\:$\\
A topological space is said to be connected if it can not be represented as the union of two (or more) disjoint open non-empty subsets. A connected manifold $M$ is a connected topological space, which in addition fulfills the properties of a manifold (see Def. \ref{def:manifold}).
\end{definition}
\noindent
Intuitively, one can think of a connected topological space as a space where the subsets do not allow a finite ''gap'' of points between them. Examples of connected manifolds are $\mathbb{R}^n$ or Lorentzian manifolds used in general relativity.

\begin{definition}[Tangent Space] \label{def:tangent-space} $\:$\\
A tangent vector is a vector, which is tangent to a curve (or space) at a given point $p$ on an $n$-dimensional manifold $M$. If $M \subset \mathbb{R}^n$, then a tangent vector is given by the directional derivative of a function $f : \mathbb{R}^n \rightarrow \mathbb{R}$ in the direction of $v \in \mathbb{R}^n$:
\begin{align}
   \left. \frac{d}{dt} f(p + tv) \right|_{t=0} = \sum_{i=1}^n v_i \left. \frac{\partial f}{\partial x^i} \left( x \right) \right|_{x=p} \: \in \: T_p M \: ,
\end{align}
where we define $T_p M$ to be the space of all tangent vectors to $M$ at the point $p \in M$. 
\end{definition}
\noindent
The notion of tangent vectors and tangent spaces can also be generalized to general smooth manifolds $M$ using a linear function called \textit{derivation} (usually denoted by $d$ or $D$). We skip further details as they are not necessary to understand the content presented in this work.

\begin{definition}[Tangent Bundle] \label{def:tangent-bundle} $\:$\\
The tangent bundle is defined as the disjoint union of all tangent spaces on an $n$-dimensional manifold M:
\begin{equation}
TM:=\bigcup\limits_{p\in M} \{p \}\times \; T_{p} M \: .
\end{equation}
\end{definition}
\noindent
For instance, the tangent bundle of the circle $S^{1}$ is diffeomorphic to a cylinder $S^{1}\times \mathbb{R}$.\\
Vector fields can be understood $W: M\rightarrow TM$ which map points on the manifold to the tangent bundle. Tangent bundles of manifolds are widely used in physics. They are populated with mathematical objects that describe physical fields such as vectors, covectors or higher rank tensors (e.g. the vector potential and the field strength tensor in electromagnetism). \\
\noindent \\
Differential forms are central in the formulations of modern physics and are closely related to gauge theories. In particular they are useful to formulate theories in a coordinate-independent way. They also provide a way of revealing the more general structure behind physical theories. We will later use them to construct our model. We will also use them to highlight the connection to existing theories.

\begin{definition}[Differential Forms] \label{def:differential-form} $\:$\\
Let $T_pM$ be the tangent space of a manifold $M$ at the point $p \in M$ and let $T^*_pM := (T_pM)^*$ be the dual vector space to the tangent space. \\
Then we can define a covector-field (or $1$-form) $\omega$ as a map which associates a covector $\omega_p \in T^\star_pM$ to every point $p \in M$ . A $k$-form is then a map $\omega$ which maps an alternating $(0,k)$-tensor $\omega_p \in A_k(T_pM)$ to every point $p \in M$. Note that for $k=1$ $A_k(T_pM) = T^\star_pM$. \\
The elements $dx^I_p := dx^{\mu_1}_p \wedge ... \wedge dx^{\mu_i}_p$ with $i = \{1, ..., k \}$ form a basis of $A_k(T_pM)$, and thus we can write every differential $k$-form as a linear combination of the elements of this basis:
\begin{align}
    \omega_p = \frac{1}{k!} \sum_I^n a_I(p) dx^I_p \: , 
\end{align}
where the factor $1/k!$ is purely conventional and prevents double-counting of terms. The coefficients $a_I(p)$ are functions on $M$. If all $a_I$ are smooth functions, we call the differential form $\omega$ a smooth differential form. The set of all smooth differential $k$-forms on $M$ is denoted by $\Omega^k(M)$.
\end{definition}
\noindent
For completeness, in Def. \ref{def:differential-form}, we defined differential forms also for non-smooth coefficients $a_I$. For the extent of this work however, we will take all functions in general to be smooth.

\begin{example}[Differential Forms] \label{exmpl:differential-form} $\:$\\
In the case $M \subset \mathbb{R}^3$, the sets of smooth differential forms can be written explicitly as
\begin{align}
    \begin{split}
        \Omega^0(M) &= C^\infty(M) \: ,\\
        \Omega^1(M) &= \left\{ a(x,y,z) dx + b(x,y,z) dy + c(x,y,z) dz \:\: | \:\: a,b,c \in C^\infty (M) \right\} \: , \\
        \Omega^2(M) &= \left\{ a(x,y,z) dy \wedge dz + b(x,y,z) dx \wedge dz + c(x,y,z) dx \wedge dy \:\: | \:\: a,b,c \in C^\infty (M) \right\} \: , \\
        \Omega^3(M) &= \left\{ a(x,y,z) dx \wedge dy \wedge dz \:\: | \:\: a \in C^\infty (M) \right\} \: , \\
        \Omega^k(M) &= \left\{ 0\right\} \:\: \forall \:\: k > 3 \: .
    \end{split}
\end{align}
This generalizes to arbitrary dimensions $n$. This opens up the possibility to write a differential $k$-form $\omega \in \Omega^k(M)$ in index notation as:
\begin{align}
    \omega = \frac{1}{k!} \omega_{\mu_1 ... \mu_k} dx^{\mu_1} \wedge ... \wedge dx^{\mu_k} \: ,
\end{align}
where summation over all $\mu_i \in \{ 1, ..., n\}$ is implied.

\end{example}

\begin{definition}[Wedge Product on Differential Forms] \label{def:wedge-product-and-differential-forms} $\:$\\
The wedge product defined in Def. \ref{def:wedge-product} generalizes also to differential forms. Let $\omega \in \Omega^k(M)$ and $\tau \in \Omega^l(M)$, then the properties a)-c) from Def. \ref{def:wedge-product} apply as well.
\end{definition}
\noindent
When computing a wedge product in index notation, one has to apply the antisymmetrization operator introduced in Eq. \ref{eq:anti-symmetrization-operator-definition} as follows:
\begin{align}
\begin{split}
    \eta = \omega \wedge \tau &= \left( \frac{1}{k!} \omega_{\mu_1 ... \mu_k} dx^{\mu_1} \wedge ... \wedge dx^{\mu_k} \right) \wedge \left( \frac{1}{l!} \tau_{\mu_{k+1} ... \mu_{k+l}} dx^{\mu_{k+1}} \wedge ... \wedge dx^{\mu_{k+l}} \right) \\
    &= \frac{1}{k! l!} \omega_{[\mu_1 ... \mu_k} \tau_{\mu_{k+1} ... \mu_{k+l}]} dx^{\mu_{1}} \wedge ... \wedge dx^{\mu_{k+l}} \\
    &= \frac{1}{(k+l)!} \eta_{\mu_1 ... \mu_{k+l}} dx^{\mu_{1}} \wedge ... \wedge dx^{\mu_{k+l}} \: . \label{def:wedge-product:index-notation-example}
\end{split}
\end{align}
As an example for $\omega \in \Omega^1(M)$ and $\tau \in \Omega^2(M)$, the expression reads:
\begin{align}
    \eta = \omega \wedge \tau = \frac{1}{2!} \omega_{[\mu_1} \tau_{\mu_2 \mu_3 ]} dx^{\mu_1} \wedge dx^{\mu_2} \wedge dx^{\mu_3} = \frac{1}{3!} \eta_{\mu_1 \mu_2 \mu_3} dx^{\mu_1} \wedge dx^{\mu_2} \wedge dx^{\mu_3} \: .
\end{align}

\begin{definition}[Exterior Derivative] \label{def:exterior-derivative} $\:$\\
For $k \in \mathbb{N}_0$ we define the exterior derivative as a mapping $d: \Omega^k(M) \rightarrow \Omega^{k+1}(M)$ acting on a differential form $\omega = \frac{1}{k!} \omega_I dx^I \in \Omega^k(M)$ with the following properties:
\begin{itemize}
    \item[a)] $d \circ d = d^2 = 0$ (nilpotent)\: ,
    \item[b)] $d(\omega \wedge \tau) = d\omega \wedge \tau + (-1)^k \omega \wedge d \tau$, for $\omega \in \Omega^k(M)$ and $\tau \in \Omega^l(M)$ \: .
\end{itemize}
Furthermore, we call differential forms with $d\omega=0$ closed, and we call them exact if there exists a $(k-1)$-form $\tau \in \Omega^{k-1}(M)$ with $d\tau = \omega$. Every exact differential form is also closed.
\end{definition}
\noindent
Explicitly, the exterior derivative $d\omega$ can be written as
\begin{align}
    d \omega := \frac{1}{k!} \sum_I d\omega_I \wedge dx^I =  \frac{1}{k!} \sum_{I,j} \frac{\partial \omega_I}{\partial x^k} dx^j \wedge dx^I  \: \in \Omega^{k+1}(M) \: .
\end{align}
\noindent
In index notation the exterior derivative of a differential form $\omega \in \Omega^k(M)$ reads
\begin{align}
    d \omega = \left( \phantom{\frac{}{}} \partial_{\mu_1} dx^{\mu_1} \right) \wedge \left( \frac{1}{k!} \omega_{\mu_2 ... \mu_{k+1}}  dx^{\mu_2} \wedge ... \wedge dx^{\mu_{k+1}} \right) = \frac{1}{k!} \partial_{[\mu_1} \omega_{\mu_2 ... \mu_{k+1} ]}  dx^{\mu_1} \wedge ... \wedge dx^{\mu_{k+1}} \: . \label{def:exterior-derivative:index-notation}
\end{align}
This mathematical structure may already seem familiar to many physicists. If we take for example the case where $\omega \in \Omega^1(M)$ is a 1-form, we obtain a structure that is identical to the definition of the electromagnetic field strength tensor $F_{\mu\nu} = \partial_\mu A_\nu - \partial_\nu A_\mu$. This is not a coincidence. In fact, the field strength tensor $F$ is formally defined as the exterior derivative of the electromagnetic vector potential $A \in \Omega^1(M)$ by $F := d A$. Using this definition, we also immediately obtain the equation of motion in the covariant formulation of electromagnetism $\partial_\mu F^{\mu\nu}=0$. This becomes apparent when taking the exterior derivative of the field strength: $dF = ddA = 0$. This is zero due to property $a)$ of the exterior derivative (see Def. \ref{def:exterior-derivative}). We will later see that all known field strength tensors follow similar relations. The difference being that the outer derivative will then be replaced by a covariant derivative.

\begin{definition}[Orientable Manifolds and Atlas'] \label{def:orientable-manifold} $\:$\\
An $n$-dimensional manifold $M$ is said to be orientable if there exists an $n$-form $\omega \in \Omega^n(M)$ with $\omega |_p \neq 0 \:\: \forall \:\: p \in M$. An atlas on $M$ is said to be orientable if there exists a collection of local charts $(U_{i},\varphi_{i})$ such that $\{U_{i}\}$ is a covering of $M$ and such that the mapping $\varphi_{i}\circ \varphi_{j}^{-1}$ has a strictly positive definite Jacobian on its domain of definition $\varphi_{i}(U_{i}\cup U_{j})$.
\end{definition}
\noindent
Simple examples for nontrivial, orientable manifolds are the $n$-Sphere $S^{n}$ and the $n$-Torus $T^{n}$. An example for a non-orientable manifold is given by the M\"obius strip. In essence, an orientable manifold allows us to define a normal vector which preserves its direction when moved on closed curves on $M$. \\

\noindent
Next, we need to add structure to our manifold, which maps two elements into a scalar.

\begin{definition}[Scalar Product between Differential Forms] \label{def:scalar-product-over-diff-forms} $\:$\\
Because every differential form is an alternating form, the scalar product between differential forms can be defined in the same way as in Def. \ref{def:scalar-product-over-alternating-forms} but $V=T_pM$. 
\end{definition}

\begin{definition}[Volume Form] \label{def:volume-form} $\:$\\
Let $\left\{\alpha^1, ..., \alpha^n \right\}$ be a basis of a positively oriented manifold $M$ (in general a vector space $V$) with a scalar product $g$ and let $\left\{\beta^1, ..., \beta^n \right\}$ be the dual basis. Then we call \\ $vol \in \Omega^n(M)$ the volume element of $M$ and $vol$ is given by
\begin{align}
    vol := \alpha^1 \wedge ... \wedge \alpha^n = \sqrt{| det(g_{ij}) |} \: \beta^1 \wedge ... \wedge \beta^n \: .
\end{align}
\end{definition}
The definition can analogously be formulated pointwise on the manifold $M$. This will later allow us to consider the volume of arbitrary curved spaces by integrating over all volume elements. \\

\noindent
Next, we define another useful mathematical operation, the Hodge star operator. It is related to the scalar product and is used to map differential forms to scalars. In some cases it is also used to map differential forms on the boundary or inside of a manifold to each other. For example in physics, one can use it to relate the charge inside of a volume to the current going through its surface.

\begin{definition}[Hodge Star Operator] \label{def:hodge-star-operator} $\:$\\
Let $M$ be an $n$-dimensional manifold (in general a vector space $V$) with a scalar product $g$ and a volume element $vol$. The Hodge star operator is defined as the unique isomorphism $\star_g \colon \Omega^k(M) \rightarrow \Omega^{n-k}(M)$ with $\omega \wedge \star_g \tau := \left< \omega, \tau \right> \cdot vol$ for all $\omega, \tau \in \Omega^k(M)$.
\end{definition}

Properties of the Hodge star operator include:
\begin{itemize}
    \item[a)] $\star_g 1 = vol$ ,
    \item[b)] $\star_g vol = (-1)^{ind_g}$ ,
    \item[c)] $\star_g \star_g \omega = (-1)^{ind_g} (-1)^{k(n-k)} \: \omega $ ,
    \item[d)] $\left< \star_g \omega, \star_g \tau\right> = (-1)^{ind_g} \left< \omega, \tau \right>$ .
\end{itemize}

\begin{example}[Hodge Star Operator] \label{example:hodge-star-operator} $\:$\\
If for example $M= \mathbb{R}^3$, then $\star_g$ acts on differential forms as:
\begin{align}
\begin{split}
    \star_g \left( a \cdot dx + b\cdot dy + c\cdot dz \right) &= a\cdot dy \wedge dz + b\cdot dz \wedge dx + c\cdot dx \wedge dy \: , \\
    \star_g \left( a \cdot dx \wedge dy \wedge dz \right) \:\:\:\:\: &= a \: ,
\end{split}
\end{align}
and $a,b,c \in C^\infty(M)$.
\noindent \\
The Hodge-Star-Operator can also be expressed in index notation for an $n$-dimensional manifold $M$. For that, we first define the Levi-Civita tensor $E$ as
\begin{align}
    E_{\mu_1 ... \mu_n} := \sqrt{| det(g_{ij}) |} \:\: \varepsilon_{\mu_1 ... \mu_n} \:\: , \:\: E^{\mu_1 ... \mu_n} := \frac{sgn(det(g_{ij}))}{\sqrt{| det(g_{ij}) |}} \varepsilon^{\mu_1 ... \mu_n} \: ,
\end{align}
where $\varepsilon$ is the general Levi-Civita symbol (or totally antisymmetric symbol). The metric determinant ensures tensoriality. The Hodge star operator acting on the wedge product of $k$ basis-1-forms can then be expressed as follows:
\begin{align}
    \star_g \left( dx^{\mu_1} \wedge ... \wedge dx^{\mu_k} \right) = \frac{1}{(n-k)!} g^{\mu_1 \nu_1} ... g^{\mu_k \nu_k} E_{\nu_1 ... \nu_n} \:  dx^{\nu_{k+1}} \wedge ... \wedge dx^{\nu_n} \: .
\end{align}
For a 1-form in $n=4$ dimensions this explicitly becomes
\begin{align}
    \star_g dx^\mu = \frac{\sqrt{| det(g_{ij}) |}}{3!} \varepsilon^\mu_{\: \alpha \beta \gamma} dx^\alpha \wedge dx^\beta \wedge dx^\gamma \: . \label{example:hodge-star-operator:1-form}
\end{align}

\end{example}

\begin{definition}[Riemannian and Semi-Riemannian Manifolds] \label{def:riemannian-and-semi-riemannian-manifolds} $\:$\\
Let $M$ be an oriented, $n$-dimensional manifold. Then the concepts mentioned above (Def. \ref{def:scalar-product-over-diff-forms}- Def. \ref{def:hodge-star-operator}) can be applied pointwise for all points $p \in M$. A map $g$ which assigns a scalar product $g_p$ on $T_pM$ to each point $p \in M$ is called a semi-Riemannian metric on $M$. If $M$ is connected, then the index $ind_{g_p}$ is independent of $p$ and is called the index of $g$ on $M$. Manifolds with $ind_{g} = 1$ are called Lorentzian manifolds, manifolds with $ind_{g} = 0$ are called Riemannian manifolds.
\end{definition}
\noindent
An example for a Lorentzian manifold is the Minkowski space (denoted $\mathbb{R}^4_1$), which is important in special relativity. $4$-dimensional Lorentzian manifolds play a crucial role in general relativity. On Lorentzian manifolds, the square-root of the absolute value of the determinant of the metric is written as the shorthand $\sqrt{| det(g_{ij}) |} = \sqrt{-g}$.

\begin{definition}[Integral over Differential Forms] \label{def:integral-over-differential-forms} $\:$\\
Let $\omega \in \Omega^n(M)$ be a differential form on an $n$-dimensional manifold $M$. Then we can write $\omega = f(x) dx^1 \wedge ... \wedge dx^n$ (= $f \cdot vol$) with a function $f: M \rightarrow \mathbb{R}$. The integral of $\omega$ over $M$ is then defined as
\begin{align}
    \int_M \omega = \int_M f(x) dx^1 \wedge ... \wedge dx^n :=  \int_M f(x) dx^1 ...\: dx^n \: ,
\end{align}
if the Lebesgue integral of the right hand side exists. 
\end{definition}

\begin{definition}[Volume of a Manifold] \label{def:volume-of-a-manifold} $\:$\\
The volume of a manifold $M$ with a scalar product $g$ is defined as $\int_M vol$. \\
\end{definition}
Using the integral, it is also possible to define a scalar product $(\cdot \, , \cdot)$ between two differential forms $\omega, \tau \in \Omega^k(M)$ on the manifold $M$ with metric $g$ as
\begin{align}
    \left( \omega,\tau \right) := \int_M \omega \wedge \star_g \tau = \int_M \left< \omega, \tau \right> vol \: .
\label{eq:scalar-product-differential-forms}
\end{align}

\begin{definition}[Stokes' Theorem] \label{theorem:strokes-theorem} $\:$\\
Let $M$ be an oriented $n$-dimensional manifold with boundary $\partial M$. Then for every differential form $\omega \in \Omega^{n-1}(M)$ with compact support, the following statement holds:
\begin{align}
    \int_M d \omega = \int_{\partial M} \imath^* \omega \: ,
\end{align}
where $\imath : \partial M \rightarrow M$ is the inclusion mapping and $\imath^* \omega$ is the pullback of $\omega$.
\end{definition}
\noindent
Using Stokes' theorem, it is possible to show a relation that will become useful later. For two differential forms $\tau \in \Omega^{k}(M)$ and $\eta \in \Omega^{k+1}(M)$ that vanish on the boundary $\partial M$ we have:
\begin{align}
    0= \int_M d \left( \tau \wedge \star_g \eta \right) = \int_M d \tau \wedge \star_g \eta + \int_M (-1)^k \tau \wedge d \star_g \omega \: . \label{eq:stokes-useful-relation-at-infty}
\end{align}
This trick of vanishing divergence on the boundary is commonly employed in physics, where the boundary $\partial M$ is set to spatial infinity and is therefore said to not affect the physical situation within the domain of interest.

\subsection*{Lie Groups and Algebras}

We leave the territory of differential geometry and turn our attention towards group theory. Later, we will use our combined knowledge about differential forms and Lie groups to explore bundle structures on manifolds and see how they can be used to construct physical field theories. We start with the basic building blocks of many physical theories with continuous symmetries, Lie groups.

\begin{definition}[Lie Groups and Lie Algebra] \label{def:lie-groups-and-lie-algebra} $\:$\\
Let $G$ be a finite dimensional group that is also a finite-dimensional smooth manifold. $G$ is then called a Lie group if the inversion mapping $G\times G \rightarrow G$ ,\; $(\sigma, \tau)\mapsto \sigma\tau^{-1}$ is smooth, where $\sigma,\tau\in G$. The tangent space $T_{e}G$ at the identity element $e$ is called the Lie algebra and is commonly denoted by $\mathfrak{g}$.
\end{definition}
Definition \ref{def:lie-groups-and-lie-algebra} provides an intuitive geometrical picture of Lie groups. They can be thought of as smooth manifolds with a smooth multiplication operation on them. The set of tangent vectors at the identity element on this manifold defines the corresponding Lie algebra to the Lie group. Although Def. \ref{def:lie-groups-and-lie-algebra} is the formally correct definition of a Lie group, it is not particularly helpful for calculations involving them. In practice, one often speaks of Lie groups in terms of their representations.

\begin{definition}[Lie Bracket] \label{def:lie-bracket} $\:$\\
Let $M$ be a manifold with corresponding tangent bundle $TM$ and let $f\in C^{\infty}(M)$ be a smooth function. Let further $X: M\rightarrow TM$ and $Y: M\rightarrow TM$ be vector fields, then the Lie bracket $[\cdot,\cdot]$ between two vector fields is defined as 
\begin{equation}
[X,Y](f)=X(Y(f))-Y(X(f)) \hspace{1.5cm} \forall f\in C^{\infty}(M) \: .
\end{equation}
\end{definition}

\begin{definition} [Representations] \label{def:representations} $\:$\\
Let $\mathfrak{g}$ be a finite-dimensional Lie algebra and V be a vector space over a field $\mathbb{K}$ and let $\mathfrak{gl}(V)$ denote the vector space of all endomorphisms of V endowed with a Lie bracket $[\cdot,\cdot]$. Then $\mathfrak{gl}(V)$ is a Lie algebra over $\mathbb{K}$. Let G be a Lie group with corresponding Lie algebra $\mathfrak{g}$ over $\mathbb{K}$. A homomorphism of $\mathfrak{g}$ into $\mathfrak{gl}(V)$ is called a representation of $\mathfrak{g}$ on $V$.
\end{definition}

\noindent
Representations are needed when performing concrete calculations involving elements of the group or algebra. When choosing a representation, the group elements assume a specific shape, for example $n \times n$ matrices. A representation $R(G)$ of a given Lie group $G$ can always be constructed with a representation $R(\mathfrak{g})$ of the algebra. This is done via the exponential mapping $\text{exp}:R(\mathfrak{g})\rightarrow R(G), \;\; x\mapsto \text{exp}(tx)$, where $x\in \mathfrak{g}$ and $t\in\mathbb{R}$. Likewise, a representation of the algebra can be constructed by taking the derivative at the identity element of a given group representation. \\
It follows from definition \ref{def:representations} that representations of a Lie group are not unique. In principle, any homomorphism that maps the algebra onto a matrix group is a valid representation. 

\begin{example}[Fundamental and Adjoint Representation] \label{example:fundamental-and-adjoint-representation} $\:$\\
In particle physics calculations, two particular matrix representations are used. One is the fundamental representation. It is defined as the representation whose basis contains the smallest possible number of elements, called generators (see Def. \ref{def:generators}). \\
The other is the adjoint representation. It describes the group transformations of gauge fields. The adjoint representation of a Lie algebra is defined as a mapping $ad:\mathfrak{g}\rightarrow \mathfrak{gl}(\mathfrak{g}), x\mapsto [x,y],\; y\in \mathfrak{g}$, where $[\cdot\; ,\; \cdot]$ is the Lie bracket defined in Def. \ref{def:lie-bracket}. The adjoint representation of a Lie group $G$ is denoted by $Ad(G)$ and simply corresponds to the exponential mapping of the adjoint representation $ad(\mathfrak{g})$ of the corresponding Lie algebra $\mathfrak{g}$.
\end{example}

\begin{example} [Representations of the Lorentz Group] \label{example:representations-of-the-lorentz-group} $\:$\\
The Lorentz group $O(3,1)$\footnote{There is some ambiguity regarding the naming conventions of the Lorentz group. In physical application one often says Lorentz group but means $SO^+(3,1)$. This is the orthochrone Lorentz group, which preserves the direction of time.} is deeply connected to the mathematical description of quantum-mechanical spin of a quantum field. A priori, these two concepts seem to be completely disentangled. It is only when studying representations of the Lorentz group that the connection between them becomes apparent. Relativistic classiscal or quantum fields can be classified into finite dimensional representations of the Lorentz group, characterized by two half-integers $(m,n)$. Bosonic fields are described by representations of integer dimensions. Fermions are described by spinorial representations. These are  representations of the spin group $Spin(3,1)$. The spin group is defined as the double covering map of $SO^{+}(3,1)$. This double covering map is an extension of the spin concept introduced in non relatvistic quantum mechanics to relativistic fields. In non relativistic quantum mechanics the $SO^{+}(3,1)$ group reduces to the rotational group $SO(3)$ and its double cover to the special unitary group $SU(2)$ which is known to describe non relativistic spin $1/2$ particles. \\
Given a representation $\rho : Spin(3,1)\rightarrow GL(V) $ with target vector space $V$, then the elements of $V$ are called spinors. Elements of the section $\Gamma(V)$ are called spinor fields. These spinor fields can be used to describe fermions of arbitrary half integer spin.\\ 
Note that the manifold $Spin(3,1)$ is not orientable due to the nature of the double cover map. This means that if we parallel transport a spinor field, the normal component of the spinor field will not point in a uniform direction but will undergo a change of orientation.
\end{example}

\begin{definition}[Generators] \label{def:generators} $\:$\\
Let G be a finite-dimensional Lie group with Lie algebra $\mathfrak{g}$. Then the elements of a set of basis vectors $\{X_{i}\}_{i=1,...,dim(\mathfrak{g})} \subset \mathfrak{g}$ of the algebra are called generators if the smallest subalgebra containing those elements is the Lie algebra itself. 
\end{definition}
In particular, one can choose a base in which the Lie bracket assumes the shape \\ $[X_{i},X_{j}]=i \sum_{k}^{} f^{k}_{\:\: ij}X_{k}$, where $f^{k}_{\:\: ij}$ are called structure coefficients. They are unique for every Lie algebra. Generators of matrix groups are usually denoted by $T_a$, $a\in\{1,...,dim(\mathfrak{g})\}$. For general Lie groups, the position of the group indices is not arbitrary and raising and lowering indices is done with the corresponding Cartan-Killing form.

\begin{definition}[Cartan-Killing Forms] \label{def:catan-killing-forms} $\:$\\
Let $\mathfrak{g}$ be a finite dimensional Lie algebra over a field $\mathbb{K}$ and $X,Y\in \mathfrak{g}$. Then the symmetric, bilinear Cartan-Killing form $K:\mathfrak{g}\times \mathfrak{g}\rightarrow \mathbb{K}$ is defined by:
\begin{equation}
K(X,Y)= Tr(ad(X) ad(Y)) \: .
\end{equation}
\end{definition}
For some basis of generators $\{T_{i}\}_{i=1,...,\text{dim}(\mathfrak{g})}$ of a Lie algebra $\mathfrak{g}$, one can express the components of the Cartan-Killing form K as

\begin{equation}
K_{il}=Tr(ad(T_i)ad(T_l))=-\sum_{j,k} f^{k}_{\:\: ij}f^{j}_{\:\: lk} \: .
\label{eq:Cartan Killing-form metric}
\end{equation}
The Cartan-Killing form is also referred to as the Cartan-Killing metric of the Lie algebra $\mathfrak{g}$. We will use those terms interchangeably. \\

\begin{definition}[Compact Lie Algebras] \label{def:compact-lie-algebras} $\:$\\
A Lie algebra is compact if and only if its corresponding Cartan-Killing form is negative definite.
\end{definition}

\begin{definition}[Semisimple Lie Algebras] \label{def:semisimple-lie-algebras} $\:$\\
A Lie algebra is semisimple if and only if its corresponding Cartan-Killing form is non-degenerate (i.e. has nonzero determinant).
\end{definition}
\noindent 
One example for a compact, semisimple Lie algebra is $\mathfrak{su}(N)$. Its structure coefficients satisfy the relation $f^{k}_{\:\: ij}f^{j}_{\:\: lk}=N\delta_{il}$ in a local basis of $\mathfrak{g}$, i.e. the Cartan-Killing metric is negative definite and the generators satisfy the relation 

\begin{align}
    Tr(T_{i}T_{j})=\lambda K(T_{i},T_{j})=\frac{1}{2}\delta_{ij} \: ,\label{eq:Cartan Killing-form metric-2}
\end{align}
with the proper normalization convention for the scale factor $\lambda\in \mathbb{R}$.

\subsection*{Principal Bundles}

We now have defined quantities and properties related to single groups. Next we consider how different groups can be used together. In physics for example, Lie groups are often used together with spacetime manifolds to describe a specific physical symmetry. The next few definitions will therefore cover mathematical structures used to embed groups actions into a manifold. One of them are fiber bundles. A fiber bundle is a space, that is \textit{locally} a product space, but can \textit{globally} have different topological properties.

\begin{definition}[Fiber Bundle] \label{def:fibre-bundle} $\:$\\
Let $E=B \times F$, $B$ and $F$ be topological spaces, and let there be be a continuous surjective projection $\pi : E \rightarrow B$. The structure containing $(E,B,F,\pi)$ is called a fiber bundle. The fiber bundle fulfills the local triviality condition, which means that locally $E$ is a product group. B is called the base space of the bundle, E the total space, and F is the fiber. 
\end{definition}
\noindent
Later, we will interpret $B$ as a manifold describing spacetime and $F$ as a Lie group, describing the desired symmetries. \\
A trivial bundle is a fiber bundle which can \textit{globally} be written as a product space $E=B \times F$. Global symmetries known from physics are trivial bundles.

\begin{definition}[G-Principal Bundle] \label{def:g-principial-bundle} $\:$\\
Let $M$ be a smooth manifold and $G$ a finite dimensional Lie group. We define a $G$-principal bundle as a fiber bundle $P$ at a point $p \in M$ together with a right action $r: P\times G \rightarrow P$ such that r preserves the fibers of $P$ and acts freely and transitively for all $g\in G$ and $p\in P$. 
\end{definition}
\noindent
This provides a way to incorporate local symmetries into a manifold. A G-principal bundle is invariant, when it is multiplied with any element from the group at each point $p \in M$. Globally, this is not the case since the given basis of the group (or Lie algebra) will be different for each point on the manifold. One can picture this as a coordinate basis that depends on the location on the manifold. 
A $G$-principal bundle therefore can be locally understood as a product space between a fiber bundle and a manifold, where the elements of both subspaces locally commute. 

\begin{example}[Minkowski Bundle] $\:$\\
One example for a $G$-principal bundle, used frequently in modern physics is the Minkowski bundle. It is a vector bundle $B$ with fibers $F$ over a semi Riemannian manifold M and structure group $G=O(n-1,1)$ acting on its fibers through a free right action in the fundamental representation. In other words, the Minkowski bundle defines a local $O(n-1,1)$ symmetry on the Manifold $M$ and locally one can write $B=M\times_{\rho} F$, with $\rho$ denoting the fundamental representation of the Lorentz group.
\end{example}

\begin{definition}[Vector Valued Differential Forms] \label{def:lie-algebra-valued-differential-forms} $\:$\\
Let $M$ be an $n$-dimensional differentiable manifold and $V$ a finite dimensional vector bundle. We define a \textit{vector valued differential form} as a smooth section of the bundle $V \tens{} \wedge^{k} ( T^{*}M )$, with $\wedge^{k}$ denoting the $k$-th exterior (wedge-) product and $T^{*}M$ being the cotangent bundle of the Manifold M. We denote the space of smooth vector valued k-forms as $\Omega^{k}(M;V)$. 
\end{definition}
We note that a Lie algebra (see Def. \ref{def:lie-groups-and-lie-algebra}) is a vector space. It is thus possible to also define Lie algebra valued differential forms in the same way. In this work we are only interested in Lie algebra valued differential forms. Therefore, we formulate the following definitions for Lie algebras even though more general definitions can be made using vector valued differential forms. \\

\noindent
If we assume local triviality then $\Omega^{k}(M;\mathfrak{g}) \simeq \Omega^{k}(M) \tens{} \mathfrak{g}$ and a given differential form  \\ $\omega \in \Omega^{k}(M;\mathfrak{g})$ can be written as $\omega = \frac{1}{k!} \omega^a_{\mu_1 ... \mu_k} \; T_a \tens{} \, ( dx^{\mu_1} \wedge ... \wedge dx^{\mu_k} )$. Here, $T_a$ are the generators of the Lie algebra $\mathfrak{g}$ and summation over the index $a$ is implied. The generators $T_a$ and the basis 1-forms $dx^\mu$ then form a common basis for the Lie algebra valued differential form. Note that in the context of standard-model particle physics, all group-indices are usually written as upper indices by convention.

\begin{definition}[Lie Bracket of Lie Algebra Valued Differential Forms] \label{def:lie-bracket-of-differential-forms} $\:$\\
Let $M$ be an $n$-dimensional manifold and G be the Lie group with algebra $\mathfrak{g}$ and let $\omega\in \Omega^{k}(M;\mathfrak{g})$ and $\eta\in\Omega^{l}(M;\mathfrak{g})$. We define the Lie bracket between these differential forms as a mapping $[\cdot,\cdot]:\Omega^{k}(M;\mathfrak{g})\times \Omega^{l}(M;\mathfrak{g})\rightarrow \Omega^{k+l}(M;\mathfrak{g})$ such that
\begin{equation}
[\omega,\eta](v_{1},...v_{k+l})=\frac{1}{(k+l)!}\; \sum_{\sigma\in S_{n}} \text{sign}(\sigma) [\omega(v_{\sigma(1)},...v_{\sigma(k)}),\eta(v_{\sigma(k+1)},...v_{\sigma(k+l)})] \: ,
\label{eq: Definition Lie bracket-Lie valued diff-froms }
\end{equation}
where $v_{i}$ are elements of the basis of the tangent space.
\end{definition}

\noindent
For $\omega\in \Omega^{k}(M;\mathfrak{g})$ and $\eta\in \Omega^{l}(M;\mathfrak{g})$, where $\mathfrak{g}$ can be represented by a matrix-algebra, the Lie bracket can be written using the group generators $T_a \in \mathfrak{g}$ as follows:
\begin{align}
    [\omega,\eta] = (\omega^a\wedge \eta^b) \tens{} [T_a,T_b] \: , \hspace{1cm} a,b \in \left\{1, \cdots dim(\mathfrak{g}) \right\} \: , \label{eq:lie-bracket-of-differential-forms}
\end{align}
where $[T_a,T_b]$ is the commutator (i.e. Lie bracket) of two matrices and summation over the indices $a$ and $b$ is implied.

\subsection*{Vielbeins and Connections}

To study explicit representations of differential forms, we need a way to describe coordinate frames. A useful tool for that are vielbeins. They can be used to construct a local frame basis. Another use of them is to map points between local and global coordinates.

\begin{definition}[Vielbeins] \label{def:Tetrads}  $\:$\\
Let $M$ be an n-dimensional differentiable semi-Riemannian manifold with a metric $g$ and let $V$ be a vector bundle. Then the isomorphisms $e:TM\rightarrow V$ are called vielbeins. Dual vielbeins are denoted by $\Bar{e}$. We call the space of vielbeins $\Omega^1(M;V)$.
\end{definition}
\noindent
It is not trivial why vielbeins are also $1$-forms, i.e. why $e\in \Omega^1(M;V)$. This can be understood in the following way: Even though vielbeins are defined on the vector bundles $TM$ and $V$, locally $TM$ and $V$ look like a vector spaces. These vector spaces will then also have a local basis. Using these bases, it is possible to express vielbeins locally as vectors. Because vectors are the same as $1$-forms this explains why $e\in \Omega^1(M;V)$. Vielbeins are essentially tools with which we can represent the tangent space using the basis of local vector spaces. \\
In particular, if the vector bundle $V$ is also endowed with a metric $K$ and if one expresses the vielbeins $e\in \Omega^1(M;V)$ in terms of a basis of the vector bundle $V$ as $e=e_{a}v^{a},a\in \{1,...,\text{dim}(V) \}$, \;then the components of the metric $g$ of the semi-Riemannian manifold and the components of the metric $K$ are related as

\begin{equation}
g_{\mu\nu}=e^{a}_{\mu}\;K_{ab}\;e^{b}_{\nu} \: . \label{eq:vielbein-definition-local-and-globals-metric}
\end{equation}

\begin{example}[Tetrads in General Relativity] \label{example:tetrads-in-GR} $\:$\\
In general relativity, the $O(n-1,1)$ group is a local isometry of some curved spacetime. We can define a bundle-isomorphism $e: TM\rightarrow W$ with W being the Minkowski bundle with a Minkowskian metric $\eta$ relating local isometry groups with the global structure of a semi-Riemannian manifold $M$ with metric $g$. In an orthonormal basis of the vector space $W$, $e$ can be expressed as $e=e^{a}w_{a}\in \Omega^{1}(M,W)$. The Minkowski metric and the metric $g$ are then related as:
\begin{equation}
g_{\mu\nu}=e^{a}_{\mu}\eta_{ab}e^{b}_{\nu} \: .
\end{equation}
The latin indices a and b are the Lorentz indices in the local frame and the greek indices are the global spacetime indices. These mappings are called frame fields or tetrads. Choosing an explicit expression for the tetrads is equivalent to choosing a basis of a local frame. 
\end{example}

\begin{definition}[Connection Forms] \label{def:connection-forms} $\:$\\
Let M be a smooth $n$-dimensional manifold and P be a $G$-Principal bundle with total space $E$. \\
A Lie algebra valued one-form $\omega\in \Omega^{1}(P;\mathfrak{g})$ is called a connection form if it satisfies:
\begin{equation}
\omega(v)=\begin{cases} \xi \: , & \text{if $v=\sigma(\xi)\:, \, \xi\in C^{\infty}(P,\mathfrak{g})$} \: , \\
0 \: , &\text{if $v$ horizontal}
\end{cases}
\end{equation}
with $v\in \Gamma(TG)$ being left-invariant vector fields and isomorphism $\sigma: \mathfrak{g}\rightarrow \Gamma(VE)$, where VE is the vertical bundle (for more see ref \cite{Tecchiolli:2019hfe}). We denote the space of connection forms as $\Omega_{\text{con}}(M;\mathfrak{g})$.\\
\end{definition}
Essentially, Lie algebra valued connection one-forms make it possible to transform objects between different tangent spaces at different points on the $G$-principal bundle. 

\begin{definition}[Spin Connection] \label{def:spin-connections}  $\:$\\
Let $M$ be a finite dimensional semi-Riemannian manifold and $P$ a G-Principal bundle with local symmetry bundle group $O(3,1)$. Then the spinorial representation $spin(M)$ of the algebra $\mathfrak{o}(3,1)$ is an associated vector bundle called the spinor bundle. A connection $\omega\in \Omega_{con}(M;\mathfrak{spin}(M))$ on this spinor bundle is called a spin connection. 
\end{definition}
The spin connection is used to define parallel transports of spinors. Since the spin group $spin(M)$ is a non-orientable manifold, the normal vector undergoes a change of direction. This notion becomes important when considering e.g. spinor fields in curved spacetime, where the existence of torsion is manifested in the covariant derivative. \\
The spin connection is induced by the affine connection. It can be interpreted as the covariant derivative of a vielbein with respect to the affine connection.

\begin{example}[Connections] \label{eample:connections} $\:$\\
The choice of a connection is in general arbitrary. In practice, however, a connection is chosen to simplify the computation of relevant quantities. Any connection induces an affine connection and a spin connection (see Def. \ref{def:spin-connections}). The affine connection is necessary to define a parallel transport. Its coefficients are called Christoffel symbols. In the standard formulation of general relativity for example, the connection chosen is the Levi-Civita connection. It is defined as the connection which has zero torsion and fulfills compatibility with the spacetime metric (covariant derivative of the metric is zero). In the language of tetrads, metric compatibility translates to the following condition for tetrads $e\in \Omega^{1}(M,V)$:
\begin{equation}
e_{a}\; (d_{\omega}e)^{a} = Q \: ,
\label{eq:metric-compatibility-tetrads}
\end{equation}
\noindent 
where $Q \in \Omega^{2}(M;\mathfrak{o}(3,1))$ is the torsion two-form. By using the condition \eqref{eq:metric-compatibility-tetrads} and imposing zero torsion, one may derive the following relation between the components of the spin connection $\omega_{\mu a}^{b}$ and the Christoffel symbols $\Gamma^{\mu}_{\nu\rho}$ 

\begin{align}
  \omega_{\mu b}^{a}=e^{\nu}_{b} \Gamma^{\alpha}_{\mu\nu} e_{\alpha}^{a} - e^{\nu}_{b} \partial_{\mu} e_\nu^a \: .
\end{align}
\end{example}

\begin{definition}[Covariant Derivative] \label{def:covariant-derivative} $\:$\\
Let $M$ be a smooth $n$-dimensional manifold and $P$ a G-Principal bundle with structure Group $G$ which has a Lie algebra $\mathfrak{g}$. Furthermore, let $\omega\in \Omega_{\text{con}}(M;\mathfrak{g})$ be a connection and $d: \Omega^{k}(M;\mathfrak{g})\rightarrow \Omega^{k+1}(M;\mathfrak{g})$ an exterior derivative of differential forms. We define the covariant derivative $d_{\omega}$ as:
\begin{equation}
d_{\omega}=d+[\omega, \cdot\;] \: ,
\label{eq: Covariant derivative definition}
\end{equation}
where $[\;\cdot\; , \cdot\;]$ is the Lie bracket (see Def. \ref{def:lie-bracket-of-differential-forms}). \\
\end{definition}
The covariant derivative is used to perform parallel transports of tensor fields between tangent spaces of the G-principal bundle, while respecting the structure of the underlying space. 
The Lie bracket term corrects for deviations from euclidean space.
In general relativity, it is commonly denoted by $\nabla_{\mu}$ and is a special case of \eqref{eq: Covariant derivative definition} when taking $\omega$ to be the Levi-Civita connection and the Group $G=O(3,1)$.

\begin{definition}[Curvature Form] \label{def:curvature-form} $\:$\\
Let $M$ be a smooth manifold and $P$ a G-Principal bundle with structure group $G$ which has a Lie algebra $\mathfrak{g}$. Furthermore, let $\omega \in \Omega_{\text{con}}(M;\mathfrak{g})$. We define the curvature form $\Omega\in \Omega^{2}(M;\mathfrak{g})$ as:
\begin{equation}
\Omega=d\omega+\frac{1}{2} [\omega,\omega] \: .
\end{equation}
\end{definition}
\noindent 
Note that this is nearly identical to the covariant derivative $d_\omega$ of the connection $\omega$. The factor one half in front of the Lie bracket ensures tensoriality.

\begin{example}[Yang-Mills Field Strength] \label{example:yang-mills-field-strength} $\:$\\
  One important example for a curvature form is the field strength of Yang-Mills theory. Mathematically, the gauge field $A$, which is the fundamental object in Yang-Mills theory, is a connection, i.e. $A\in \Omega_{\text{con}}(M;\mathfrak{su}(N))$ (see Def. \ref{def:connection-forms}). The corresponding curvature form $F\in \Omega^{2}(M;\mathfrak{su}(N))$ is the Yang-Mills field strength and can be written as
  \begin{equation}
    F=dA+\frac{1}{2}[A,A] \: .
  \end{equation}
  This mathematical equivalence further emphasizes the identical mathematical structure behind the concepts of differential geometry used in general relativity and the concepts used in gauge theories to describe the standard model of particle physics. 
  \label{ex:yang-mills-field-strength}
\end{example}

\begin{example}[Riemann Curvature Tensor]   $\:$\\
The Riemann curvature tensor for a generic group $G$ can be constructed from a curvature $2$-form. Let $\omega\in \Omega_{\text{con}}(M;\mathfrak{g})$ be a spin connection, $\Omega\in \Omega^{2}(M;\mathfrak{g})$ the corresponding curvature $2$-form, $V$ a vector bundle and let $e\in \Gamma(TM,V)$ be the corresponding vielbeins as defined in Def. \ref{def:Tetrads}. Let further $K:\mathfrak{g}\times \mathfrak{g}\rightarrow \mathbb{R}$ be the Cartan-Killing metric of the Lie algebra $\mathfrak{g}$, then the components of the Riemann tensor can be expressed as
\begin{equation}
R_{\mu\nu\rho}^{\sigma}=e_{a}^{\sigma}\;\Omega_{\mu\nu}^{ab}\;K_{bc} \; e^{c}_{\rho} \: ,
\end{equation}
with $a$,$b$ and $c$ denoting local structure group indices.
In the case of general relativity, the structure group is the Lorentz group $O(3,1)$. The corresponding Cartan-Killing metric is the Minkowski metric $\eta=diag(-1,+1,+1,+1)$. It describes flat spacetime. If $G= O(3,1)$, then the Riemann curvature tensor corresponds to the known expression from general relativity. 
\end{example}

\newpage
\section{Formulation and Motivation of Action} \label{sec:formulation-and-motivation-of-action}
                                                        
\subsection{Review of Classical Yang-Mills Theory}

In classical and quantum field theories, the starting point usually is an action functional from which equations of motions can be derived. In Yang-Mills theory for example, the action is given by a combination of field strength tensors $F$. The action is constructed such that it transforms like a scalar under Lorentz and $SU(N)$ group transformations. The simplest such combination, ignoring topological terms, is a contraction over all spacetime-indices of two field strength tensors. \\
It is possible to write the action functional in a coordinate-independent manner using differential forms as
\begin{align}
    S_{YM} :=- \frac{1}{2} \int_M Tr_{\mathfrak{su}(N)} \left[ F \wedge \star_\eta F \right] \: , \label{eq:yang-mills-pure-form-notation}
\end{align}
where the trace is taken over the Lie algebra $\mathfrak{su}(N)$ of the $SU(N)$ symmetry group. $M$ is an $n$-dimensional manifold. The Hodge star $\star_\eta$ used here is defined with respect to the Minkowski metric $\eta$. \\
The field strength $F\in \Omega^2(M;\mathfrak{su}(N))$ is the curvature $2$-form associated to a connection form $A\in \Omega_{\text{con}}(M;\mathfrak{g})$:

\begin{equation}
F = dA + \frac{1}{2} [A,A] \: . \label{eq:yang-mills--field-strength-pure-form-notation}
\end{equation}
The only dynamical field in Yang-Mills theory is the connection $A\in \Omega_{\text{con}}(M;\mathfrak{g})$. The total space of fields is hence given by
\begin{align}
    \mathcal{F}_{YM} = \Omega_{\text{con}}(M;\mathfrak{su}(N)) \: .
\end{align}
Studying Yang-Mills theory then means that we are investigating certain types of connections, which are always $1$-forms. Hence, Yang-Mills theory does not permit to study higher order $k$-forms as their dynamical field -- because they would not be connection forms. \\
\noindent
The action defined in \eqref{eq:yang-mills--field-strength-pure-form-notation} translates to index based notation as

\begin{align}
   S_{YM} = - \frac{1}{4} \int F_{\mu\nu}^a F^{\mu\nu}_{a} d x^4 \: 
   \label{eq:Yang-Mills-action-index-notation}.
\end{align}
The field strength tensor $F_{\mu\nu}^a$ itself is composed out of terms that include the connection form components $A_{\mu}^a$:

\begin{align}
    F_{\mu\nu}^a := \partial_\mu A_\nu^a - \partial_\nu A_\mu^a + g f^a_{\:\, bc} A_\mu^b A_\nu^c \: .
\end{align}
Here, $g \in \mathbb{R}$ is arbitrary and is usually interpreted as a coupling constant. \\
As a final note of this review, we comment on the transformation behaviour of the connection $A$. Let $U \in SU(N)$ be a group element, then the connection $A$ transforms under $SU(N)$ transformations as
\begin{equation}
A\rightarrow A^{'}=UAU^{-1}-(dU)U^{-1} \: . \label{eq:transformation-property-connection-form}
\end{equation}
This is the typical transformation behaviour of a connection form (for more details see \cite{Zucchini:2019pbv}). In physics, this type of transformation is called a gauge transformation.

\newpage 
\subsection{Constructing a Field Theory for Arbitrary $k$-Forms} \label{subsection:generalizing-yang-mills-theory-using-differential-forms}

The main goal of this work is to construct a classical field theory (with similar building blocks as Yang-Mills theory) which has arbitrary $k$-form fields as the dynamical field and is applicable to a wide range of local symmetries, described by Lie groups. We call this theory $k$-form theory (KFT). To construct KFT, we use Yang-Mills theory as a motivational starting point. \\

\noindent
We start by stating a few elementary assumptions about our field theory.

\begin{assumption} 
We construct our field theory on a G-principal bundle $P: E\rightarrow M$, where $E$ is the total space (see Def. \ref{def:fibre-bundle}). $M$ is an $n$-dimensional semi-Riemannian manifold equipped with a metric $g$. This entails the choice of coordinate systems on $M$.
\label{def:group-properties}
\end{assumption}

\begin{assumption} 
The symmetry group $G$ is a Lie group with a semisimple, compact Lie algebra $\mathfrak{g}$.
\label{def: compact-group-assumption}
\end{assumption}
In contrast to Yang-Mills theory, we take the dynamical field $A\in \Omega^{k}(M;\mathfrak{g})$ of our theory to be a Lie algebra valued differential form of degree $k$, instead of a connection form.

\begin{assumption} 
The dynamical field of our theory is a Lie algebra valued differential $k$-form $A\in \Omega^k (M;\mathfrak{g})$.
\end{assumption}
It directly follows that the field $A$ can in general \textit{not} function as the connection, because connections are \textit{always} $1$-forms. To solve this issue, we need to introduce a new connection form $B$. 

\begin{assumption} 
We introduce the connection form $B\in \Omega_{con} (M;\mathfrak{g})$ to the theory. In general it is also a dynamical field, but we will later choose it in a way that is is non-dynamical.
\end{assumption}
Using this connection, we can now define a covariant derivative $d_B: \Omega^k (M;\mathfrak{g}) \rightarrow \Omega^{k+1} (M;\mathfrak{g})$. In general, the connection $B$ is another degree of freedom of the theory besides $A$. \\
We now define the field strength $F$ as the covariant derivative of the differential form field $A\in \Omega^k (M;\mathfrak{g})$ with respect to the connection $B\in \Omega_{con} (M;\mathfrak{g})$:

\begin{equation}
F := d_B A = dA + [B,A] \: . \label{eq:field-strength-differential-form-notation}
\end{equation}
Note that the field strength $F\in \Omega^{k+1} (M;\mathfrak{g})$ is a Lie algebra valued differential $(k+1)$-form.
This sets a mathematical constraint on the $k$-form field $A$. For any $n$-dimensional space, any outer derivative and/or wedge product of two differential forms must at most be an $n$-form, otherwise it is zero by definition. It follows that $k \leq n-1$ because otherwise, $F$ would be zero. \\

\noindent
There also exists a curvature form $H \in \Omega^2 (M;\mathfrak{g})$ associated to the connection $B$ (see Def. \ref{def:curvature-form} for the definition of the curvature form). \\

\noindent
We now construct an action for our field theory, such that it symbolically resembles classical Yang-Mills theory. We do this by choosing the simplest action that includes a Hodge star and is invariant under group transformations of $G$. The simplest choice, ignoring topological terms, for an action for the field $A$ and the connection $B$ is to take the scalar product (see Def. \ref{def:integral-over-differential-forms}) of the field strength $F$ with itself and the scalar product of the curvature form $H$ with itself, respectively:

\begin{equation}
S_{KFT} := S[A,B] = - \alpha \int_{M} Tr_{\mathfrak{g}}[F\wedge \star_\eta F] - \beta \int_{M} Tr_{\mathfrak{g}}[H\wedge \star_\eta H] \: . \label{eq:general-k-form-action}
\end{equation}
Here $\alpha \in \mathbb{R}$ and $\beta \in \mathbb{R}$ are arbitrary scalar constants. The subscript $\mathfrak{g}$ on the trace indicates that the trace is taken over the Lie algebra $\mathfrak{g}$. The negative sign in front of the action integral is purely conventional. \\
The first term of the action \eqref{eq:general-k-form-action} can be interpreted as a kinetic term for $A$. The second term can likewise be interpreted as a kinetic term for the connection $B$. We note that our theory reduces to Yang-Mills theory if we set the field $A=0$ or $\alpha=0$. If we choose $\beta=0$, we ignore the dynamics of the connection through its curvature term. The curvature term can also be set to zero by choosing $B$ such that the curvature form $H$ vanishes. Note that this is not equivalent to setting $\beta=0$ because $B$ also contributes to the field strength $F$. Different choices for $\alpha$ and $\beta$ can thus be used to study different subsets and limiting cases of KFT. \\
The total space of fields of the action \eqref{eq:general-k-form-action} corresponds to 

\begin{align}
\mathcal{F}_{KFT} =\Omega^{k}(M;\mathfrak{g}) \times \Omega_{\text{con}}(M;\mathfrak{g})  \: ,
\end{align}
Similarly to the standard Yang-Mills theory, it is possible to write the action \eqref{eq:general-k-form-action} in index based notation. It can then be written as 

\begin{align}
S_{KFT} = -\frac{\alpha}{(k+1)!} \int_{M}  Tr_{\mathfrak{g}} \left( F_{\mu_1 ... \mu_{k+1}} F^{\mu_1 ... \mu_{k+1}} \right) \: \sqrt{-g} \: dx^n \:- \frac{\beta}{2} \int_{M} Tr_{\mathfrak{g}} \left( H_{\mu_1\mu_2} H^{\mu_1 \mu_2} \right) \: \sqrt{-g} \: dx^n \: . \label{eq:general-k-form-action-index-notation}
\end{align}
\noindent
Index notation is useful to perform explicit calculations. Since explicit calculations are one of the main goals of this work, we will primarily consider the form \eqref{eq:general-k-form-action-index-notation} for most calculations performed with the action. \\

\noindent 
To close this section, we consider the transformation properties of the involved fields and the action \eqref{eq:general-k-form-action}. The field $A\in \Omega^{k}(M;\mathfrak{g})$ transforms like an element of the Lie algebra $\mathfrak{g}$. Let $U \in G$ be a group element, then the field transforms as

\begin{equation}
A \rightarrow A^{'} = UAU^{-1} \: . \label{eq:transformation-property-field}
\end{equation}
\noindent
Note that this transformation property is different from the transformation of a connection form (see Eq. \eqref{eq:transformation-property-connection-form}). The connection form $B \in \Omega_{\text{con}}(M;\mathfrak{g})$ in turn transforms like

\begin{equation}
B \rightarrow B^{'} = UBU^{-1} - (dU)U^{-1} \: . \label{eq:transformation-property-connection}
\end{equation}
Using these transformation rules, we can show the invariance of the action \eqref{eq:general-k-form-action} under group transformations: The second term has the same transformation properties as the Yang-Mills action \eqref{eq:yang-mills-pure-form-notation} and is thus known to be invariant under transformations in $G$. The first term is non-trivially invariant under group transformations. We show the invariance explicitly in Appendix \ref{sec:appendix C}. The entire action is thus well-behaved with regards to invariance under group transformations.

\newpage
\subsection{Review of Palatini-Cartan Gravity} \label{subsec:review-of-einstein-hilbert-gravity}

In this section, we give a brief review of the coordinate-independent formulation of gravity. It uses vielbeins and connection forms. For a more detailed discussion, we refer to \cite{Tecchiolli:2019hfe} and references therein. \\
The most general action without topological terms in four dimensions which is 
\begin{itemize}
\itemsep0em
    \item [$a)$] polynomial in the vielbeins $e \in \Omega^1(M;W)$, where $W$ is the Minkowski bundle, and the connection $\omega \in \Omega_{con}(M;\mathfrak{so}(3,1))$,
    \item [$b)$] is invariant under diffeomorphisms (coordinate transformations),
    \item [$c)$] is invariant under \textit{local} Lorentz transformations,
    \item [$d)$] and does not use the Hodge star,
\end{itemize}
is the Palatini–Cartan (PC) action:

\begin{align}
    S_{PC} = \frac{1}{2 \kappa} \int_M Tr_{\mathfrak{so}(3,1)} \left[  e \wedge e \wedge \Omega_\omega \right] \: , \label{eq:palatini-cata-action}
\end{align}
where $\Omega_\omega = d\omega + \frac{1}{2} [\omega,\omega]$ is the curvature form associated to the connection form \\ $\omega \in \Omega_{con}(M;\mathfrak{so}(3,1))$ (see \eqref{def:curvature-form}). $\kappa$ is a factor which depends on the chosen system of units (in SI units $\kappa = 8\pi G c^{-4}$). The trace maps the wedge product of $e$ and $\Omega_{\omega}$ onto a $4$-form and is normalized in such a way that a specific orientation of the manifold is chosen.  \\
A higher-dimensional description of the Palatini–Cartan action for $n>4$ dimensions is also possible. This can be done by adding additional wedge-products of the vielbeins $e$ (see e.g. \cite{Tecchiolli:2019hfe,Montesinos:2019bkc}). The symmetry group then becomes $SO(n-1,1)$. \\
The Palatini-Cartan action can be interpreted as a functional of the connection $\omega$ and the vielbeins $e$. The space of fields is therefore
\begin{align}
\mathcal{F}_{PC} = \Omega^{1}(M;W) \times \Omega_{con}(M;\mathfrak{so}(3,1)) \: .
\end{align}
Alternatively, the action can also be interpreted as a functional of the connection and the spacetime metric $g \in \mathcal{PR}(M)$ (here $\mathcal{PR}(M)$ denotes the set of pseudo-Riemannian metrics on the manifold $M$). This is possible since the vielbeins are uniquely determined by the spacetime metric $g$ and vice-versa (see Def. \ref{def:Tetrads} and Ex. \ref{example:tetrads-in-GR}). It is therefore arbitrary whether one considers the vielbeins or the metric to be the dynamical field. In the standard formulation of general relativity, the metric is chosen as the dynamical field. \\

\noindent
In contrast to general relativity, the action (\ref{eq:palatini-cata-action}) does not make any assumption about the curvature or torsion properties of the theory. In fact, the torsion (or torsion-freeness) of spacetime is solely linked to the choice of the connection $\omega$. In the case of Palatini–Cartan, both torsion and curvature are allowed. The possibility to have non-metricity is not included in PC, but generalizations exist. This makes the Palatini–Cartan action a generalized version of the more widely known Einstein-Hilbert (EH) action, which describes gravity in a torsion-free spacetime (i.e the connection is chosen as the unique metric-compatible Levi–Civita connection). \\
In coordinate representation, the Einstein-Hilbert action in four dimensions is written as

\begin{align}
    S_{EH} = \frac{1}{2\kappa}  \int_M \sqrt{-g} \: R \: dx^4 \: , \label{eq:einstein-hilbert-action}
\end{align}
where $R$ is the Ricci curvature scalar and $\kappa$ is the same constant as defined above. The space of dynamical fields in Einstein-Hilbert gravity is

\begin{align}
\mathcal{F}_{EH} = \mathcal{PR}(M) \: .
\end{align}
Similar to Yang-Mills theory, the actions \eqref{eq:palatini-cata-action} and \eqref{eq:einstein-hilbert-action} are invariant under \textit{local} transformations with respect to a Lie group. In the case of the Palatini–Cartan/Einstein-Hilbert actions in four dimensions, the local symmetry group is the Lorentz group $SO(3,1)$. Recalling special relativity, the idea of the non-uniqueness of local frames is mathematically manifested through the global $SO(3,1)$ symmetry of the theory. If the global $SO(3,1)$ symmetry is now promoted to a local symmetry, one directly obtains general relativity. Physically, this corresponds to the strong equivalence principle that states that at any point in (curved) spacetime, special relativity is recovered.


\newpage
\subsection{Minimally Coupling $k$-Form Theory to Gravity}\label{sec:minimal-coupling-of-extended-YM}

In the previous sections, we constructed a field theory for arbitrary $k$-forms (see section \ref{subsection:generalizing-yang-mills-theory-using-differential-forms}). We also reviewed the coordinate independent formulations of gravity in section \ref{subsec:review-of-einstein-hilbert-gravity}. The goal of this section is to combine the $k$-form theory (KFT) with gravity in a common framework, which we call gravitational $k$-form theory (GKFT). \\
There are two possibilities to couple additional fields to gravity:

\begin{itemize}
\itemsep0em
    \item [$a)$] \textbf{minimal coupling:} Treatment of KFT as a self-gravitating matter field. This means that the $k$-form fields do not couple directly to the spacetime curvature, but only indirectly through their matter-energy content present in spacetime and the back-reaction of the curvature on the motion of the fields.
    \item [$b)$] \textbf{non-minimal coupling:} Treatment of the $k$-form field as a modification of the spacetime curvature. This means that there are explicit coupling terms between KFT terms and the spacetime curvature.
\end{itemize}
In this work, we assume minimal coupling between KFT and gravity.

\begin{assumption} 
We couple $k$-form theory minimally to gravity.
\end{assumption}
Minimal coupling is achieved by adding the Lagrangians describing gravity \eqref{eq:palatini-cata-action} and $k$-form theory \eqref{eq:general-k-form-action}. In $n$ dimensions, the combined action is:

\begin{align}
    S_{GKFT} = \int_M \frac{1}{2 \kappa} Tr_{\mathfrak{so}(n-1,1)} \left[ \underbrace{e\wedge e \wedge ... \wedge e\, \wedge}_{n-2 \text{ times } "e\wedge "} \Omega_\omega \right] - \alpha \int_{M} Tr_{\mathfrak{g}}[F\wedge \star_g F] - \beta \int_{M} Tr_{\mathfrak{g}}[H\wedge \star_g H] \: . \label{eq:total-action}
\end{align}
Here $\alpha \in \mathbb{R}$ and $\beta \in \mathbb{R}$ are scalar constants. $\kappa$ is a scalar constant defined in the same way as in Eq. \eqref{eq:palatini-cata-action}. The traces are taken over the Lie algebras $\mathfrak{so}(n-1,1)$ and $\mathfrak{g}$ respectively. The Hodge stars $\star_g$ are taken with respect to the spacetime metric $g$. \\
Due to the coupling to gravity, all partial derivatives $d$ need to be changed to covariant derivatives $d_\omega$ with respect to the spacetime connection $\omega \in \Omega_{\text{con}}(M;\mathfrak{so}(n-1,1))$. This is necessary to ensure invariance of the GKFT terms under local $SO(n-1,1)$ transformations. The definitions of the $k$-form field strength $F$ and the curvature form of the connection $B$ are then changed to

\begin{align}
    F = d_\omega A + [B,A] \: , \hspace{1cm} H = d_\omega B + \frac{1}{2}[B,B] \: . \label{eq:field-strangth-and-curvature-form-with-spacetime-corrections}
\end{align}
Note that the field strength $F$ and the curvature form $H$ do now have sightly different interpretations compared to their flat-space counterparts (see Eq. \eqref{eq:field-strength-differential-form-notation} and below). In a global sense, $F$ and $H$ are no longer simply a covariant derivatice or a curvature form in the Lie algebra. Rather, they can be seen as a covariant derivative/curvature form with correction terms accounting for the curved background spacetime. Locally, both $F$ and $H$ will however still retain their meaning as a covariant derivative/curvature form. \\
We now comment on the transformation behaviour of the action \eqref{eq:total-action}. The Palatini-Cartan terms are already known to be invariant under local $SO(n-1,1)$ transformations (see \cite{Montesinos:2019bkc}). The $k$-form terms are invariant under $SO(n-1,1)$ transformations as well. This is because the field $A$ transforms as a $k$-form and the connection $B$ transforms like a $1$-form under $SO(n-1,1)$. The field strength $F$ and the curvature form $H$ then transform as a $(k+1)$-form and as an anti-symmetric $(0,2)$-tensor, respectively. We elobarate on this further in Appendix \ref{sec:appendix C}. The invariance of \eqref{eq:total-action} under internal group transformations in $G$ is also fulfilled. The KFT terms are invariant by construction. The Palatini-Cartan terms are invariant because they transform trivially with respect to the group $G$ (they do not take values in the corresponding Lie algebra $\mathfrak{g}$). The total symmetry behaviour of GKFT is thus as follows: The Palatini-Cartan term is invariant under local $SO(n-1,1)$ transformations and trivially invariant under transformations in $G$. The KFT term is non-trivially invariant under local transformations in both $SO(n-1,1)$ and $G$. \\

\noindent
In the action \eqref{eq:total-action}, there is a total of four dynamical fields. From the KFT part there is the $k$-form field $A \in \Omega^k(M;\mathfrak{g})$ and the connection form $B \in \Omega_{con}(M;\mathfrak{g})$. Both are used to construct the field strength $F\in \Omega^{k+1}(M;\mathfrak{g})$ and the curvature form $H\in \Omega^2(M;\mathfrak{g})$ (also see the total action \eqref{eq:total-action}). The Palatini–Cartan part also contributes two dynamical fields. These are the vielbeins $e\in \Omega^{1}(M;W)$ and the spacetime connection $\omega \in \Omega_{con}(M;\mathfrak{so}(n-1,1))$. The corresponding space of fields to the full GKFT action \eqref{eq:total-action} is:

\begin{align}
    \begin{split}
    \mathcal{F}_{GKFT} &= \Omega^{k}(M;\mathfrak{g}) \times \Omega_{con}(M;\mathfrak{g}) \times \Omega^{1}(M;W) \times \Omega_{con}(M;\mathfrak{so}(n-1,1)) \: . \label{eq:total-space-of-fields-GKFT}
    \end{split}
\end{align}
Instead of the vielbeins $e$, it is equivalently possible to regard the spacetime metric $g\in \mathcal{PR}(M)$ as the dynamical field. \\
The total space of fields (\ref{eq:total-space-of-fields-GKFT}) will be reduced in practice when choosing specific connections for the spacetime connection $\omega$ and the group connection $B$. There are many possible choices for the connections. Starting with the spacetime connection $\omega$, one popular choice is the Levi-Civita connection. It is the unique connection which has zero torsion and zero non-metricity. It is often chosen because it greatly simplifies explicit calculations. We will later also choose this connection for this reason. Other connections which include torsion (or non-metricity) are also possible choices. The connection $B$ is a free choice of the theory as well. We will later choose the connection $B$ in a way that it only depends on the group structure coefficients $f^{a}_{\;bc}$ and the background spacetime metric $g$. In this case, $B$ has the primary function to provide coefficients for the covariant derivative $d_B$. In particular, the interpretation of $B$ as a physical field would change. If it is chosen as stated above, it will not introduce a new physical degree of freedom. When $B$ is non-dynamical, the term with the curvature forms of $B$ in \eqref{eq:total-action} could be regarded as a source term. \\

\noindent
We now find an expression of the total action \eqref{eq:total-action} in local coordinates. To do this, in index-based notation, we exchange the Palatini-Cartan term in \eqref{eq:total-action} for the Einstein-Hilbert action. However this is only strictly valid in vacuum. In the presence of bosonic matter fields -- which is the case here -- the Palatini-Cartan term can be written as the Einstein-Hilbert-Cartan action and the curvature scalar $R$ will thus  also include torsion terms in the general case. Torsion can later be set to zero by choosing the Levi-Civita connection. \\
The total action becomes

\begin{align}
    S_{GKFT} = \int_{M} \sqrt{-g} \left( \frac{1}{2\kappa} R - \frac{\alpha}{2 (k+1)!} F_{\mu_1 ... \mu_{k+1}}^a F^{\mu_1 ... \mu_{k+1}}_a - \frac{\beta}{4}\, H_{\mu_1 \mu_2}^a H^{\mu_1 \mu_2}_a \right) dx^n \: . \label{eq:total-action-coordinate-representation}
\end{align}
Summation over repeated indices $\mu_i \in \{1,...,4\}$ and $a \in \{1,...,dim(\mathfrak{g})\}$  is understood. This representation of the total action is convenient to perform explicit calculations. We will use this form later in this work. \\

\noindent
A few mathematical constraints arise directly from the action \eqref{eq:total-action}. The Palatini-Cartan term sets a lower bound on the spacetime dimension of $n\geq 2$. The current form of the action also necessitates to use a metric and coordinates. This is because of two reasons: (1.) The usage of Hodge star operators necessitates the choice of a specific metric and (2.) the usage of the vielbeins $e$ is also equivalent to choosing a metric. Note that the metric used for the Hodge star and the metric from the vielbeins are the same. The necessity of choosing a metric can thus be seen as either a necessary condition to use the Hodge star or as a consequence of using vielbeins to construct the action. This, however, is of no concern for us since a metric must be chosen anyways to do explicit calculations. Additionally, we will also choose coordinates whenever concrete calculations are performed. \\
The total action \eqref{eq:total-action} also inherits the same constraints on the $k$-form field $A$ as present in the original KFT action \eqref{eq:general-k-form-action}. Due to the field $A$ being a $k$-form, the degree $k$ of the form must fulfill $k \leq n-1$ ($k \leq 3$ in four dimensions) because otherwise, the field-strength $F \in \Omega^{k+1}(M;\mathfrak{g})$ would be zero (see Ex. \ref{exmpl:differential-form}). \\
As discussed in Example \ref{example:representations-of-the-lorentz-group}, the spin of a field (describing e.g. a particle) is related to representations of the Lorentz group. The interpretation of the physical of the field $A$ will thus change depending on $k$. For $k=0$, $A$ is a scalar field. This can be interpreted with regards to the representations of the Lorentz group as a spin-0 field. For $k\geq 1$, $A$ can be interpreted as a spin-$1$ field. However, note that the spin-representations can be different in different spacetime dimensions $n\neq 4$. \\

\noindent
To close this section, we summarize the building blocks of gravitational $k$-form theory \eqref{eq:total-action} below:

\begin{summary}
The building blocks of GKFT are:
\begin{itemize}
    \item A dynamical $k$-form field $A\in \Omega^{k}(M;\mathfrak{g})$ \: ,
    \item A connection $B\in \Omega_{\text{con}}(M;\mathfrak{g})$ \: ,
    \item A covariant derivative $d_{B}:\Omega^{k}(M;\mathfrak{g})\rightarrow \Omega^{k+1}(M;\mathfrak{g})$ which acts only on the group part \: ,
    \item A spacetime metric $g\in \mathcal{PR}(M)$ or equivalently vielbeins $e \in \Omega^1(M;W)$ \: ,
    \item A metric compatible spacetime connection $\omega \in \Omega_{\text{con}}(M;\mathfrak{so}(n-1,1))$ \: ,
    \item A covariant derivative $d_{\omega}:\Omega^{k}(M;\mathfrak{so}(n-1,1))\rightarrow \Omega^{k+1}(M;\mathfrak{so}(n-1,1))$ which acts only on the spacetime part \: .
\end{itemize}
\end{summary}

\newpage
\subsection{Possible Extensions} \label{subsection:possible-extensions}

The field theory presented in the last section allows to study Lie algebra valued differential $k$-form fields on arbitrary spacetime manifolds and with internal symmetries described by semisimple compact Lie groups. \\
Hereafter, we explore possible extensions to the $k$-form field theory action \eqref{eq:general-k-form-action} and also to the combined theory action \eqref{eq:total-action}. Most of these involve adding terms to the respective actions. All additional terms to the action \eqref{eq:general-k-form-action} must be invariant under local group transformations in $G$. In addition, they must also be invariant under local $SO(n-1,1)$ transformations, if coupling to gravity is considered (as is the case in Eq. \eqref{eq:total-action}).

\subsubsection*{Terms Involving Only the Field $A$}

In general, the additional terms involving solely the field $A$ can be written as the action of a "potential form" $V_A(A) \in \Omega^{n}(M)$:

\begin{equation}
S_{A} = \int_{M} V_A(A) \: .
\end{equation}
Here, $V_{A}(A)\in \Omega^{n}(M)$ is the image of the mapping $V_{A}:\Omega^{k}(M;\mathfrak{g})\times\cdots \times\Omega^{k}(M;\mathfrak{g})\rightarrow \Omega^{n}(M)$ that maps multiple fields $A$ onto an $n$-form which is invariant under $G$ and $SO(n-1,1)$. To ensure invariance, one can construct the terms in the following way: Because $A$ transforms like an algebra element (see section \ref{subsection:generalizing-yang-mills-theory-using-differential-forms} and appendix \ref{sec:appendix C}), any trace $Tr_\mathfrak{g}$ over any combination of $A$-fields, which results in an $n$-form, will ensure invariance under $G$. To ensure invariance under $SO(n-1,1)$, the resulting term should also be a Lorentz scalar. A simple example for $V_A(A)$ is a mass term for the $k$-form field $A$, given by

\begin{equation}
V_A(A) = m_A^2 \; Tr_\mathfrak{g}[A\wedge \star_g A] \: ,
\label{eq:potential-mass-term}
\end{equation}
where $m_{A}$ is the mass of the field $A$.
In a similar manner, also potential forms that take more than two fields as arguments can be constructed. We give additional examples for potential forms $V_{A}(A)$ in Appendix \ref{sec:appendix D}.

\subsubsection*{Terms Involving Only the Connection $B$}

Due to the transformation properties of the connection $B$ \eqref{eq:transformation-property-connection} under transformations in $G$, building invariant terms is more involved compared to the case with the fields $A$. One possibility is to consider combinations of the curvature form $H=dB+\frac{1}{2}[B,B]$ instead. This combination must be chosen in a way that the resulting potential form $V_B(B)$ is an $n$-form and invariant under $G$ and $SO(n-1,1)$. We give some examples in Appendix \ref{sec:appendix D}. \\
Alternatively, it is also possible to add a mass term for the connection $B$ by using a generalized Stückelberg form (see \cite{Barros:2021tnn,PhysRevD.80.103509,Koivisto2009InflationFN}). This however only works if the group $G$ is also Abelian (i.e., the group elements commute). But we do not consider this case in this work.

\subsubsection*{Terms Involving the Field $A$ and the Connection $B$}

Such a term can be described by a potential form $V_{AB}(A,B)\in \Omega^{n}(M)$, which is invariant under local transformation in both $G$ and $SO(n-1,1)$. If the dimension $n=4$ and $A \in \Omega^{1}(M;\mathfrak{g})$, a valid term would be a combination of the field strength $F$ and the curvature form $H$:

\begin{equation}
V_{AB}(A,B) = Tr_\mathfrak{g}[F \wedge \star_g H] \: , \hspace{1cm} F = dA + [B,A] \in \Omega^{2}(M;\mathfrak{g}) \: . \label{eq:potential-form-AB-1}
\end{equation}
If $A$ is a $2$-form, a term of the following shape becomes valid:

\begin{equation}
V_{AB}(A,B) = Tr_\mathfrak{g}[A \wedge \star_g H] \: , \hspace{1cm} A \in \Omega^{2}(M;\mathfrak{g}) \: .
\label{eq:potential-form-AB-2}
\end{equation}
There is a large -- and possibly infinite -- amount of viable terms for the potential form $V_{AB}(A,B)$. The possibility of a specific term will also depend on the spacetime dimension $n$ and the degree of the $k$-form field $A$. We give some additional examples in the Appendix \ref{sec:appendix D}. \\

\noindent
The most general action for the $k$-form field theory (without gravitational terms) including potential terms can finally be written as
\begin{equation}
\begin{split}
S_{GKFT,AB}&= - \alpha \int_{M} Tr_{\mathfrak{g}}[F\wedge \star_g F] - \beta \int_{M} Tr_{\mathfrak{g}}[H\wedge \star_g H] - \int_M V_A(A) - \int_M V_B(B) - \int_M V_{AB}(A,B) \: .
\end{split}
\label{eq:most-general-action-potential}
\end{equation}
where $V_A(A)$, $V_B(B)$, $V_{A,B}(A,B)$ are the potential forms defined above.

\subsubsection*{Considering More Complex Symmetry Groups}

As stated in Assumption \ref{def: compact-group-assumption}, the Lie group $G$, which describes the local symmetry, must be a compact semisimple group. However, it is possible that the group $G$ can be constructed from a product of different groups, as long as the resulting group fulfills the aforementioned properties. One could thus consider multiple local symmetry groups (e.g. product groups of the form $SU(3) \times SU(2) \times (...) $. Physically, this setup can be used to study theoretical models for classical fields, which are sensitive to multiple local symmetries. \\
In addition to local symmetries, global symmetries may also be added to the theory. For example a global $U(1)$ symmetry could be included by making the field $A$ a complex valued differential form.

\subsubsection*{Modifying the Underlying Theory of Gravity}

Since the choice of the spacetime connection $\omega \in \Omega_{\text{con}}(M;\mathfrak{so}(n-1,1))$ is in principle arbitrary, it is possible to consider spacetimes which include torsion effects (see for example Einstein-Cartan gravity \cite{HehlReview76,Tecchiolli:2019hfe,Cattaneo:2023cnt}). One could also add terms of higher order in the curvature $R$, $R_{\mu\nu}$ and $R_{\mu\nu\rho\sigma}$ via $f(R)$ or Gauss-Bonnet gravity \cite{DeFelice:2010aj,Lobo:2008sg,Nojiri:2006ri,Nojiri:2010wj,Tsujikawa:2010zza}. A treatment in higher or lower dimensions is also possible by adding or removing wedge products of the vielbeins $e$ in the Palatini-Cartan term in equation \eqref{eq:total-action}. Note that in the Palatini-Cartan formulation, the lowest possible spacetime dimension is two, since the curvature form $\Omega_\omega$ is a $2$-form.

\subsubsection*{Matter Terms}

A matter action $S_{matter}$ usually describes some kind of fluid, for example an ideal fluid of nuclear matter (as is regularly done in the case of neutron stars). The action itself can be written as
\begin{align}
    S_{matter} = \int_M \sqrt{-g} \: \mathscr{L}_{matter} \: dx^n \: , \label{eq:matter-action}
\end{align}
with $\mathscr{L}_{matter}$ being the Lagrangian density of a matter fluid. When considering such matter terms in our framework, we note that $\mathscr{L}_{matter}$ must be chosen so that it respects both the spacetime symmetry $SO(n-1,1)$ and the internal group symmetry $G$. \\
It is possible to couple a (fermionic) matter term \eqref{eq:matter-action} minimally to the (bosonic) $k$-form action \eqref{eq:total-action}. This action can describe self-gravitating hybrid objects which consist of both fermions and bosons with internal symmetries. Some of these mixed systems have been studied in the past and are called fermion boson stars (see for example \cite{Henriques:1989ez,Kouvaris:2010vv,Liebling:2012fv,Bertone:2007ae,Zurek:2013wia,Kouvaris:2013awa,Diedrichs:2023trk,DiGiovanni:2021ejn,Jockel:2023rrm}). Although most of the models studied feature bosonic fields with global $U(1)$ symmetries, versions with gauge bosons have also been investigated, see \cite{daRocha:2020jdj,Bartnik:1988am,Schunck:2003kk,Brihaye:2004nd,Dzhunushaliev:2006sy,Brihaye:2009hf,Soni:2016gzf,Martinez:2022wsy,Jain:2022kwq,Soni:2016yes}. It is also conceivable to take excitations of branes from type IIA superstring theory \cite{Polchinski_1995,Polchinski_1996,Witten_1996,Brodie_1997}, described by $k$-form fields, as the bosonic part of the fermion boson star.

\newpage
\section{Derivation of Equations of Motion} \label{sec:derivation-of-EOM}
In this section, we derive the equations of motion that emerge from the KFT actions \eqref{eq:general-k-form-action} and \eqref{eq:total-action}. We do this by computing the variation of the action integral and imposing that the variation vanishes. We give the concrete form of the resulting field-equations in differential-form notation as well as in index notation. Thereafter, we analyze their mathematical properties. \\
We also explore the case where the $k$-form action is coupled to gravity (see \eqref{eq:total-action}). The coupling to gravity introduces a new set of equations of motion, the Einstein equations. We then discuss the steps to derive an energy-momentum tensor associated to the $k$-form field $A$ and connection $B$.

\subsubsection*{Computing the Variation}

We start by defining the variation with respect to a differential $k$-form $A \in \Omega^k (M; \mathfrak{g})$ on an $n$-dimensional semi-Riemannian manifold $M$. To do this we use the fact that the space in which the fields live is locally trivial. This is due to the G principal bundle structure of our theory (see Def. \ref{def:g-principial-bundle}). The total space of the G-principial bundle factorizes locally to a product space of $SO(n-1,1)$ and $G$. We can make use of this property to write the field $A$ and the connection $B$ as a linear combination of generators $T_a \in \mathfrak{g}$:

\begin{equation}
A(x)=A^a(x)\tens{} T_a \: , \hspace{1.0cm} \text{for every point $x\in M$} \: . \label{eq:k-form-decompositon-with-generators}
\end{equation}
Summation over repeated indices is implied and the sum runs over the dimension of the Lie algebra so that $a \in \{1,\cdots, dim(\mathfrak{g})\}$. It is now possible to define the variation pointwise with respect to the $k$-form field components $A^{a}\in \Omega^{k}(M)$ as:

\begin{equation}
\delta_A S = \lim_{\epsilon\rightarrow 0} \frac{ S[A^a + \epsilon \, \delta A^a] - S[A^a]}{\epsilon} \: . \label{eq:variation-of-a-k-form}
\end{equation}
In this definition of the variation, we only vary the field $A$ and treat the connection $B$ as an independent field with a separate equation of motion. This is similar to the first order formalism in general relativity (see e.g. \cite{Corichi:2016zac,osti_6608582}). It is therefore possible to analogously define a variation with respect to $B$. \\

\noindent
We now present the computation of the variation of the KFT part of the the total GKFT action \eqref{eq:total-action}. For all explicit calculations presented here, we also ignore additional potential terms (see section \ref{subsection:possible-extensions}) to the action \eqref{eq:total-action}. But the calculations presented in this section can easily be generalized to incorporate arbitrary potential terms of the field $A$ and connection $B$. To simplify notation, we hereafter write $d$ instead of $d_\omega$ for the covariant derivative relative to the spacetime connection $\omega \in \Omega_{con}(M;\mathfrak{so}(n-1,1))$. \\
We first express the KFT part of the action \eqref{eq:total-action} in terms of the field components $A^a(x)$ and the components of the connection $B^b(x)$. This leads to:

\begin{align}
\begin{split}
    S_{AB} = &- \alpha \int_M Tr_\mathfrak{g} \left(d A^a \wedge \star_g d A^b \tens{} T_aT_b + 2(B^b \wedge A^c)\wedge \star_g d A^a \tens{} [T_b,T_c]T_a \right. \\
     &+ \left. (B^a \wedge A^b) \wedge \star_g (B^c \wedge A^d)\tens{} [T_a,T_b][T_c,T_d] \right) \\
    &- \beta\int_M Tr_\mathfrak{g} \left(d B^a \wedge \star_g d B^b \tens{} T_a T_b + (B^b \wedge B^c)\wedge \star_g d B^a \tens{} [T_b,T_c] T_a \right. \\
     &+ \left. \frac{1}{4} (B^a \wedge B^b) \wedge \star_g (B^c \wedge B^d)\tens{} [T_a,T_b][T_c,T_d] \right) \: , \label{eq:k-form-action-ausgeschrieben}
\end{split}
\end{align}
where the trace $Tr_\mathfrak{g}$ only acts on the Lie algebra elements $T_a \in \mathfrak{g}$. The Hodge star $\star_g$ is defined with respect to the metric $g$ associated to the manifold $M$. We also used that the scalar product (see Def. \ref{def:integral-over-differential-forms}) is symmetric to simplify the expression. We then perform the variation with respect to $A^a$ as defined in Equation \eqref{eq:variation-of-a-k-form} and use Stokes theorem (see Def. \ref{theorem:strokes-theorem}, Eq. \eqref{eq:stokes-useful-relation-at-infty}) to simplify the integrals. Note that we explicitly assume that the field components $A^a(x)$ vanish at the boundary $\partial M$, which we set to be positioned at infinity. \\
After taking the limit $\epsilon \rightarrow 0$ and cancelling the remaining terms, we obtain the following expression for the variation 
$\delta_A$ of the action 

\begin{equation}
\begin{split}
    \delta_A S_{AB} &= -2 \alpha (-1)^{k} \int_M Tr_\mathfrak{g} \left[  \delta A^b \wedge \left\{ - \left( d\star_g dA^a \right) \tens{} T_a T_b \right. \right. \\ 
    &\left. - \left( d \star_g (B^a \wedge A^c) \right) \tens{} T_b[T_a,T_c] + \left( B^c \wedge \star_g d A^a   \right) \right. \tens{} T_b[T_c,T_a] \\
    &+ \left( B^a \wedge \star_g ( B^d \wedge A^c ) \right) \left. \left. \tens{} \, [T_a,T_b][T_d,T_c] \right\} \right] \: .
\end{split}
\label{eq:variation-wrt-to-A}
\end{equation}
For a step by step version of this derivation, we refer to Appendix \ref{sec:Appendix B}.

\subsubsection*{Equations of Motion}

By enforcing equation \eqref{eq:variation-wrt-to-A} to vanish for arbitrary variations of the fields (i.e. $\delta_A S_{AB} = 0$ for all $\delta A^a \neq 0$), we obtain a set of field equations for the field $A$:

\begin{align}
Tr_\mathfrak{g}[ \, (d \star_g F+[B,\star_g F]) T_a \, ] = 0 \: . \label{eq:field-equation-for-A}
\end{align}
Note that $F$ and $B$ are matrix-valued in the Lie algebra $\mathfrak{g}$ (see Eq. \eqref{eq:k-form-decompositon-with-generators}). Multiplying them with the generators $T_a \in \mathfrak{g}$ is therefore equivalent to a matrix multiplication. We then take the trace over the result. One can also write the trace as a sum over matrix components:

\begin{align}
 ( d \star_g F_{ij} + [B, \star_g F]_{ij} )\, (T_a)_{ji} = 0 \: .
\end{align}
The indices $i$ and $j$ denote the rows and columns of some matrix representation of the Lie algebra $\mathfrak{g}$ (see Def. \ref{def:representations}). Summation over both indices is implied. \\
When also considering general potentials, the equations of motion become:

\begin{equation}
Tr_\mathfrak{g}\left[\,(d\star_g F+[B,\star_g F])T_{a}\right] = \sum_{i}^{} \xi_{A}^{(i)}\; \frac{\delta V_{A}^{(i)}(A)}{\delta A^a} + \sum_{i}^{} \xi_{AB}^{(i)}\; \frac{\delta V_{AB}^{(i)}(A,B)}{\delta A^{a}} \: .
\end{equation}
The parameters $\xi^{(i)}_A$ and $\xi^{(i)}_{AB}$ are constants. \\

\noindent
In addition to the field equations for the field $A$, we also obtain field equations for the connection $B$ by varying the action \eqref{eq:k-form-action-ausgeschrieben} with respect to $B^{a}(x)$. The variation is defined in the same way as for the field $A$ in \eqref{eq:variation-of-a-k-form}. The resulting field equation for $B$ is given by:

\begin{equation}
Tr_\mathfrak{g}\left[ \left(\,\beta d\star_g H + \beta [B,\star_g H]\right) T_a \right] = Tr_{\mathfrak{g}}\left(\,\alpha [A,\star_g F]\, T_a\right) \: . \label{eq:field-equation-for-B}
\end{equation}
When writing the trace using index notation, we obtain:

\begin{equation}
 \left(\,\beta d\star_g H_{ij} + \beta [B,\star_g H]_{ij}\right) (T_a)_{ji} = \alpha [A,\star_g F]_{ij} (T_a)_{ji} \: .
\end{equation}
Here also we can consider general potentials. The equations of motion become:

\begin{equation}
Tr_\mathfrak{g}\left[ \left(\,\beta d\star_g H + \beta [B,\star_g H]\right) T_a \right] = Tr_{\mathfrak{g}}\left(\,\alpha [A,\star_g F]\, T_a\right) + \sum_{i}^{} \zeta_{B}^{(i)}\; \frac{\delta V_{B}^{(i)}(B)}{\delta B^a} + \sum_{i}^{} \zeta_{AB}^{(i)}\; \frac{\delta V_{AB}^{(i)}(A,B)}{\delta B^{a}} \: .
\end{equation}
The parameters $\zeta^{(i)}_B$ and $\zeta^{(i)}_{AB}$ are constants.

\subsubsection*{Equations of Motion Including Coupling to Gravity}

As of now, we have derived a set of equations, which describe the dynamics of the $k$-form field $A$ and the connection $B$. To describe the dynamics of $k$-form field theory coupled to gravity, the full system of equations \eqref{eq:total-action} must be considered. It includes minimal coupling terms between gravity and the $k$-form field. This gives rise to an additional set of equations, the Einstein equations. They describe the gravitational interaction between matter and spacetime. \\
The Einstein equations with a matter term are described using an energy-momentum tensor $T$. The effective energy momentum tensor of the field $A$ can be calculated by varying the action (\ref{eq:general-k-form-action}) with respect to the metric tensor $g$ (or alternatively w.r.t. the vielbein fields). The general form of the components of an energy-momentum tensor is

\begin{equation}
T_{\mu\nu}^{AB} := - \frac{2}{\sqrt{-g}}\frac{\delta S_{AB}}{\delta g^{\mu\nu}} \: . \label{eq:definition-energy-momentum-tensor}
\end{equation}
This energy momentum tensor is used on the right hand side of the Einstein equations. \\ 
The complete set of equations to describe the gravitating $k$-form field with the Levi-Civita connection for the spacetime connection is:

\begin{subequations}
    \begin{align}
G_{\mu\nu} &= \kappa \; T_{\mu\nu}^{AB} \: , \label{eq:complete-k-form-star-equation-1} \\
 (d \star_g F_{ij} + [B, \star_g F]_{ij} )\,  (T_a)_{ji} &= 0 \: , \label{eq:complete-k-form-star-equation-2} \\
  \left(\,\beta d\star_g H_{ij} + \beta [B,\star_g H]_{ij}\right) (T_a)_{ji} &= \alpha [A,\star_g F]_{ij}\, (T_a)_{ji} \: . \label{eq:complete-k-form-star-equation-3}
\end{align}
\end{subequations}
Equation \eqref{eq:complete-k-form-star-equation-1} describes the curvature of spacetime, whose energy-momentum content is described by $T_{\mu\nu}^{AB}$. Equations (\ref{eq:complete-k-form-star-equation-2}) and (\ref{eq:complete-k-form-star-equation-3}) describe the dynamics of the $k$-form field $A$ and the connection $B$ respectively. \\

\noindent
The equations of motion of $k$-form theory discussed above are essentially covariant derivatives set equal to a source term. This is similar to e.g. electrodynamics where the derivative of the field strength tensor is set equal to a charge current. This analogy opens up the possibility to interpret the right hand sides of the equations of motion as source currents.

\subsubsection*{Degrees of Freedom of the Equations of Motion}

We now analyze the full system of field equations \eqref{eq:complete-k-form-star-equation-1}-\eqref{eq:complete-k-form-star-equation-3} regarding their degrees of freedom. The components of the fields involved are the metric components $g_{\mu\nu}$ (or the vielbeins $e$), the components of the $k$-form field $A^a_{\mu_1 ... \mu_k}$ and the components of the connection $B^a_\mu$. In $n$ dimensions, the symmetric rank-2 tensor $g$ has a total of $\frac{n(n+1)}{2}$ independent components. A general $k$-form has $\begin{pmatrix} n \\ k \end{pmatrix}$ independent components. A Lie algebra valued $k$-form then also has additional degrees of freedom for each generator of the Lie algebra (because the generators form a basis of the Lie algebra). This leads to $\text{dim}(\mathfrak{g})$ independent components for each component of the $k$-form. The connection form has a number of $n \cdot \text{dim}(\mathfrak{g})$ independent components. This leads to a total number of free components $N_\mathrm{free}$ of

\begin{align}
    N_\mathrm{free} = \frac{n(n+1)}{2} + \left[ \begin{pmatrix} n \\ k \end{pmatrix} + n \right] \cdot \text{dim}(\mathfrak{g}) \: . \label{eq:number-of-free-components}
\end{align}
When counting the total number of equations (\ref{eq:complete-k-form-star-equation-1} - \ref{eq:complete-k-form-star-equation-3}) one obtains $\frac{n(n+1)}{2}$ equations for Eq. (\ref{eq:complete-k-form-star-equation-1}), $\begin{pmatrix} n \\ n-k \end{pmatrix} \cdot \text{dim}(\mathfrak{g})$ equations for Eq. (\ref{eq:complete-k-form-star-equation-2}) and $\begin{pmatrix} n \\ n-1 \end{pmatrix} \cdot \text{dim}(\mathfrak{g})$ equations for Eq. (\ref{eq:complete-k-form-star-equation-3}). Using the properties of the binomial coefficient, it is easy to show that the number of equations is equal to the number of independent components (\ref{eq:number-of-free-components}). \\

\noindent
We summarize the amounts of degrees of freedom in $n=4$ dimensions for different degrees of the $k$-form field and different symmetry groups $G$:

\begin{table}[h!]
\centering
\adjustbox{max height=\dimexpr\textheight-0.2cm\relax,
          max width=\textwidth}{
\renewcommand{\arraystretch}{1.2}
\begin{tabular}{|p{2cm}||p{3cm}|p{3cm}| p{3cm}|p{3cm}|} 
 \hline
 $G$& $k=0$ & $k=1$ & $k=2$ & $k=3$ \\ 
 \hline
     $SU(2)$ & $25$      & $34$ & $40$ & $34$ \\ 
     $SU(3)$ & $50$      & $74$ & $90$ & $74$ \\ 
     $SU(N)$ & $5(N^2+1)$ & $2 + 8N^2$ & $10 N^2$ & $2 + 8N^2$ \\
     $SO(N)$ & $10 + 5(N^2-N)/2$ & $10 + 4(N^2-N)$ & $10 + 5(N^2-N)$ & $10 + 4(N^2-N)$ \\
 \hline
\end{tabular}
}
\caption{Total number of equations for gravitational $k$-form theory in $n=4$ dimensions with a given inner group symmetry $G$. Note that $k$ is restricted by the dimension and fulfills $k \leq n-1$.}
\label{table:1}
\end{table}

\noindent
Also note that the number of degrees of freedom can decrease drastically when choosing a concrete ansatz for the spacetime and the $k$-form fields. The choice of the connection (e.g. whether it can be derived from other quantities) also has an influence on the total number of degrees of freedom. This will be addressed in more detail in sections \ref{sec:connection-for-Lie-groups} and \ref{sec:solving-the-equations-of-motion}.

\newpage
\section{Fixing the Group Connection Using a Spacetime Connection} \label{sec:connection-for-Lie-groups}
In sections \ref{sec:formulation-and-motivation-of-action} and \ref{sec:derivation-of-EOM}, we discussed $k$-form theory and derived the equations of motion for the $k$-form field $A\in \Omega^k(M;\mathfrak{g})$ and the connection $B\in \Omega_{\text{con}}(M;\mathfrak{g})$. The field $A$ is a degree of freedom but the connection $B$ needs yet to be chosen. The choice of the connection is in general arbitrary, but usually we choose a connection with properties that simplify calculations. One choice is to regard $B$ as a dynamical field with separate equations of motion (see Eq. \eqref{eq:field-equation-for-B}). This allows us to interpret $B$ as another degree of freedom alongside the field $A$. \\
In this work however, we are interested in the case where the connection $B$ is not a dynamical field. We do this because our aim is mainly to study $k$-form fields with internal symmetries and not connections. We do this with future applications in mind, for example to describe boson stars of a single dynamical $k$-form field $A$ with internal symmetry. \\
We here choose the connection $B$ such that it is solely determined by the background spacetime and the internal symmetry group. We develop the general procedure to compute the connection coefficients of $B$. Then, we compute the coefficients explicitly for a static spherically symmetric spacetime and an internal $SU(2)$ internal symmetry. We further choose the Levi-Civita connection as the spacetime connection in this example.

\subsection{General Procedure}

To develop an algorithm for computing the connection coefficients, we make use of a mathematical analogy between the connection $B$ and the spin connection introduced in (Def. \ref{def:spin-connections}). We derive this analogy by considering the torsion tensor $Q = \overline{e} \cdot (d_\omega e) = \overline{e}_a (d_\omega e)^a$, with respect to a connection $\omega$ and (dual) vielbeins ($\Bar{e}$) $e$ (see Def. \ref{def:Tetrads}), is given by:

\begin{align}
    \begin{split}
        \overline{e}_a (d_\omega e)^a &= \overline{e}_a^\sigma \left( \partial_{[\mu} e^a_{\nu ]} + \omega^a_{[\mu c} e^c_{\nu ]} \right) dx^\mu \wedge dx^\nu \tens{} \partial_\sigma \: , \\
    &= \frac{1}{2} \left( \Gamma^\sigma_{\mu\nu} - \Gamma^\sigma_{\nu\mu} \right) \, dx^\mu \wedge dx^\nu  \tens{} \partial_\sigma  \: , \label{eq:torsion-tensor-christoffel-symbols}
    \end{split}
\end{align}
where $\Gamma^\sigma_{\mu\nu} = \overline{e}_a^\sigma ( \partial_{\mu} e^a_{\nu} + \omega^a_{\mu c} e^c_{\nu} )$ are the affine connection coefficients (Christoffel symbols) induced by the connection $\omega$. The coefficients $\omega^a_{\mu c}$ are the coefficients of the corresponding spin connection. \\
One can now perform the same calculation as above, but for the connection $B$ of $k$-form theory, and obtain:

\begin{align}
    \begin{split}
        \overline{e}_a (d_B e)^a &= \overline{e}_a^\sigma \left( \partial_{[\mu} e^a_{\nu ]}+ B^b_{[ \mu} e^c_{\nu ]} [T_b,T_c]^a \right) dx^\mu \wedge dx^\nu \tens{} \partial_\sigma \: , \\
        &= \frac{1}{2} \overline{e}_a^\sigma \left( \partial_{\mu} e^a_{\nu } - \partial_\nu e^a_\mu  + i f^a_{\:\, bc} \left( B^b_{\mu} e^c_{\nu} - B^b_\nu e^c_\mu \right)  \right) \, dx^\mu \wedge dx^\nu \tens{} \partial_\sigma \: . \label{eq:torsion-tensor-christoffel-symbols-B-connection}
    \end{split}
\end{align}
We here used the commutator property of the group generators (see the paragraph below Def. \ref{def:generators}) and have expanded the anti-symmetrization operator. In the same way as the connection $\omega$, the connection $B$ also induces an affine connection and a spin connection. This allows us to define the affine connection coefficients in a similar manner as in \eqref{eq:torsion-tensor-christoffel-symbols}: $\Gamma^\sigma_{\mu\nu} := \overline{e}_a^\sigma ( \partial_{\mu} e^a_{\nu} + i f^a_{\:\, bc} B^b_{\mu} e^c_{\nu} )$. When comparing equations (\ref{eq:torsion-tensor-christoffel-symbols}) and (\ref{eq:torsion-tensor-christoffel-symbols-B-connection}), we can see that the spin connection coefficients $\omega^a_{\mu c}$, induced by the connection $B$, and the coefficients $B_\mu^a$ are related by:

\begin{align}
    \omega^a_{\mu c} \equiv i f^a_{\:\, bc} B^b_{\mu} \: . \label{eq:analogy-B-spin-connection}
\end{align}
Note that the connection coefficients are non-tensorial and can be zero or nonzero solely based upon the choice of the local frame and coordinate system. Once the coefficients $\omega^a_{\mu b}$ are known, one can also compute the connection coefficients $B_\mu^a$ using the relation \eqref{eq:analogy-B-spin-connection}. We stress that \eqref{eq:analogy-B-spin-connection} is just another way of expressing the connection coefficients of the group connection $B$ via a spin-connection. Later, we choose the connection $B$ in a way so that the affine connection-coefficients induced by $B$ correspond to the affine connection coefficients induced by the spacetime connection. Through this prescription, we effectively fix the group connection using the spacetime connection. This is a choice we make in this work to ensure that the connection $B$ is not a degree of freedom and only depends on the background spacetime and the local symmetry group. \\

\noindent
To compute the coefficients $\omega^a_{\mu b}$ we can invert the relationship $\Gamma^\sigma_{\mu\nu} = \overline{e}_a^\sigma ( \partial_{\mu} e^a_{\nu} + \omega^a_{\mu c} e^c_{\nu} )$ and obtain:

\begin{align}
\omega_{\mu b}^{a} \: = \bar{e}_{\: b}^{\nu} \: \Gamma^{\alpha}_{\mu \nu} \: e_{\alpha}^{\: a} - \bar{e}_{\: b}^{\nu} \: \partial_{\mu} \: e_{\nu}^{\: a} \: . \label{eq:inverted-expression-spin-connection}
\end{align}
If the connection has no torsion, the spin connection coefficients can be computed using the relation
\begin{align}
    \omega_{\mu}^{\: ab} = \frac{1}{2} e^{\nu a} (\partial_\mu e_\nu^{\: b} - \partial_\nu e_\mu^{\: b}) - \frac{1}{2} e^{\nu b}(\partial_\mu e_\nu^{\ a}-\partial_\nu e_\mu^{\ a}) - \frac{1}{2} e^{\rho a} e^{\sigma b}(\partial_\rho e_{\sigma c} - \partial_\sigma e_{\rho c})e_\mu^{\: c} \: .
\end{align}
Alternatively, one can also derive the spin connection coefficients in \eqref{eq:inverted-expression-spin-connection} using the first Cartan structure equation (see \cite{Tecchiolli:2019hfe}). Both approaches have been shown to give equivalent results. \\
Note that in \eqref{eq:inverted-expression-spin-connection}, the spacetime indices run over $\mu,\nu,\alpha \in \{0,1,...,n-1 \}$ and the group indices assume values $a,b \in \{1, ... , dim(\mathfrak{g}) \}$. \\

\noindent
We now discuss a crucial point regarding the vielbein fields $e$ that appear in the equations above. Vielbeins $e$ are mappings between a tangent bundle and a vector bundle. In our case, the tangent bundle is the tangent bundle of the spacetime manifold and the vector bundle is the Lie algebra of the symmetry group. This is why they can be thought of as a mapping between the global spacetime and the local symmetry space. \\
Given an $n$-dimensional semi-Riemannian manifold, vielbeins can be defined using the global metric $g$ and the local metric $K$ (see Eq. \ref{eq:vielbein-definition-local-and-globals-metric}). In component notation:

\begin{align}
    g_{\mu\nu} = e^a_\mu \, K_{ab} \, e^b_\nu \: , \label{eq:def-tetrads-in-liealgebra-local-basis}
\end{align}
where $g_{\mu\nu}$ and $K_{ab}$ are the coefficients of the metric of the global and local space, respectively. Also $\mu,\nu \in \{0,1,...,n-1 \}$ and $a,b \in \{1, ... , dim(V) \}$. In the case of general relativity (see Ex. \ref{example:tetrads-in-GR}), both the local space and the global space have the same number of dimensions. Vielbeins can then be represented using $n \times n$ matrices. \\
However, it is also possible that the global space and the local space $V$ have a different dimension. The vielbeins then would have $n \times dim(V)$ elements and are thus representable by rectangular matrices. This introduces the problem that in general, the inverse vielbeins $\overline{e}$ do not exist and that the vielbeins in general do not form a complete orthonormal basis of the local frame.
Nevertheless, it is still possible to formally define the coefficients of these "rectangular vielbeins" using equation (\ref{eq:def-tetrads-in-liealgebra-local-basis}). \\

\noindent
The non-invertibility of the vielbeins $e$ and $\Bar{e}$ can be resolved by applying a technique that is similar to the one used by the author of \cite{Volovik:2022dxw}: Depending on whether $dim(V)$ is smaller or larger than the dimension $n$ of the global space, one can complete the basis by adding dimensions to either the global or the local space, so that the vielbeins have a quadratic shape. These "extra dimensions" are a mathematical trick to make the vielbeins well-behaved. They should have no effect on the physics of the system. In particular, this reasoning can also be applied to Lie groups and their algebras, which can have different dimensionality than the spacetime manifold in which they are embedded. \\

\noindent
The last remaining step is to find the connection-coefficients of the affine connection $\Gamma^\sigma_{\mu\nu}$. The exact form and shape of the connection coefficients will depend on the connection chosen for the spacetime. A connection must be chosen for both the spacetime manifold and for the internal symmetry group. In this work, we make a choice for the connection of the inner symmetry space so that it depends on the tetrads (vielbeins) and the connection coefficients of spacetime, i.e. equation \eqref{eq:inverted-expression-spin-connection}. \\

\noindent
For example, the Levi-Civita connection may be chosen as the spacetime connection. In that case, this choice for the connection of the spacetime manifold entails vanishing torsion and compatibility with the spacetime metric. This way, the well known expression
\begin{align}
    \Gamma^\sigma_{\mu\nu} = \frac{1}{2} g^{\sigma \alpha} \left( \partial_\mu g_{\nu \alpha} + \partial_\nu g_{\mu \alpha} - \partial_\alpha g_{\mu \nu} \right) \label{eq:christoffel-symbols-levi-civita}
\end{align}
may be used to compute the Christoffel symbols. The coefficients $\omega_{\mu b}^a$ are then computed through equation \eqref{eq:inverted-expression-spin-connection}. We finally compute the connection coefficients $B_\mu^a$ through equation \eqref{eq:analogy-B-spin-connection}. Throughout the remainder of this work, we will assume the Levi-Civita connection for the spacetime manifold so that the expression \eqref{eq:christoffel-symbols-levi-civita} may be used. \\

\noindent
Before we move on to compute the connection coefficients $B_\mu^a$ explicitly, we consider the interpretation of the connection $B$. The connection $B$ can be thought of as a connection in the Lie group space that also respects the curved spacetime background. When computing e.g. a parallel transport or a covariant derivative of a Lie valued differential form, this connection acts on both the spacetime and the group components of the differential form.

\subsection{Spacetime Connection for SU(2)} \label{sec:connection-for-SU(2)}
In this section, we use the previously discussed procedure to compute the connection coefficients $B_\mu^a$ explicitly for the symmetry group $SU(2)$ and a spherically symmetric spacetime. The metric components depend solely on the radial coordinate $r$:

\begin{equation}
    ds^{2}=g_{tt}(r)  dt^{2} + g_{rr}(r) dr^{2} + r^{2} (d\theta^2 + \sin^2 \theta d\phi^2) \: .
    \label{eq: spherical metric}
\end{equation}
The vielbein components $e_\mu^a$ and the dual vielbein components $\overline{e}^\mu_a$ are determined by their definition:

\begin{equation}
g_{\mu\nu}=e_{\mu}^{a}K_{ab} e_{\nu}^{b} \:\: ,\:\: g^{\mu\nu}=\overline{e}^{\mu}_{a}K^{ab} \overline{e}^{\nu}_{b} \: .
\end{equation}
Here $K_{ab} = - N \delta_{ab}$ (with $N=2$) is the Cartan-Killing metric tensor of the Lie algebra $\mathfrak{su}(2)$. $K^{ab} = - \frac{1}{N} \delta^{ab}$ is the inverse metric. The choice of vielbeins is not unique and depends upon the chosen local basis. In this work we pick an orthonormal basis (i.e. $e_\mu^a \, \overline{e}^\mu_b = \delta^a_b$), where only the spatial components of the tetrads are non-vanishing. Since $dim(\mathfrak{su}(2)) = 3$, we complete the basis by the additional vielbein group-components $e^0_\mu$ and $\overline{e}^\mu_0$. The components are given by:

\begin{equation}
e^{\:a}_{j}=i\: \sqrt{\frac{g_{jj}}{N}}\delta^{\: a}_{j},\hspace{2cm} e_{\mu}^{\:0}=i\: \sqrt{\frac{g_{tt}}{N}}\delta^{0}_{\mu} \: , \label{eq:tetrad-su2-zero-components}
\end{equation}
where the indices are in the range $j \in \{ r, \theta,\phi\}$, $\mu \in \{ t, r, \theta,\phi\}$ and $a \in \{1,2,3\}$. Note that there is no summation performed over $j$ in Eq. \eqref{eq:tetrad-su2-zero-components}. We here choose the ranges of the indices $j,\mu,a$ such that the spatial spacetime components are mapped to the $\mathfrak{su}(2)$ Lie algebra components. The non-zero components of the Kronecker-deltas therefore are $\delta_t^0, \delta_r^1, \delta_\theta^2, \delta_\phi^2$. The corresponding dual tetrads are given by:

\begin{equation}
\overline{e}^{j}_{\: a}=-i\: \sqrt{\frac{N}{g_{jj}}}\delta^{j}_{\:a},\hspace{2cm} \overline{e}_{\: 0}^{\mu}=-i\: \sqrt{\frac{N}{g_{tt}}}\delta^{\mu}_{\:0}
\end{equation}
The spatial spin connection components $\omega_{j\: b}^{a}$ can be calculated accordingly as:

\begin{equation}
\omega_{j\: b}^{a} \: = \overline{e}_{\: b}^{k} \: \Gamma^{i}_{j k} \: e_{i}^{\: a} - \overline{e}_{\: b}^{k} \: \partial_{j}\:e_{k}^{\: a}
\end{equation}
\\
The only non-vanishing components are:

\begin{align}
    \omega_{\theta 2}^{1}\: = -\omega_{\theta 1}^{2} = -\frac{1}{\sqrt{g_{rr}}} \:\:\: , \:\:\:
    \omega_{\varphi 3}^{2}\: = -\omega_{\varphi 2}^{\:3} = - \cos \theta \:\:\: , \:\:\:
    \omega_{\varphi 3}^{\:1}\: = -\omega_{\varphi 1}^{\:3} = - \frac{\sin\theta}{\sqrt{g_{rr}}} \: . 
\end{align}
Note that all the group factors cancel out. When considering groups with $dim(\mathfrak{g}) > 3$ the coefficients of the spin connection will be more involved because one would need to expand the spacetime by additional extra dimensions. It is essential to note that this procedure is merely a mathematical trick and should first and foremost not impact any physical results. Moreover, the extension of  spacetime dimensions can be realized in the most simplistic geometry which for spherical symmetry is a $\mathbb{S}^{dim(\mathfrak{g})-n}$ sphere. The spacetime metric then acquires additional $\text{dim}(\mathfrak{g})-n$ multiple non-zero components. For spherically symmetric spacetimes this dimensional extension can be parameterized by $dim(\mathfrak{g})-n$ additional angles $\theta_{i}$ (see Appendix \ref{sec:appendix a} for an example where the symmetry group is $SU(3)$). \\
It is now possible to use equation (\ref{eq:analogy-B-spin-connection}) to compute the connection coefficients $B_\mu^a$:

\begin{subequations}
    \begin{align}
\omega_{\theta 2}^{1} &= i f^{1}_{\:\: a 2} \: B_{\theta}^{a}=i f^{1}_{\:\: 3 2} \: B_{\theta}^{3} = -i B_{\theta}^{3}, \hspace{2cm} B_{\theta}^{3} = -\frac{i}{\sqrt{g_{rr}}} \: , \\
\omega_{\varphi 3}^{2} &= i f^{2}_{\:\: a 3} \: B_{\varphi}^{a}=if^{2}_{\:\: 1 3} \: B_{\varphi}^{1} = -i B_{\varphi}^{1} , \hspace{2cm} B_{\varphi}^{1} = -i \cos\theta \: , \\
\omega_{\varphi 1}^{3} &= i f^{3}_{\:\: a 1} \: B_{\varphi}^{a} = i f^{3}_{\:\: 2 1} \: B_{\varphi}^{2} = -i B_{\varphi}^{2} , \hspace{2cm} B_{\varphi}^{2} = i \frac{\sin \theta}{\sqrt{g_{rr}}} \: .
\end{align}
\end{subequations}
All other coefficients of $B_\mu^a$ are zero.

\newpage
\section{Solving the Equations of Motion} \label{sec:solving-the-equations-of-motion}
In this chapter, we apply the general equations of motion (\ref{eq:complete-k-form-star-equation-1})-(\ref{eq:complete-k-form-star-equation-3}), derived in the previous section, to fixed $k$-form degrees. We first translate the equations from differential form notation to equations in full index notation. The aim is to make the resulting equations more intuitive to understand. We explicitly derive the equations of motion in index notation for the $0$-form field and highlight central algebraic steps. Further, we derive the corresponding energy-momentum tensor components (see Eq. \ref{eq:definition-energy-momentum-tensor}). Explicit derivations for higher order $k$-form fields are omitted and we show only the equations of motion in index notation. For the case $k=0$, we choose a concrete ansatz for the spacetime and the $0$-form field. From that, we derive a system of partial differential equations, which can then be solved numerically.

\subsection{0-Form Field} \label{sec:derivation-of-0-form-equations}

In this subsection, we translate the equations of motion from section \ref{sec:derivation-of-EOM}, for both the Lie algebra valued differential form field $A$ and for the connection $B$, to index notation. Here, we consider only the case where the field is a $0$-form (i.e. a scalar) and thus $A = \phi^a \tens{} T_a$. \\
Throughout the derivations, we show an adequate amount of in-between algebra steps. We do this so that these calculations can serve as a reference for future scientists. We go into more detail here because in later sections, we will omit detailed calculations. The general steps are essentially the same for any $k$-form field an can be applied analogously. We do not consider potential form terms in this derivation, but they can easily be handled using analogous calculations. We adopt the same notation, $d$ instead of $d_\omega$, for the covariant derivative relative to the spacetime connection $\omega \in \Omega_{con}(M;\mathfrak{so}(n-1,1))$ as in section \ref{sec:derivation-of-EOM}. \\
\noindent
We start by expressing the field strength $F$ in index notation. Since $A$ is a $0$-form, $F$ will be a $1$-form:

\begin{subequations}
\begin{align}
    F :=& d_B A = d A + [B,A] = \nabla_\mu (A) dx^\mu + [B_\mu dx^\mu, A] \: ,  \\
    =& \left( \nabla_\mu \phi^a T_a + B^b_\mu \phi^c [T_b, T_c] \right) dx^\mu \: , \\
    =& \left( \nabla_\mu \phi^a T_a + i B^b_\mu \phi^c f^a_{\:\,bc} T_a \right) dx^\mu \: , \\
    =& F_\mu^a T_a dx^\mu \: .
\end{align}
\end{subequations}
Here, we have used the commutator relation for two Lie algebra valued differential forms (see Eq. \eqref{eq:lie-bracket-of-differential-forms}). We also evaluated the commutator between two generators $[T_a,T_b] = i f^c_{\:\,ab} T_c$. \\
Therefore, the components of the field strength can be compactly written as

\begin{align}
     F_\mu^a  = \nabla_\mu \phi^a + i f^a_{\:\,bc} B_\mu^b \phi^c \: . \label{eq:field-strength-0-form-index-notation}
\end{align}

\subsubsection*{Equations of Motion for the Field $A$}

We first consider the field equations for $A$ \eqref{eq:field-equation-for-A}. We rewrite the terms using their differential-form-basis and the generators of the symmetry group:
\begin{subequations}
\begin{align}
    0 &= d \left( \star_g F_{ij} \right) (T_a)_{ji} + [B, \star_g F]_{ij} (T_a)_{ji} \: , \\
      &= d \left( \star_g F_\mu^b T^b_{ij} dx^\mu \right) (T_a)_{ji} + B^b \wedge \star_g F^c [T_b, T_c]_{ij} (T_a)_{ji} \: , \\
      &= d \left( \star_g F_\mu^b (T_b)_{ij} dx^\mu \right) (T_a)_{ji} + B_\nu^b dx^\nu \wedge \left(\star_g F_\mu^c dx^\mu \right) [T_b, T_c]_{ij} (T_a)_{ji} \: .
\end{align}
\end{subequations}
We now use the trace relation for the generators, valid for $\mathfrak{su}(N)$: $tr(T_aT_b) = \frac{1}{2} \delta_{ab}$ (see \eqref{eq:Cartan Killing-form metric-2}) and the relation for the Lie bracket of differential forms to arrive at the expression (note that the generators and the basis $1$-forms commute):

\begin{align}
   0 = d \left(  F_\mu^a  (\star_g dx^\mu) \right) + B_\nu^b dx^\nu \wedge  F_\mu^c (\star_g dx^\mu) i f^a_{\:\,bc} \: . \label{eq:eom-for-field-A-1}
\end{align}
We now write out the Hodge dual of the 1-form (see Eq. \ref{example:hodge-star-operator:1-form}):

\begin{align}
    \star_g dx^\mu = \frac{g^{\mu\rho}}{3!} E_{\rho \alpha \beta \gamma} dx^\alpha \wedge dx^\beta \wedge dx^\gamma \: .
\end{align}
Let us only consider the first term of equation \eqref{eq:eom-for-field-A-1} for now. When applying the Hodge star, it becomes:

\begin{align}
    d \left(  F_\mu^a  (\star_g dx^\mu) \right) = d \left(  F_\mu^a  \frac{g^{\mu\rho}}{3!} E_{\rho \alpha \beta \gamma} dx^\alpha \wedge dx^\beta \wedge dx^\gamma \right) \: .
\end{align}
As a shorthand notation, we define the term in the brackets as a new $3$-form $G^a_{\alpha \beta \gamma}$. We then take the exterior derivative (see Eq. \eqref{def:exterior-derivative:index-notation}) (note that one has to take care of the antisymmetrization terms in index-notation):

\begin{subequations}
\begin{align}
     d \left(  F_\mu^a  (\star_g dx^\mu) \right)&= \nabla_\nu dx^\nu \wedge (  G^a_{\alpha \beta \gamma} dx^\alpha \wedge dx^\beta \wedge dx^\gamma) \: , \\
    &= \nabla_{[\nu}  G^a_{\alpha \beta \gamma ]} dx^\nu \wedge dx^\alpha \wedge dx^\beta \wedge dx^\gamma \: . \label{eq:0-form-EOM-derivation-wedge-product-calc-start}
\end{align}
\end{subequations}
To convert this $4$-form into a $0$-form, we again apply the Hodge operator:

\begin{align}
      \star_g d \left(  F_\mu^a  (\star_g dx^\mu) \right)= \nabla_{[\nu}  G^a_{\alpha \beta \gamma ]} E^{\nu \alpha \beta \gamma} \: .
\end{align}
Now, one can write out the anti-symmetrization operator of this resulting $0$-form explicitly. Since $G^a_{\alpha \beta \gamma}$ is totally antisymmetric, many terms will be equal to each other and can therefore be simplified:

\begin{align}
     \star_g d \left(  F_\mu^a  (\star dx^\mu) \right)= \frac{1}{4!} \left( 3! \nabla_\nu G^a_{\alpha\beta\gamma} - 3! \nabla_\alpha G^a_{\nu\beta\gamma} + 3! \nabla_\beta G^a_{\nu\alpha\gamma} - 3! \nabla_\gamma G^a_{\nu\alpha\beta} \right)  E^{\nu \alpha \beta \gamma} \: .
\end{align}
Inserting the definition of $G^a_{\alpha \beta \gamma}$ from above (note that the metric determinant $\sqrt{-g}$ will cancel out), one obtains:

\begin{align}
     \star_g d \left(  F_\mu^a  (\star_g dx^\mu) \right)=- \frac{3!}{4!} \frac{1}{3!} \left(  \nabla_\nu F^{\mu a} \varepsilon_{\mu\alpha\beta\gamma} - \nabla_\alpha F^{\mu a} \varepsilon_{\mu\nu\beta\gamma} +  \nabla_\beta F^{\mu a} \varepsilon_{\mu\nu\alpha\gamma} - \nabla_\gamma F^{\mu a} \varepsilon_{\mu\nu\alpha\beta} \right)  \varepsilon^{\nu \alpha \beta \gamma} \: .
\end{align}
Now we can use the contraction relation: $\varepsilon_{ijkl} \varepsilon^{ijkm} = -3! \delta_l^m$ to get

\begin{subequations}
\begin{align}
     \star_g d \left(  F_\mu^a  (\star_g dx^\mu) \right)&= \frac{1}{4!} \left( \nabla_\nu F^{\mu a} 3! \delta_\mu^\nu + \nabla_\alpha F^{\mu a} 3! \delta_\mu^\alpha + \nabla_\beta F^{\mu a} 3! \delta_\mu^\beta + \nabla_\gamma F^{\mu a} 3! \delta_\mu^\gamma  \right) \\
    &= \frac{3!}{4!} \left( 4 \nabla_\mu F^{\mu a}   \right) \\
    &= \nabla_\mu F^{\mu a}
\end{align}
\end{subequations}
The calculations for the second term can be done in a similar manner:

\begin{align}
    i f^a_{\:\,bc} B_\nu^b dx^\nu \wedge  F_\mu^c (\star_g dx^\mu)  = i f^a_{\:\,bc} B_\nu^b dx^\nu \wedge  \left(  F_\mu^c \, \frac{g^{\mu\rho}}{3!} E_{\rho \alpha \beta \gamma} dx^\alpha \wedge dx^\beta \wedge dx^\gamma \right) \: .
\end{align}
Again, define the new $3$-form $G^c_{\alpha\beta\gamma}$, as above, such that

\begin{align}
    i f^a_{\:\,bc} B_\nu^b dx^\nu \wedge  F_\mu^c (\star_g dx^\mu)= i f^a_{\:\,bc} B_\nu^b dx^\nu \wedge  \left( G^c_{\alpha\beta\gamma} dx^\alpha \wedge dx^\beta \wedge dx^\gamma \right) \: .
\end{align}
Then we evaluate the wedge-product while keeping in mind the anti-symmetrizations:

\begin{align}
    i f^a_{\:\,bc} B_\nu^b dx^\nu \wedge  F_\mu^c (\star_g dx^\mu)= i f^a_{\:\,bc} B_{[\nu}^b G^c_{\alpha\beta\gamma ]} dx^\nu \wedge  dx^\alpha \wedge dx^\beta \wedge dx^\gamma \: .
\end{align}
The following calculation steps are analogous to the steps performed in Eq. \eqref{eq:0-form-EOM-derivation-wedge-product-calc-start} onwards. The only difference being that $B_\mu^b$ takes the place of $\nabla_\mu$. After straightforward algebraic simplifications we obtain

\begin{align}
    \star_g \left(i f^a_{\:\,bc} B_\nu^b dx^\nu \wedge  F_\mu^c (\star_g dx^\mu)\right)= i f^a_{\:\,bc} B_\mu^b F^{\mu c} \: .
\end{align}
The full equation of motion thus becomes

\begin{align}
    0 = \nabla_\mu F^{\mu a} + i f^a_{\:\,bc} B_\mu^b F^{\mu c} \: ,
    \label{eq: Field-equation-0-form-index-notation}
\end{align}
with $F^{\mu a} =  \nabla^\mu \phi^a  + i f^a_{\:\,bc} B^{\mu b} \phi^c$, as obtained above in Eq. \eqref{eq:field-strength-0-form-index-notation}. Note that this expression is also equivalent to taking the gradient $D_{\mu c}^a F^{\mu c}$ of the field-strength $F$ with respect to the covariant group derivative $D_{\mu c}^a := \delta^a_c \nabla_\mu +i f^a_{\:\,bc} B_\mu^b$ in coordinate notation. We thus have derived the full equation of motion for a $0$-form field with an inner group-symmetry with a semisimple compact Lie algebra.

\subsubsection*{Equations of Motion for the Connection $B$}

We now show the calculation steps to derive the equation of motion for the connection form $B$ \eqref{eq:field-equation-for-B}:
\begin{align}
    \beta \left\{ d \left( \star_g H_{ij} \right) (T_a)_{ji} + [B, \star_g H]_{ij} (T_a)_{ji} \right\} &= \alpha A^b \wedge (\star_g F^c) [T^b, T^c]_{ij} T^a_{ji} \: . \label{eq:solving-the-equations-of-motion:eom-for-connection-b-full}
\end{align}
Where $\alpha$ and $\beta$ are scalar constants. The terms proportional to $\beta$ can be simplified in the same way as presented previously. Hereafter, we therefore only show the algebra steps of the terms proportional to $\alpha$ (right hand side of Eq. \eqref{eq:solving-the-equations-of-motion:eom-for-connection-b-full}). The curvature form $H$ associated to $B$ in index notation can be written as
\begin{subequations}
\begin{align}
    H :=&\, d B + \frac{1}{2} [B,B] = \nabla_{[\mu} B_{\nu ]} \; dx^\mu \wedge dx^\nu + \frac{1}{2} [B_\mu dx^\mu, B_\nu dx^\nu] \: ,  \\
    =& \left( \nabla_{[\mu} B^a_{\nu ]} T_a + \frac{i}{2} B^b_{[\mu} B^c_{\nu ]} f^a_{\:\,bc} T_a \right) dx^\mu \wedge dx^\nu \: , \\
    =& \frac{1}{2} H_{\mu\nu}^a T_a dx^\mu \wedge dx^\nu \: .
\end{align}
\end{subequations}
The computation of the equations of motion is in large parts analogous to the calculation presented from equation \eqref{eq:field-strength-0-form-index-notation} onwards. We use the commutator and the trace relations for the group generators to write the parts proportional to $\alpha$ (we omit the factor $\alpha$ in the following) as

\begin{subequations}
\begin{align}
    A^b \wedge (\star_g F^c) [T^b, T^c]_{ij} T^a_{ji} = i \frac{1}{2} f^a_{\:\,bc} A^b \wedge  F_\mu^c (\star_g dx^\mu) \: .
\end{align}
\end{subequations}
Note that for a $0$-form $A^a=\phi^a$. We expand the Hodge-star term and rewrite the whole term again in terms of the fully antisymmetric $3$-form $G^c_{\alpha\beta\gamma}$:

\begin{subequations}
\begin{align}
    \cdots &= -\frac{i}{2} f^a_{\:\,bc} \phi^b \wedge  F_\mu^c \left( \frac{g^{\mu\rho}}{3!} E_{\rho \alpha \beta \gamma} dx^\alpha \wedge dx^\beta \wedge dx^\gamma \right) \: , \\
     &= -\frac{i}{2} f^a_{\:\,bc} \phi^b \wedge G^c_{\alpha\beta\gamma} dx^\alpha \wedge dx^\beta \wedge dx^\gamma \: . 
\end{align}
\end{subequations}
We then compute the wedge product, which in this case is trivial because $A^a=\phi^a$ is a $0$-form:

\begin{align}
    \cdots &= -\frac{i}{2} f^a_{\:\,bc} \phi^b G^c_{[\alpha\beta\gamma]} dx^\alpha \wedge dx^\beta \wedge dx^\gamma \: .
\end{align}
The antisymmetrization operator on $G$ can now be neglected due to it being fully antisymmetric. Then, we apply the Hodge operator again and write out $G^c_{\alpha\beta\gamma}$ to get

\begin{subequations}
\begin{align}
    \cdots &= -\frac{i}{2} f^a_{\:\,bc} \phi^b G^c_{\alpha\beta\gamma} E^{\alpha\beta\gamma}_{\:\:\:\:\:\:\: \mu} dx^\mu \: , \\
     &= -\frac{i}{2} \frac{1}{3!} f^a_{\:\,bc} \phi^b F_\nu^c  \varepsilon^\nu_{\: \alpha \beta \gamma} \varepsilon^{\alpha\beta\gamma}_{\:\:\:\:\:\:\: \mu} dx^\mu \: .
\end{align}
\end{subequations}
This can be written as:
\begin{align}
    \cdots &= -\frac{i}{2} \frac{1}{3!} f^{a}_{\:\,bc} \phi^b F^{\nu c} g_{\mu \lambda} \varepsilon_{\nu \alpha \beta \gamma} \varepsilon^{\alpha\beta\gamma\lambda} dx^\mu
\end{align}
Now, a contraction relation of the Levi-Civita symbol $\varepsilon_{\nu \alpha \beta \gamma} \varepsilon^{\alpha\beta\gamma\lambda}  = -3! \delta^\lambda_\nu$ can be used to obtain

\begin{align}
    \cdots &= \frac{i}{2} f^a_{\:\,bc} \phi^b F_\mu^c  dx^\mu \: .
\end{align}
We now combine this expression with the terms proportional to $\beta$ to obtain the total equation of motion:

\begin{align}
   \beta \left\{ \nabla_{\mu}H^{\mu\nu a}+i\, f^{a}_{\;bc}B_{\mu}^{b}H^{\mu\nu c} \right\} = \alpha \; i f^a_{\:\,bc} \phi^b g^{\mu\nu} F_\mu^c \: . \label{eq: Field-equation-0-form-index-notation-B}
\end{align}

\subsubsection*{Energy Momentum Tensor}

Next, we derive the explicit expression for the energy-momentum tensor of the $0$-form field. The action (Eq. \eqref{eq:total-action-coordinate-representation}) can be written in index notation as

\begin{align}
    S &=-\frac{\alpha}{2}\sum_{a=1}^{\text{dim}(\mathfrak{g})} \int \sqrt{-g} F_{\mu}^{a}\; F^{\mu}_{a} \: dx^{4}-\frac{\beta}{4}\sum_{a=1}^{\text{dim}(\mathfrak{g})} \int \sqrt{-g} H_{\mu\nu}^{a}\; H^{\mu\nu}_{a} \: dx^{4} \: .
\end{align}
It is now possible to compute the variation with respect to the spacetime metric tensor $\delta g^{\mu\nu}$ to obtain the energy-momentum tensor $T_{\mu\nu}$, as defined in equation \eqref{eq:definition-energy-momentum-tensor}. We obtain
\begin{align}
    T_{\mu\nu}\; = \alpha \sum_{a=1}^{\text{dim}(\mathfrak{g})} \left( F_{\mu}^{a} F_{\nu a}-\frac{1}{2} g_{\mu\nu} F_{\alpha}^{a} F^{\alpha}_{a} \right) + \beta \sum_{a=1}^{\textrm{dim}(\mathfrak{g})} \left(\; H_{\mu}^{\; \alpha a}H_{\nu\alpha a}-\frac{1}{4}g_{\mu\nu}H_{\alpha\beta}^{\;\; a} H^{\alpha\beta}_{a} \right) \:
    \label{eq: Energy momentum tensor k=0}
\end{align}

\subsubsection*{Finding Explicit Solutions}

Hereafter, we focus on deriving explicit solutions to the equations of motion \eqref{eq: Field-equation-0-form-index-notation}. For this, we choose an explicit ansatz for the spacetime metric and for the components of the differential $k$-form field. \\ 
Although equation \eqref{eq: Field-equation-0-form-index-notation} holds for any compact, semisimple Lie group, for all following calculations we will fix the Lie group to be $SU(2)$. All of the following calculations may also be extended to other Lie groups by following the procedure discussed in section \ref{sec:connection-for-Lie-groups}, if a suitable basis for the connection coefficients $B_{\mu}^{a}$ is constructed. The following calculation is meant to be a proof of principle calculation. It can be used as a reference when considering more complex fields in future works. \\

\noindent
We choose an ansatz for a static and spherically symmetric spacetime with the following line element:

\begin{equation}
ds^2=g_{tt}(r)dt^{2}+g_{rr}(r)dr^{2}+r^{2} (d\theta^{2}+\sin^{2}(\theta) d\varphi^2) \: . \label{eq:metric-ansatz-static-spherically-symmetric}
\end{equation}
The field components $\phi^a$ are chosen such that they depend on the radial distance $r$ from the origin and the polar angle $\theta$:

\begin{equation}
\phi^a(t,\Vec{x})=\phi^a(r,\theta) \:\:,\hspace{1.5cm} a=1,2,3 \: .
\label{eq:Ansatz-Field-0-form}
\end{equation}
We insert the ansatz \eqref{eq:metric-ansatz-static-spherically-symmetric} and \eqref{eq:Ansatz-Field-0-form} into \eqref{eq:field-strength-0-form-index-notation} to obtain the field strength $1$-form. Its components are given by:

\begin{align}
F_{\mu}^{\; 1} &= \left( \; 0,\; \partial_{r}\phi^1  , \; \partial_\theta \phi^1 -\frac{1}{\sqrt{g_{rr}}}\phi^2, \; -\frac{\sin\theta}{\sqrt{g_{rr}}} \phi^3\right)\: , \hspace{0.5cm} F_{\mu}^{\; 2}=\left( \; 0,\; \partial_{r}\phi^2  , \; \partial_\theta \phi^2 +\frac{1}{\sqrt{g_{rr}}}\phi^1,-\cos\theta \phi^3\right) \: , \nonumber \\
F_{\mu}^{\;3} &= \left(\; 0,\; \partial_{r}\phi^3,\; \partial_\theta \phi^3,\; \frac{\sin\theta}{\sqrt{g_{rr}}}\phi^1+\cos\theta\phi^2\right) \: 
\label{eq:field-strength-1-form-components}
\end{align}
The field-strength $1$-form and the connection coefficients computed in section \ref{sec:connection-for-Lie-groups} may now be plugged in into \eqref{eq: Field-equation-0-form-index-notation}. The resulting set of equations of motions can be compiled into three partial differential equations:

\begin{subequations}
    \begin{align}
\Box \phi^1-\frac{2}{r^{2}g_{rr}}\phi^1-\frac{2 \cot\theta}{r^{2}\sqrt{g_{rr}}}\phi^2 -\frac{2\partial_\theta \phi^2}{r^2 \sqrt{g_{rr}}}\; &=\; 0 \: , \label{eq:ODE-0-form-field-explicit-1} \\
\Box \phi^2-\left(\frac{\cot^{2}\theta}{r^2}+\frac{1}{r^{2}g_{rr}} \right)\phi^2 +\frac{2\partial_\theta \phi^1}{r^2 \sqrt{g_{rr}}}\; &=\; 0 \: , \label{eq:ODE-0-form-field-explicit-2} \\
\Box \phi^3-\left(\frac{\cot^{2}\theta}{r^2}+\frac{1}{r^{2}g_{rr}} \right)\phi^3 \; &=\; 0 \: , \label{eq:ODE-0-form-field-explicit-3}
\end{align}
\end{subequations}
where $\Box = \nabla^\mu \nabla_\mu$ is the d'Alembert operator. These equations can be interpreted as describing different components of the $0$-form field that have different relative amplitudes depending on the location in spacetime. The equations \eqref{eq:ODE-0-form-field-explicit-1} and \eqref{eq:ODE-0-form-field-explicit-2} are coupled in a non-trivial way. In the limit of large radii $r\rightarrow \infty$ the system of equations reduces to three independent wave equations. \\

\noindent
With knowledge of the field-strength $1$-form \eqref{eq:field-strength-1-form-components}, we can compute the components of the energy momentum tensor via Equation \eqref{eq: Energy momentum tensor k=0}:

\begin{subequations}
\begin{align}
T_{tt}&=-\frac{\alpha}{2} g_{tt} \left[ \frac{1}{g_{rr}}\left((\partial_{r}\phi^1)^2+(\partial_{r}\phi^2)^2+(\partial_{r}\phi^3)^2\right)+\frac{1}{r^{2}}\left((\partial_\theta \phi^1)^2+(\partial_\theta \phi^2)^2+(\partial_\theta \phi^3)^2  \right) \right. \nonumber \\ &\;+ \left. \frac{1}{r^{2}g_{rr}}\left(2(\phi^1)^2+(\phi^2)^2+(\phi^3)^2  \right)
 +\frac{\cot^2\theta}{r^{2}}((\phi^2)^2+(\phi^3)^2)+\frac{2\cot\theta}{r^{2}\sqrt{g_{rr}}}\phi^1\phi^2 \right. \nonumber \\
 &\;- \left. \frac{2}{r^2\sqrt{g_{rr}}}\left((\partial_\theta \phi^1)\phi^2-(\partial_\theta \phi^2)\phi^1\right) \right] \: + \frac{\beta}{4} g_{tt} \left[ \frac{(\partial_r g_{rr})^2}{r^2 g_{rr}^4} + \frac{1}{r^4} \left( 1 - \frac{1}{g_{rr}}\right)^2 \right] \\
T_{rr}&=\frac{\alpha}{2} \left[\left(\partial_r \phi^1\right)^2+(\partial_r \phi^2)^2+(\partial_r \phi^3)^2 \right] -\frac{\alpha}{2} g_{rr} \left[\frac{1}{r^{2}}\left((\partial_\theta \phi^1)^2+(\partial_\theta \phi^2)^2+(\partial_\theta \phi^3)^2  \right) \right. \nonumber \\ &\;+ \left. \frac{1}{r^{2}g_{rr}}\left(2(\phi^1)^2+(\phi^2)^2+(\phi^3)^2  \right)
 +\frac{\cot^2\theta}{r^{2}}((\phi^2)^2+(\phi^3)^2)+\frac{2\cot\theta}{r^{2}\sqrt{g_{rr}}}\phi^1\phi^2 \right. \nonumber \\
 &\;- \left. \frac{2}{r^2\sqrt{g_{rr}}}\left((\partial_\theta \phi^1)\phi^2-(\partial_\theta \phi^2)\phi^1\right) \right] +\frac{\beta}{4} g_{rr} \left[ - \frac{(\partial_r g_{rr})^2}{r^2 g_{rr}^4} + \frac{1}{r^4} \left( 1 - \frac{1}{g_{rr}}\right)^2 \right]  \: .
\end{align}
\end{subequations}
The equations of motion for the metric components are derived from the Einstein equations

\begin{equation}
    G_{\mu\nu}\; =\; 8\pi T_{\mu\nu} \: .
    \label{eq: Einstein equations k=0}
\end{equation}
\noindent
On the left hand side of equation \eqref{eq: Einstein equations k=0}, only functions of the radial coordinate appear when choosing the radially symmetric metric ansatz \eqref{eq:metric-ansatz-static-spherically-symmetric}. However, the energy momentum tensor components explicitly depend the radial distance $r$ and the polar angle $\theta$. This indicates a poloidal deformation of spacetime since the energy-momentum distribution encoded in $T_{\mu\nu}$ is anisotropic. For more general considerations, the components of the $0$-form field $\phi^a$ \textit{and} metric components will have to depend on the polar angle $\theta$. For the scope of this work, the simple ansatz chosen in equations \eqref{eq:Ansatz-Field-0-form} and \eqref{eq:metric-ansatz-static-spherically-symmetric} is acceptable, since it was only meant as a proof of concept to illustrate the application of our formalism for a concrete physical case. \\
For numerical applications, one could substitute the simple ansatz \eqref{eq:metric-ansatz-static-spherically-symmetric} with an axially symmetric ansatz for a metric that depends on both $r$ and $\theta$. Another option would be to take an average of $T_{\mu\nu}$ over the polar angle $\theta$. The tensor $T_{\mu\nu}$ on the r.h.s of equation \eqref{eq: Einstein equations k=0} would then have to be replaced by its poloidal average $\left<T_{\mu\nu}\right>_{\theta\in [0;\pi]}$. The equations \eqref{eq: Einstein equations k=0} then reduce to two ordinary differential equations in the radial distance $r$. A future numerical study is necessary to assess the quality of the averaging prescription $\left<T_{\mu\nu}\right>_{\theta\in [0;\pi]}$ compared to using a fully axially symmetric ansatz for the field and the metric.

\subsection{Numerical Implications}

To solve the $k$-form equations numerically, it is important to know the type of equations and initial conditions. We discuss these points for the case of the $0$-form field introduced in the previous subsection.

\subsubsection*{Analyzing the Equations of Motion}

We start by analyzing the type of the partial differential equations \eqref{eq:ODE-0-form-field-explicit-1}-\eqref{eq:ODE-0-form-field-explicit-3}. All three of the equations are linear second order PDEs. There exists a unique classification for linear second order PDEs with two variables. Let $f(r,\theta)$ be a smooth and (at least) twice differentiable function of two variables $r$ and $\theta$. Then, every second order linear PDEs can be written as

\begin{equation}
A\partial_{r}^2 f(r,\theta)+C\partial_{\theta}^2 f(r,\theta)+2B\partial_{r}\partial_{\theta}f(r,\theta)+D\partial_{r}f(r,\theta)+E\partial_{\theta} f(r,\theta)+F=0 \: ,
\end{equation}
with smooth functions $A,B,C,D,E,F$ that depend on $r$ and $\theta$. All second order linear partial differential equations can now be categorized into hyperbolic, elliptic and parabolic. A second order linear PDE is called hyperbolic if $B^2-AC>0$ for all $r$ and $\theta$. It is called elliptic if $B^2-AC<0$ and parabolic  if $B^2-AC=0$ for all $r$ and $\theta$. In our case, for the equations \eqref{eq:ODE-0-form-field-explicit-1}- \eqref{eq:ODE-0-form-field-explicit-1}, one obtains 
\begin{equation}
B^2-AC=-\frac{1}{r^2 g_{rr}} \: .
\end{equation}
Since $g_{rr}>0$ for all radial distances $r\in (0,\infty)$ it follows that $B^2-AC<0$ for all $r$ and $\theta$. Thus, equations \eqref{eq:ODE-0-form-field-explicit-1}-\eqref{eq:ODE-0-form-field-explicit-3} are a system of elliptic partial differential equations. In a numerical analysis, standard solution methods for elliptic partial differential equations can be applied to find an explicit solution. One of the inputs needed for this methods are appropriate boundary conditions.

\subsubsection*{Boundary Conditions}

For the ansatz chosen in this work, there are four boundaries at which appropriate boundary conditions must be applied. These four boundaries are located at $(r,\theta)=(0,\theta)$, $(r,\theta)=(R,\theta)$, $(r,\theta)=(r,0)$ and $(r,\theta)=(r,\pi)$, where $R$ denotes the largest radial position of a two dimensional polar grid. For the boundary located at $(r,\theta)=(R,\theta)$, a consistent boundary condition is given by 
\begin{equation}
\phi^a(R,\theta) = \phi^a(R) \: .
\label{eq:no-outflow-boundary-condition}
\end{equation}
Here $\phi^a(R)$ is independent of the angle $\theta$ and must be chosen depending on the physical scenario one wants to study. One possible choice is to let the field vanish at large distances $R$. \\
This boundary condition is valid for sufficiently large $R$ since for $r\rightarrow\infty$ the equations \eqref{eq:ODE-0-form-field-explicit-1}-\eqref{eq:ODE-0-form-field-explicit-3} reduce to decoupled free wave equations. For the coordinate poles $\theta=\{0,\pi\}$, typical pole fixing boundary conditions can be applied, for example reflective or absorption boundary conditions. \\
For the remaining boundary condition at the center of the grid at $(r,\theta)=(0,\theta)$ it is instructive to investigate the proportionality behaviour of the field for small radii. Evidently, equations \eqref{eq:ODE-0-form-field-explicit-1}-\eqref{eq:ODE-0-form-field-explicit-3} all contain terms proportional to $1/r^{n},n\in\mathbb{N}$ which are singular at $r=0$. Therefore, one can not simply choose a set of constant values $\phi^a(0,\theta)=\phi_{0}^a(\theta) = const$. Instead, one needs to expand the field components $\phi^{a}$ in a polynomial series in the radial distance $r$
\begin{equation}
\phi^{a}(r,\theta)=\sum_{n=0}^{\infty} \phi^{(n),a}(\theta)\, r^{n} \: .
\label{eq:boundary-at-r-0}
\end{equation}
By inserting the boundary condition \eqref{eq:boundary-at-r-0} into equations \eqref{eq:ODE-0-form-field-explicit-1}-\eqref{eq:ODE-0-form-field-explicit-3} and imposing regularity at $r=0$ one can obtain a set of constraints for the radius-independent functions $\phi^{(n),a}(\theta)$. We subsequently obtain the following boundary condition 
\begin{equation}
\phi^{a}(r,\theta) = \phi^{(3),a}(\theta)\: r^{3} + \mathcal{O}(r^4) \: , \hspace{1.5cm} r \rightarrow 0 \: .
\end{equation}
The yet unknown functions $\phi^{(3),a}(\theta)$ can be obtained by inserting the boundary condition into equations \eqref{eq:ODE-0-form-field-explicit-1}-\eqref{eq:ODE-0-form-field-explicit-3}. The resulting equations are three second order ordinary differential equations in $\theta$:
\begin{subequations}
    \begin{align}
&\partial_{\theta}^{2}\,\phi^{(3),1}(\theta)+\cot\theta\, \partial_{\theta}\,\phi^{(3),1}(\theta)+10\;\phi^{(3),1}(\theta)-2\cot\theta\, \partial_\theta\, \phi^{(3),2}(\theta)\; =\; 0\; \label{eq:ODE-theta-1} \\
&\partial_{\theta}^{2}\,\phi^{(3),2}(\theta)+\cot\theta\, \partial_{\theta}\,\phi^{(3),2}(\theta)-(\cot^2\theta-11)\, \phi^{(3),2}(\theta)+2\partial_{\theta}\, \phi^{(3),1}(\theta)\; =\; 0 \: , \label{eq:ODE-theta-2} \\
&\partial_{\theta}^{2}\,\phi^{(3),3}(\theta)+\cot\theta\, \partial_{\theta}\,\phi^{(3),3}(\theta)+10\;\phi^{(3),1}(\theta)-(\cot^2\theta-11)\, \phi^{(3),3}(\theta) \; =\; 0 \: . 
\end{align}
\label{eq:ODE-theta-3}
\end{subequations}
\\
Here, the asymptotic flatness conditions $g_{tt}(0)=g_{rr}(0)=1$ and $\partial_{r}g_{tt}(0)=\partial_{r}g_{rr}(0)=0$ were assumed. This is only valid for the spherically symmetric metric ansatz. Coupling to matter can now be realized by either solving the equation fully self-consistently (with the metric depending on $r$ and $\theta$), or by using the averaging procedure for the energy-momentum tensor, as explained in the last paragraph of section \ref{sec:derivation-of-0-form-equations}. \\
We note that the polar grid has a singularity at $r=0$. In practice, one has to remove this point from the domain to ensure regularity at all points of the grid. 
Numerically the set of equations \eqref{eq:ODE-theta-3} can be treated as a set of constraint equations which must be satisfied for sufficiently small $r=\epsilon$. These equations constitute consistent boundary conditions that have to be fulfilled at the boundary $(r,\theta)=(\epsilon,\theta)$. With this, we conclude our discussion on boundary conditions.

\subsection{Higher Order $k$-Form Fields}

In the following, we briefly address the steps to study $1$-form and $2$-form fields in a similar fashion as laid out for the $0$-form field in section \ref{sec:derivation-of-0-form-equations}. In this work, we refrain from studying $3$-form fields. However we note that the process to extend the description to $3$-form fields is in principle straightforward. Following similar steps as for the $0$-form field, one can derive the field strengths and the field equations for the $1$-form field from the most general form of equation \eqref{eq:field-equation-for-A}. If we let $A\in \Omega^{1}(M;\mathfrak{g})$ be a Lie algebra valued $1$-form field and $B\in \Omega^{1}_{con}(M;\mathfrak{g})$ be the connection computed in chapter \ref{sec:connection-for-Lie-groups}, then the corresponding $2$-form field strength $F\in \Omega^{2}(M;\mathfrak{g})$ is given by:

\begin{equation}
F=dA +[B,A]\; =\frac{1}{2} F_{\mu\nu}^{a} \left(dx^\mu \wedge dx^\nu \right)\; \tens{} T_{a} \: .
\label{eq:1-Form-field-strength-definition}
\end{equation}
where the $F_{\mu\nu}^{a}$ denote the components of the Lie algebra valued $2$-form $F$. \\
Similarly, for the $2$-form field $A\in \Omega^{2}(M;\mathfrak{g})$, one can derive the following expression for the $3$-form field-strength $F\in \Omega^{3}(M;\mathfrak{g})$:

\begin{equation}
F=dA +[B,A]\; =\frac{1}{3!} F_{\mu\nu\rho}^{a} \left(dx^\mu \wedge dx^\nu \wedge dx^\rho \right)\; \tens{} T_{a} \: .
\label{eq:2-Form-field-strength-definition}
\end{equation}
The components of the field strengths defined in \eqref{eq:1-Form-field-strength-definition} and \eqref{eq:2-Form-field-strength-definition} can also be expressed in terms of the field components $A_{\mu}^a$, $A_{\mu\nu}^{a}$ and the connection coefficients components $B_{\mu}^a$ as

\begin{subequations}
\label{eq: Fieldstrenghts for k=1,k=2}
\begin{align}
F_{\mu\nu}^a&=2\nabla_{[\mu}A_{\nu ]}^{a}+2if^{a}_{\;bc}\; B_{[\mu}^b \; A_{\nu ]}^{c} \: , \label{eq: Fieldstrength for k=1} \\
F_{\mu\nu\rho}^a&=3\nabla_{[\mu}A_{\nu\rho ]}^{a}+3if^{a}_{\;bc}\; B_{[\mu}^b \; A_{\nu\rho ]}^{c}  \: . \label{eq: Fieldstrength for k=2}
\end{align} 
\end{subequations}
The squared brackets denote the usual total antisymmetrization with respect to the Lorentz indices (see paragraph below Def. \ref{def:alternating-linear-forms}). To derive the full equations of motion for the $1$-form field, we follow the same steps as in the previous section, where the derivation of the $0$-form equations of motion was laid out in detail. In coordinate based notation, the resulting field equations have the same structure as for the $0$-form field. They can be written for the $1$-form field using the components of the $2$-form field strength defined in equation \eqref{eq: Fieldstrength for k=1}

\begin{equation}
\nabla_{\mu}F^{\mu\nu a}+i\, f^{a}_{\;bc}B_{\mu}^{b}F^{\mu\nu c} = 0 \: .
\label{eq: Explicit field equation k=1}
\end{equation}
With similar steps we obtain the field equations for the $2$-form field $A\in \Omega^{2}(M;\mathfrak{g})$ using the 3-form field strength defined in equation \eqref{eq: Fieldstrength for k=2}

\begin{equation}
\nabla_{\mu}F^{\mu\nu\rho a}+i\, f^{a}_{\;bc}B_{\mu}^{b}F^{\mu\nu\rho c} = 0 \: .
\label{eq:2-Form-field-equations}
\end{equation}
The corresponding energy momentum tensor for the 1-form field $A\in \Omega^1(M,\mathfrak{g})$ is given in terms of the corresponding field strength

\begin{equation}
T_{\mu\nu}\; = \alpha \sum_{a=1}^{\textrm{dim}(\mathfrak{g})} \left(\; F_{\mu}^{\; \alpha a}F_{\nu\alpha a}-\frac{1}{4}g_{\mu\nu}F_{\alpha\beta}^{\;\; a} F^{\alpha\beta}_{a} \right) + \beta \sum_{a=1}^{\textrm{dim}(\mathfrak{g})} \left(\; H_{\mu}^{\; \alpha a}H_{\nu\alpha a}-\frac{1}{4}g_{\mu\nu}H_{\alpha\beta}^{\;\; a} H^{\alpha\beta}_{a} \right) \: .
\end{equation}
In an identical fashion one can derive the following expression for the energy momentum tensor for the $2$-form field $A\in \Omega^{2}(M;\mathfrak{g})$

\begin{equation}
T_{\mu\nu}\; = \alpha \sum_{a=1}^{\textrm{dim}(\mathfrak{g})} \left(\; \frac{1}{2} F_{\mu\rho\kappa}^{a}F_{\nu a}^{\; \rho\kappa} - \frac{1}{12} g_{\mu\nu} F_{\rho\kappa\lambda }^{a}F^{\rho\kappa\lambda}_{a}\right) + \beta \sum_{a=1}^{\textrm{dim}(\mathfrak{g})} \left(\; H_{\mu}^{\; \alpha a}H_{\nu\alpha a}-\frac{1}{4}g_{\mu\nu}H_{\alpha\beta}^{\;\; a} H^{\alpha\beta}_{a} \right) \:  \: .
\end{equation}
Both energy momentum tensors are hence completely determined by the corresponding field strengths. From this point on, to perform any specific calculation, one must choose a concrete ansatz for the $1$-form/$2$-form field $A$. Future work might use approaches based upon studies of $SU(2)$ Yang-Mills stars, such as \cite{VOLKOV19991,chen2022stable}, as a potential ansatz for the $k$-form fields.

\newpage
\section{Conclusion and Outlook} \label{sec:conclusions-and-outlook}
In this work, we developed a rigorous way to incorporate differential $k$-form fields $A$ with a local gauge symmetry $G$ into curved spacetimes. The symmetry of the fields was modelled using a compact, semisimple Lie group. In section \ref{sec:formulation-and-motivation-of-action}, we constructed a generalized action for Lie algebra valued differential $k$-form fields in $n$ dimensions. For this, we introduced a covariant derivative with corresponding connection $B$, which respects both the inner group structure of the fields and the spacetime curvature. We showed that the action is self-consistent and that it is invariant under local symmetry transformations in $G$ and $SO(n-1,1)$. From this action, we derived equations of motion for the field $A$ and the connection $B$ and discussed their degrees of freedom. \\
In section \ref{sec:connection-for-Lie-groups}, we explored possible choices for the connection $B$. We showed that it can be related to the spin-connection of spacetime. As an example, we explicitly studied the case of a spherically symmetric spacetime with a local symmetry group $G=SU(2)$. In this case, we were able to fully express the connection coefficients of $B$ using the spacetime metric and group structure coefficients. We also found that in some cases, the dimension of the local group space and of spacetime can be different form each other. This makes is necessary to complete the basis using "extra dimensions" and rectangular vielbein basis vectors. Calculations of the vielbeins for the group $SU(3)$ are shown in Appendix \ref{sec:appendix a}. \\
In section \ref{sec:solving-the-equations-of-motion}, we rewrote the equations of motion for the $k$-form field and the connection in index notation and used this to derive differential equations for a $0$-form field (i.e. a  scalar function) with internal $SU(2)$ symmetry in spherical symmetry. We also discussed possible applications and numerical solution strategies. \\

\noindent
Our work can be applied in a variety of scenarios. One is boson stars, which are a class of self-gravitating bosonic field solutions of the Einstein equations. In most recent studies (see citations in the introduction), the bosonic field is usually taken to be a classical scalar (or vector) field with a global $U(1)$ symmetry. They are often motivated through light dark matter particles or theories of modified gravity. An extension to internal symmetries of the fields could thus allow to probe different types of dark matter or extensions of gravity. Gravitational condensates of pions -- and other particles with internal symmetry groups -- could also be modelled using our approach because in a classical picture they can be viewed as scalar particles with an internal $SU(2)$ symmetry. \\
Another field of interest is string theory, which features various differential $k$-form fields with different degrees $k$ and internal symmetry structures. In certain superstring theories, differential $k$-form fields appear either as low energy effective modes or as excitations of $D$-branes. In particular $D$-branes would be interesting to study with the formalism in this work. The group index of the field would then correspond to the numbering of multiple branes in specific orientations. \\

\noindent
Our formalism also has some limitations. It is valid for $k$-form fields $A$ with internal symmetries. These symmetries are described by semisimple compact Lie groups. It also features a connection $B$, which is independent of the field $A$. The description of the fields in terms of Lie algebra valued differential forms allows us to study them with various internal symmetries. These groups cover a large amount of symmetry groups widely used in physics. However, it restricts us to model physical bosonic fields with spin $0$ and spin $1$. Also, the allowed Lie groups do not permit to study all possible groups that might be interesting for physical applications. \\

\noindent
Our work can be further be expanded on in two main ways: modifying the symmetry groups and modifying the fields. \\
Regarding the groups, one could consider more complicated internal symmetry groups such as product groups. Relaxing the assumption of semisimplicity and compactness of the Lie groups might also be worthwhile to study -- although relaxing semisimplicity leads to a degenerate Cartan–Killing form. \\
Another option is to add additional types of symmetries such as global symmetries. This is interesting in the context of boson stars where the bosonic fields are usually modeled as complex fields with a global $U(1)$ symmetry (see \cite{Liebling:2012fv} for an extensive review). Using internal symmetries for boson stars might be a way to circumvent Derrick's theorem \cite{Derrick:1964ww,Diez-Tejedor:2013sza,Carloni:2019cyo}, which forbids scalar boson stars without conserved Noether currents. \\ 
Regarding the fields, another avenue of study would be higher order tensor fields with internal symmetries. This would open the way to model arbitrary integer spin fields. These would not be described using differential forms. The action would thus need to be modified. \\ 
Another interesting option might be to incorporate fermionic fields into the formalism. This amounts to adding half-integer differential $k$-form fields. They are described by the $k+\frac{1}{2}$ representations of the double cover of the Lorentz group $SO(n-1,1)$. \\
Modifications to the connection $B$ are also interesting.  Promoting $B$ to a fully dynamical field might allow different physical interpretations. Considering different choices for the connection as a non-dynamical object could also be interesting. \\

\noindent
But even without further modifications, there are already a large number of cases that can be studied using our formalism. It would be interesting to formulate solutions to the equations of motion for various $k$-form fields and internal symmetry groups. \\
In this work, we made a choice for the vielbeins where we matched specific dimensions of the local symmetry space and the global spacetime. Alternatively, one could also consider to expand the rectangular vielbeins in a different manner -- e.g. matching the $\mathfrak{su}(2)$ algebra space to the spacetime components ($0,1,2$) instead of the spatial spacetime components ($1,2,3$). \\
Regarding concrete computations of solutions to the equations of motion, one might use a more complete ansatz for the fields and the metric such that both depend on the coordinates $(r,\theta)$. \\
Other extensions include the use of different potential forms in the action, adding coupling terms to matter or modifying the underlying theory of gravity. We have summarized a number of other extensions in more detail in chapter \ref{subsection:possible-extensions}. \\
The large amount of options aside, the next steps should be to study the solutions evaluated in this work numerically. This entails solving differential equations to obtain the values of the field $A$ and the spacetime. We leave the detailed analysis for a future work.

\section*{Acknowledgements}
We thank Leon Menger for his invaluable insights, discussions and proof-reading the manuscript. We also thank Laura Sagunski for providing support during the initial stages of the project.

\newpage
\appendix
\section*{Appendix}
\section{Vielbeins for SU(3)} \label{sec:appendix a}

In the following section, we briefly outline the basic computational steps to calculate vielbein components for the case when the dimension of the symmetry group $G$ is larger than the dimension $n$ of the spacetime manifold $M$. For the sake of this example, we choose $G=SU(3)$. Let further $g$ be the spacetime metric. We also assume that $g$ it is static and spherically symmetric. Let $K$ be the Cartan--Killing metric of the Lie algebra $\mathfrak{su}(3)$ with components $K_{ab}=-N\delta_{ab}$, where $N=3$. The vielbeins $e$ may then be determined using the relation

\begin{equation}
g_{\mu\nu}=e_{\mu}^{a}K_{ab}e_{\nu}^{b} \: .
\label{eq: Tetrads for SU(3)}
\end{equation}
From equation \eqref{eq: Tetrads for SU(3)} it is apparent that the vielbeins are not quadratic. The reason is that the dimension of the algebra $dim(\mathfrak{g})=8$ is larger than the dimension of the spacetime manifold $n=4$. In section \ref{sec:connection-for-SU(2)}, we discussed the procedure to calculate vielbeins for cases where $dim(\mathfrak{g})< n$. This was done by completing the basis of vielbeins using additional ''extra dimensions'' of the group. When $dim(\mathfrak{g})>n$, we may employ the same procedure by adding additional spacetime dimensions instead. There are many possible choices on how to choose the manifold that incorporates additional dimensions. In this example we use 6-dimensional spherical coordinates $\{r,\theta_{j} \}_{\; j\in [1;6]}$ with the corresponding metric line element

\begin{equation}
ds^2=g_{tt}(r)dt^2+g_{rr}(r)dr^2+r^2 \left(d\theta_{1}^2+\sum_{m=2}^{6} \left(\prod_{j=1}^{m-1} \sin^{2}\theta_{j}\right)\; (d\theta_{m})^2\right) \: .
\label{eq: metric for 6-sphere}
\end{equation}
Using the metric \eqref{eq: metric for 6-sphere}, it is straightforward to compute the vielbein components (because both $g$ and $K$ are symmetric in the chosen basis). We list all resulting non-zero vielbein components $e_{\mu}^a$ for $SU(3)$ below:
\begin{align}
e_{t}^1& = i\;\sqrt{\frac{g_{tt}}{3}} \hspace{1.5cm} e_{r}^2=i\; \sqrt{\frac{g_{rr}}{3}} \hspace{1.5cm} e_{\theta_{1}}^3=i\; \frac{r}{\sqrt{3}} \hspace{1.5cm} e_{\theta_{2}}^4=i\; \frac{r\sin\theta_{1}}{\sqrt{3}} \nonumber \\ \nonumber \\
e_{\theta_{3}}^5&=i\; \frac{r\sin\theta_{1} \sin\theta_{2}}{\sqrt{3}}\hspace{1.4cm} e_{\theta_{4}}^6=i\; \frac{r\sin\theta_{1} \sin\theta_{2}\sin\theta_{3}}{\sqrt{3}} \hspace{1.4cm}
e_{\theta_{5}}^7=i\; \frac{r\sin\theta_{1} \sin\theta_{2}\sin\theta_{3}\sin\theta_{4}}{\sqrt{3}} \nonumber \\ e_{\theta_{6}}^8&=i\; \frac{r\sin\theta_{1} \sin\theta_{2}\sin\theta_{3}\sin\theta_{4}\sin\theta_{5}}{\sqrt{3}} \: .
\end{align}
\\
Using these vielbeins one can compute the connection coefficients $B_{\mu}^a$ in an identical manner as was shown for $SU(2)$ in section \ref{sec:connection-for-SU(2)}. The connection coefficients can be computed using the structure coefficients $f^a_{\; b c}$ of $\mathfrak{su}(3)$ and equation \eqref{eq:analogy-B-spin-connection}. \\
Due to the artificial extension of the spacetime dimension from $dim(M_{4})=4$ to $dim(M_{8})=8$, additional components will be introduced to all tensorial and non-tensorial objects. This includes the connection, spacetime metric and also the equations of motion. But the the additional dimensions were added as a mere mathematical tool to make the vielbeins well-defined. These extra dimensions are thus unphysical and must be removed before evaluating any physically relevant expressions. This is formally achieved by taking the following intersection:
\begin{equation}
M_{4}=\bigcap_{\theta_{j} = 0\; \forall \;j\in [3;6]} \; M_{8} \: .
\end{equation}
This effectively reduces all final expressions to the physical four-dimensional spacetime and sets any contributions from the ''extra dimensions'' to zero.

\newpage

\section{Derivation of the Equations of Motion} \label{sec:Appendix B}

Equation \eqref{eq:variation-of-a-k-form} provides a formal definition of a variation for an action that contains Lie algebra valued differential $k$-form fields $A\in \Omega^{k}(M;\mathfrak{g})$. It is defined by the limit 

\begin{equation}
\delta  S_{KFT}=\lim_{\epsilon\rightarrow 0} \frac{S_{KFT}[A+\epsilon\delta A]-S_{KFT}[A]}{\epsilon} \: .
\label{eq:variation-appendix}
\end{equation}
Before performing the variation, we rewrite the KFT part of the action (\ref{eq:total-action}) using the $k$-form field $A$. By inserting the field strength, defined in equation \eqref{eq:field-strangth-and-curvature-form-with-spacetime-corrections}, the action can be expressed solely in terms of the field $A$ and the connection $B$. For the remainder of this derivation, summation over repeated group indices is assumed. We also adopt the same notation, $d$ instead of $d_\omega$, for the covariant derivative relative to the spacetime connection $\omega \in \Omega_{con}(M;\mathfrak{so}(n-1,1))$ as in section \ref{sec:derivation-of-EOM}:
\begin{equation}
\begin{split}
S_{KFT}&=\int_M \left(dA^a \wedge \star_g dA^b \tens{} Tr(T_aT_b)+2(B^b \wedge A^c)\wedge \star_g dA^a \tens{} Tr([T_b,T_c]T_a) \right. \\
&+ \left.(B^a \wedge A^b) \wedge \star_g (B^c \wedge A^d)\tens{} Tr([T_a,T_b][T_c,T_d]) \right) \: .
\label{eq:action-in-fields}
\end{split}
\end{equation}
We now turn to the calculation of the variation of the action \eqref{eq:action-in-fields}.
\noindent 
First, we write out the terms from the variation defined in equation \eqref{eq:variation-appendix} explicitly 
\begin{equation}
\begin{split}
& S_{KFT}[A^a+\epsilon \delta A^a] =
 \int_M Tr \left[ d(A^a+\epsilon\delta A^a)\wedge \star_g d(A^b+\epsilon \delta A^b)\tens{} T_a T_b \right.  \\ 
& +\left. 2 d(A^a+\epsilon \delta A^a) \wedge \star_g (B^b\wedge (A^c+\epsilon \delta A^c))  \tens{}  T_a[T_b,T_c] \right. \\ 
& \left. + (B^b\wedge (A^a+\epsilon \delta A^a))\wedge \star_g (B^d\wedge (A^c+\epsilon \delta A^c))\tens{} [T_b,T_a][T_d,T_c] \right] \: .
\end{split}
\end{equation}
All terms of the form $B^{b}\wedge (A^{c}+\epsilon \delta A^{c})$ can be rewritten as

\begin{equation}
B^b\wedge (A^c+\epsilon \delta A^c) = B^b\wedge A^c + \epsilon B^b\wedge \delta A^c \: .
\end{equation}
We further use that the scalar product (see equation \eqref{eq:scalar-product-differential-forms}) is symmetric.

\begin{equation}
\begin{split}
&S_{KFT}[A^a+\epsilon \delta A^a]= S_{KFT}[A] + \epsilon \int_M Tr_{\mathfrak{g}} \left[ 2 dA^a \wedge \star_g d(\delta A^b) \tens{}T_aT_b \right. \\
& +\left. 2 \left( d \delta A^a \wedge \star_g (B^b \wedge A^c) + d A^a\wedge \star_g (B^b \wedge \delta A^c) \right) \tens{} T_a[T_b,T_c] \right. \\
& + \left. \left( (B^b\wedge \delta A^a) \wedge \star_g (B^d \wedge A^c) + (B^d\wedge \delta A^c ) \wedge \star_g ( B^b \wedge A^a )\right) \tens{} [T_b,T_a][T_d,T_c] \right] \\
&+\epsilon^2 \int_M Tr_{\mathfrak{g}} \left[ d(\delta A^a)\wedge \star_g d(\delta A^b) \tens{} T_aT_b +2B^b\wedge \delta A^c\wedge \star_g d(\delta A^a) \tens{} [T_b,T_c]T_a \right. \\ 
&+ \left. B^a\wedge \delta A^b \wedge \star_g B^c \wedge \delta A^d \tens{} [T_a,T_b][T_c,T_d]\right] \: .
\end{split}
\end{equation}
\noindent
Now, we again use the symmetry of the scalar product as stated in Def. \ref{def:scalar-product-over-diff-forms} as well as Stokes' theorem for differential $k$-forms. After applying these relations, the following terms remain

\begin{equation}
\begin{split}
&S_{KFT}[A^a+\epsilon \delta A^a]= S_{KFT}[A^a] + 2 \epsilon \int_M Tr_{\mathfrak{g}} \left[ d(\delta A^b) \wedge \star_g dA^a \tens{} T_aT_b \right. \\ &+ \left. \left( d (\delta A^a) \wedge \star_g (B^b \wedge A^c) + (B^d \wedge \delta A^c) \wedge \star_g d A^a   \right) \tens{} T_a[T_b,T_c] \right. \\ 
&+\left. (B^b \wedge \delta A^a) \wedge \star_g ( B^d \wedge A^c  )\tens{} [T_b,T_a][T_d,T_c] \right]\\
&+ O(\epsilon^2) \: .
\end{split}
\end{equation}
\noindent 
We again make use of Stokes theorem(see Def.\eqref{eq:stokes-useful-relation-at-infty}) and also make use of the anti-symmetry of the wedge-product to shift the terms $\delta A$ to the left. We obtain
\begin{equation}
\begin{split}
&S_{KFT}[A^a+\epsilon \delta A^a]= S_{KFT}[A^a] + 2 \epsilon \int_M Tr_{\mathfrak{g}} \left[ (-1)^{k+1} \delta A^b \wedge (d \star_g dA^a) \tens{} T_aT_b \right. \\ 
&+ \left. \left( (-1)^{k+1} \delta A^a \wedge (d \star_g (B^b \wedge A^c)) + (-1)^{k}\delta A^c \wedge B^b \wedge \star_g d A^a   \right) \tens{} T_a[T_b,T_c] \right. \\ 
&+ \left. (-1)^{k}  \delta A^a \wedge B^b \wedge \star_g ( B^d \wedge A^c  )  \tens{} [T_b,T_a][T_d,T_c] \right]\\
&+ O(\epsilon^2) \: .
\end{split}
\end{equation}
\noindent 
Now, we can factorize the $dA^a$-term and rename a few indices, if necessary
\begin{equation}
\begin{split}
&= S_{KFT}[A^a] + 2 \epsilon (-1)^{k} \int_M Tr_{\mathfrak{g}} \left[  \delta A^b \wedge \left\{ - (d \star_g dA^a) \tens{} T_aT_b \right. \right. \\ 
&-  \left. \left. d \star_g (B^a \wedge A^c) \tens{} T_b[T_a,T_c] + (B^c \wedge \star_g d A^a)    \tens{} T_b[T_c,T_a] \right. \right. \\
&+ \left. \left. ( B^a \wedge \star_g ( B^d \wedge A^c )) \tens{} [T_a,T_b][T_d,T_c] \right\} \right]\\
&+ O(\epsilon^2) \: .
\end{split}
\end{equation}
If we impose that the variation must vanish for arbitrary values of $\delta A^{a}$, it follows 
that the term in the curly brackets must be zero.
Now, we can make use of the trace identity $Tr([A,B][C,D]) = Tr([[C,D],A]B )$ for matrices $A,B,C,D\in \mathbb{C}^{n\times n}$ to rewrite the term in the curly brackets as 

\begin{equation}
\begin{split}
    \hspace*{-3cm} 0 &= Tr_{\mathfrak{g}} \left[ \left\{ d \star_g dA^b T_b + d\star_g(B^b \wedge A^c) [T_b,T_c] + B^b \wedge (\star_g dA^c)  [T_b, T_c] + B^b \wedge \star_g (B^d \wedge A^c) [T_b,[T_d,T_c]]  \right\} T_a \right] \\
    &= Tr_{\mathfrak{g}} \left[ \left\{ d \star_g dA + d \star_g [B,A] + [B, \star_g dA] + [B, \star_g [B,A]]  \right\} T_a \right] \: .
\end{split}
\end{equation}
To shorten notation, we dropped all tensor products. Now we can identify all terms corresponding to the field strength $F$.
This results in the final equations of motion (one for each open index $a$):

\begin{align}
    Tr_{\mathfrak{g}} \left( (d \star_g F + [B, \star_g F])  T_a \right) = 0 \: .
\end{align}
\noindent 
The internal structure of the equation can be made more explicit by writing out the matrix indices of the field strength $F$ and the generators $T_{a}$:
\begin{align}
 (d \star_g F_{ij} + [B, \star_g F]_{ij} )  (T_a)_{ji} = 0 \: ,
\end{align}
where $(F_{\mu_1 ... \mu_{k+1}})_{ij} \hat{=} (F_{\mu_1 ... \mu_{k+1}})^a\tens{} (T_a)_{ij}$.

\newpage

\section{Transformation Properties of the Fields and Equations of Motion}\label{sec:appendix C}

\subsubsection*{Invariance Under Local $G$ Group Transformations}

The field strength $F\in \Omega^{k+1}(M;\mathfrak{g})$ for a general $k$-form $A\in \Omega^{k}(M;\mathfrak{g})$ and a general connection $B\in \Omega_{\text{con}}(M;\mathfrak{g})$ can be defined as
 \begin{equation}
    F=dA + [B,A] \: .
\end{equation}
The following computations are equally valid when $d=d_{\omega}$ is the spacetime covariant derivative. \\
The field $A$ and the connection $B$ transform under Lie group transformations with group element $U\in G$ as
\begin{equation}
A\rightarrow A^{'}=\text{adj}(U)=UAU^{-1},\hspace{2cm} B\rightarrow B^{'}=\text{adj}(U)-(dU)U^{-1}=UBU^{-1}-(dU)U^{-1} \: .
\end{equation}
The transformed field strength can be calculated as 
\begin{equation}
\begin{split}
F^{'}&=d(A^{'})+[B^{'},A^{'}] \\
   &= d(UAU^{-1})+[UBU^{-1}-(dU)U^{-1},UAU^{-1}] \\
   &=U(dA)U^{-1}+(dU) AU^{-1}+UA(dU^{-1})+[UBU^{-1},UAU^{-1}]-[(dU)U^{-1},UAU^{-1}] \\
   &=U(dA)U^{-1}+(dU) AU^{-1}+UA(dU^{-1})+(UBU^{-1})\wedge (UAU^{-1})-(UAU^{-1})\wedge (UBU^{-1}) \\
   &-(dU)AU^{-1}+UAU^{-1} (dU)U^{-1} \\
   &=U(dA)U^{-1}+(dU)AU^{-1}+UA(dU^{-1})+U(B\wedge A)U^{-1}-U(A\wedge B)U^{-1}-(dU)AU^{-1}-UA(dU^{-1}) \\
   &=U(dA)U^{-1}+U(B\wedge A)U^{-1}-U(A\wedge B)U^{-1} \\
   &=U(dA+[B,A])U^{-1} \\
   &=UFU^{-1}\in \Omega^{k+1}(M;\mathfrak{g}) \: ,
\end{split}
\end{equation}
where we used the relation $(dU)U^{-1}=-UdU^{-1}$.

\subsubsection*{Invariance Under Local $SO(n-1,1)$ Transformations}

The total action \eqref{eq:total-action} is invariant under local $SO(n-1,1)$ transformations if the individual terms are. The Palatini--Cartan term is known to be invariant, see \cite{Montesinos:2019bkc}. We now show the invariance of the remaining terms individually. \\
Lorentz scalars are invariant under local $SO(n-1,1)$ transformations if they are constructed from a contraction of tensors. It is therefore sufficient to show that the field strength $F$ and the curvature form $H$ behave like tensors under local $SO(n-1,1)$ transformations. \\
The field $A$ and the connection $B$ transform like a $k$-form and like a $1$-form (vector) under local $SO(n-1,1)$ transformations, respectively. Note that the connection $B$ is only a connection with respect to $G$ and not $SO(n-1,1)$. Therefore, contractions and products of $A$ and $B$ will also be tonsorial with respect to $SO(n-1,1)$. The only non-trivial parts remaining in the definitions of $F$ and $H$ are the covariant derivatives $d_\omega A$ and $d_\omega B$, with respect to the spacetime connection $\omega \in \Omega_{con}(M,\mathfrak{so}(n-1,1))$. These expressions will be covariant because covariant derivatives of tensors are also tensors.

\newpage

\section{Additional Potential Forms}\label{sec:appendix D}

We define a potential form $V$ as an $n$-form, which is invariant under transformations in $G$ and $SO(n-1,1)$. There are many -- potentially infinite -- different ways to construct valid potential forms. This depends on e.g. the spacetime dimension $n$ and the degree of the $k$-forms, which are used to construct $V$. In this section we give a brief overview of valid choices and we give instructions how to construct certain classes of potential forms.

\subsubsection*{Potential Forms Involving Only the Fields $A$}

Apart from the mass term mentioned in section \ref{subsection:possible-extensions}, any term involving wedge products of $A$ with itself is a valid potential form. This is the case if a trace is taken over the whole expression and if the resulting form is an $n$-form. For example, contracting a $1$-form $A\in \Omega^1(M;\mathfrak{g})$ $n$ times with itself will be a valid term. It will have the form
\begin{align}
    V_A(A) = Tr_\mathfrak{g} \left[ \underbrace{A \wedge \cdots \wedge A}_{n \text{ times}} \right] \hspace{1cm} \text{if} \hspace{0.5cm} A\in \Omega^1(M;\mathfrak{g}) \: .
\end{align}
Another possibility is to combine terms including Hodge stars, because $(\text{any term}) \wedge \star (\text{any term})$ will always be an $n$-form, provided that the given term is at most an $n$-form. Thus, terms of the following form will always be possible:

\begin{align}
    V_A(A) = Tr_\mathfrak{g} \left[ A \wedge \star_g A \right] \hspace{1cm} \text{for all} \hspace{0.5cm} A\in \Omega^k(M;\mathfrak{g}) \: .
\end{align}
Incidentally, this is the mass term, which was mentioned in the main text. Such terms can also be nested like this:

\begin{align}
    V_A(A) = Tr_\mathfrak{g} \left[ \left(A \wedge \star_g A\right) \wedge \star_g \left(A \wedge \star_g A\right) \right] \hspace{1cm} \text{for all} \hspace{0.5cm} A\in \Omega^k(M;\mathfrak{g}) \: . \label{Appendix D: infinitely-nesting-A-field-potential-form}
\end{align}
This construction can be iterated an arbitrary number of times, allowing for an infinite amount of valid potential forms. \\
For certain spacetime dimensions $n$, terms that are quartic in the field $A$ can also be constructed like this:
\begin{align}
    V_A(A) = Tr_\mathfrak{g} \left[ \left(A \wedge A\right) \wedge \star_g \left(A \wedge A\right) \right] \hspace{1cm} \text{for} \hspace{0.5cm} A\in \Omega^k(M;\mathfrak{g}) \: , \: k \leq n/2 \: .
\end{align}
Terms of this form with the wedge product of three or more fields are also possible, depending on the degree of the field $A$. Interestingly, for $0$-forms, an arbitrary amount of wedge products can be taken and crafted into an n-form using a Hodge star:

\begin{align}
    V_A(A) = Tr_\mathfrak{g} \left[ \left( \underbrace{A \wedge \cdots \wedge A}_{\text{arbitrary amount}} \right) \wedge \star_g \left( \underbrace{A \wedge \cdots \wedge A}_{\text{arbitrary amount}} \right) \right] \hspace{1cm} \text{if} \hspace{0.5cm} A\in \Omega^0(M;\mathfrak{g}) \: .
\end{align}
The above examples should be enough to enable the interested reader to construct further terms as needed.

\subsubsection*{Potential Forms Involving Only the Connection $B$}

The connection $B$ does not transform like an algebra element under transformations in $G$. This excludes potential forms which simply are wedge products of $B$ with itself or with an algebra element. However, the corresponding field strength $H = dB + \frac{1}{2}[B,B]$ does transform like an algebra element and can easily be combined into valid terms for a potential form $V_B(B)$. A difference to the $k$-form field is that $H$ will always be a $2$-form, since $B$ is a connection. This limits the possible amount of non-trivial terms (apart from the infinitely nesting configuration discussed in \eqref{Appendix D: infinitely-nesting-A-field-potential-form}). In $n=4$ dimensions, the only valid non-trivial potential forms using only the connection $B$ are

\begin{subequations}
\begin{align}
    V_B(B) &= Tr_\mathfrak{g} \left[ H \wedge \star_g H \right] \: , \\
    V_B(B) &= Tr_\mathfrak{g} \left[ H \wedge H \right] \: .
\end{align}
\end{subequations}
The first term is the kinetic term for the connection $B$ and the second term is a topological term. The kinetic term is already included in the full KFT action. The topological term is optional, depending on whether one wants to study topological terms.

\subsubsection*{Potential Forms Involving the Field $A$ and the Connection $B$}

Combining the field $A$ and the connection $B$ is more involved due to the different transformation properties of $A$ and $B$. For example, terms that are simply wedge products between $A$ and $B$ will not work because they break invariance under transformations in $G$. This leaves the possibility to combine the field strengths $F$ and $H$ with the fields $A$. This is possible since all these objects individually transform like an algebra element. Apart from the terms discussed in equations \eqref{eq:potential-form-AB-1} and \eqref{eq:potential-form-AB-2}, other possible terms include:

\begin{subequations}
\begin{align}
V_{AB}(A,B) &= Tr_\mathfrak{g} \left[ (A \wedge A) \wedge \star_g H \right] \hspace{1cm} \text{for} \hspace{0.5cm} A\in \Omega^1(M;\mathfrak{g}) \: , \forall\, n\geq 2  \: , \\
V_{AB}(A,B) &= Tr_\mathfrak{g} \left[ (A \wedge A) \wedge \star_g F \right] \hspace{1cm} \text{for} \hspace{0.5cm} A\in \Omega^1(M;\mathfrak{g}) \: , \forall\, n \geq 2 \: , \\
V_{AB}(A,B) &= Tr_\mathfrak{g} \left[ (F \wedge F) \wedge \star_g H \right] \hspace{1cm} \text{for} \hspace{0.5cm} A\in \Omega^0(M;\mathfrak{g}) \: , \forall\, n\geq 2 \: , \\
V_{AB}(A,B) &= Tr_\mathfrak{g} \left[ (F \wedge F) \wedge \star_g (F \wedge F)) \right] \hspace{1cm} \text{for} \hspace{0.5cm} A\in \Omega^k(M;\mathfrak{g}) \: , k \leq n/2 \: , \\
V_{AB}(A,B) &= Tr_\mathfrak{g} \left[ (F \wedge F\wedge F\wedge F) \wedge \star_g (F \wedge F\wedge F\wedge F) \right] \hspace{1cm} \text{for} \hspace{0.5cm} A\in \Omega^0(M;\mathfrak{g}) \: , \forall\, n\geq 4 \: , \\
V_{AB}(A,B) &= Tr_\mathfrak{g} \left[ (H \wedge \star_g H ) \underbrace{\wedge A \wedge \cdots \wedge A}_{\text{arbitrary amount}}\right] \hspace{1cm} \text{for} \hspace{0.5cm} A\in \Omega^0(M;\mathfrak{g}) \: , \forall\, n \geq 2 \: .
\end{align}
\end{subequations}
This is only a small selection of the infinite amount of possible terms. In practical applications, some terms might also be excluded for different reasons. For example, one might exclude topological terms or terms above a certain order in the field or the connection.

\newpage
\printbibliography

\end{document}